\documentclass[journal]{IEEEtran}
\usepackage[cmex10]{amsmath}
\ifCLASSINFOpdf
\else
   \usepackage[dvips]{graphicx}
\fi
\usepackage{graphicx,cite}
\usepackage{color}
\usepackage[dvipsnames]{xcolor}
\usepackage{wasysym} 
\usepackage{amssymb}
\usepackage{dsfont}
\usepackage[caption=false,font=footnotesize]{subfig}
\usepackage{mathrsfs}
\usepackage{url}
\usepackage[printonlyused]{acronym}
\usepackage{datetime}
\usepackage[inline,shortlabels]{enumitem}
\usepackage{algorithm} 
\usepackage{algpseudocode} 
\usepackage[official]{eurosym}
\usepackage{tcolorbox}
\usepackage{tabu} 
\usepackage[normalem]{ulem}
\usepackage{multirow}
\usepackage{eucal} 
\usepackage[colorlinks=true, citecolor=blue,linkcolor=blue]{hyperref}
\hypersetup{
pdftitle={Positive Semidefinite Matrix Factorization},
pdfsubject={Positive Semidefinite Matrix Factorization},
pdfauthor={Dana Lahat},
pdfkeywords={Positive Semidefinite Matrix Factorization}
}
\usepackage{amsthm} 
\usepackage[capitalize,nameinlink]{cleveref}
\usepackage{balance}
\usepackage{mathtools} 
\usepackage{thmtools}

\tcbset{colframe=blue!50!black,colback=white, colbacktitle=blue!2!white, coltitle=blue!50!black,fonttitle=\bfseries}
\tcbuselibrary{breakable}
\newtcolorbox{danabox}[2][]{colbacktitle=red,coltitle=white,colframe=red,colback=red!2!white,coltext=red,title={#2},fonttitle=\bfseries,#1,breakable}
\crefname{appsec}{Appendix}{Appendices}

\crefformat{equation}{\textup{#2(#1)#3}}
\crefrangeformat{equation}{\textup{#3(#1)#4--#5(#2)#6}}
\crefmultiformat{equation}{\textup{#2(#1)#3}}{ and \textup{#2(#1)#3}}{, \textup{#2(#1)#3}}{, and \textup{#2(#1)#3}}
\crefrangemultiformat{equation}{\textup{#3(#1)#4--#5(#2)#6}}%
{ and \textup{#3(#1)#4--#5(#2)#6}}{, \textup{#3(#1)#4--#5(#2)#6}}{, and \textup{#3(#1)#4--#5(#2)#6}}

\Crefformat{equation}{#2Equation~\textup{(#1)}#3}
\Crefrangeformat{equation}{Equations~\textup{#3(#1)#4--#5(#2)#6}}
\Crefmultiformat{equation}{Equations~\textup{#2(#1)#3}}{ and \textup{#2(#1)#3}}
{, \textup{#2(#1)#3}}{, and \textup{#2(#1)#3}}
\Crefrangemultiformat{equation}{Equations~\textup{#3(#1)#4--#5(#2)#6}}%
{ and \textup{#3(#1)#4--#5(#2)#6}}{, \textup{#3(#1)#4--#5(#2)#6}}{, and \textup{#3(#1)#4--#5(#2)#6}}

\crefname{assumption}{assumption}{assumptions}
\Crefname{assumption}{Assumption}{Assumptions}
\Crefname{figure}{Fig.}{Fig.} 
\crefname{defi}{Definition}{Definitions}
\Crefname{defi}{Definition}{Definitions}
\crefname{theorem}{Theorem}{Theorems}
\Crefname{theorem}{Theorem}{Theorems}
\algrenewcommand\algorithmicrequire{\textbf{Input:}}
\algrenewcommand\algorithmicensure{\textbf{Output:}}
\newcommand{\bzero}{{\mathbf{0}}}
\newcommand{\ba}{{\mathbf{a}}}
\newcommand{\bb}{{\mathbf{b}}}
\newcommand{\bc}{{\mathbf{c}}}
\newcommand{\bD}{{\mathbf{D}}}
\newcommand{\bd}{{\mathbf{d}}}

\newcommand{\bu}{{\mathbf{u}}}
\newcommand{\bv}{{\mathbf{v}}}
\newcommand{\bx}{{\mathbf{x}}}
\newcommand{\by}{{\mathbf{y}}}
\newcommand{\bA}{{\mathbf{A}}}
\newcommand{\bB}{{\mathbf{B}}}
\newcommand{\bC}{{\mathbf{C}}}
\newcommand{\bG}{{\mathbf{G}}}
\newcommand{\bH}{{\mathbf{H}}}
\newcommand{\bI}{{\mathbf{I}}}
\newcommand{\bL}{{\mathbf{L}}}
\newcommand{\bM}{{\mathbf{M}}}
\newcommand{\bN}{{\mathbf{N}}}
\newcommand{\bP}{{\mathbf{P}}}
\newcommand{\bQ}{{\mathbf{Q}}}
\newcommand{\bU}{{\mathbf{U}}}
\newcommand{\bV}{{\mathbf{V}}}
\newcommand{\bW}{{\mathbf{W}}}
\newcommand{\bX}{{\mathbf{X}}}
\newcommand{\bY}{{\mathbf{Y}}}
\newcommand{\balpha}{{\boldsymbol{\alpha}}}
\newcommand{\bbeta}{{\boldsymbol{\beta}}}
\newcommand{\bLambda}{{\boldsymbol{\Lambda}}}

\newcommand{\indrange}[2]{#1=1,\ldots,#2}
\newcommand{\vect}[1]{{\mathrm{vec}}\{{#1}\}}
\newcommand{\vectt}[1]{{\mathrm{vec}}\transpose\{{#1}\}}

\newcommand{\trace}[1]{{\mathrm{tr}}\{{#1}\}}

\newcommand{\Hardpsd}[2]{{\mathrm{H}}_{\mathbb{S}_+^K,#1}(#2)}
\newcommand{\Her}{^\mathsf{H}}  
\newcommand{\adj}{^\dagger}

\newcommand{\inv}{^{-1}}

\newcommand{\transpose}{^\mathsf{T}}
\newcommand{\itranspose}{^{-\top}}
\newcommand{\had}{\ast}    
\newcommand{\defn}{\triangleq}
\DeclareMathOperator{\rank}{rank}
\DeclareMathOperator*{\argmin}{arg\,min}

\newcommand\etal{et~al.}
\newcommand{\TolFun}{\mathtt{TolFun}}

\newcommand{\TolRMFE}{\mathtt{TolRMFE}}
\newcommand{\new}[1]{{\color{black} #1}}
\acrodef{ABG}{alternating block gradient descent}
\acrodef{ADMM}{alternating direction method of multipliers}
\acrodef{ARM}{affine rank minimization}
\acrodef{ARMP}{affine rank minimization problem}
\acrodef{CC}{computational complexity}
\acrodef{CD}{coordinate descent}
\acrodef{CG}{conjugate gradient}
\acrodef{CGIHT}{conjugate gradient iterative hard thresholding}
\acrodef{d.o.f.}{degrees of freedom}
\acrodef{EDM}{Euclidean distance matrix}
\acrodef{EVD}{eigenvalue decomposition}
\acrodef{FISTA}{fast iterative shrinkage-thresholding algorithm}
\acrodef{FPGM}{fast projected gradient method}
\acrodef{FSVP}{fast singular value projection}
\acrodef{GKLD}{generalized Kullback-Leibler divergence}
\acrodef{GS}{Gauss-Southwell}
\acrodef{IHT}{iterative hard thresholding}
\acrodef{KLD}{Kullback-Leibler divergence}
\acrodef{L-BFGS}{limited-memory Broyden–Fletcher–Goldfarb–Shanno}
\acrodef{LHS}{left-hand side}
\acrodef{LRMR}{low-rank matrix recovery}
\acrodef{MC}{Monte Carlo}
\acrodef{ML}{maximum likelihood}
\acrodef{MLE}{maximum likelihood estimator}
\acrodef{MM}{majorization-minimization}
\acrodef{NIHT}{normalized iterative hard thresholding}
\acrodef{NMF}{nonnegative matrix factorization}
\acrodef{p.d.f.}{probability density function}
\acrodef{pdf}{probability density function}
\acrodef{PGM}{projected gradient method}
\acrodef{PMF}{probability mass function}
\acrodef{POVM}{positive operator valued measure}
\acrodef{psd}{positive semidefinite}
\acrodef{PSDMF}{positive semidefinite matrix factorization}
\acrodef{PSM}{Planck Sky Model}
\acrodef{PR}{phase retrieval}
\acrodef{QMFE}{quadratic model fit error}
\acrodef{RMFE}{relative model fit error}
\acrodef{SDP}{semidefinite programming}
\acrodef{SM}{supplemental material}
\acrodef{SVD}{singular value decomposition}
\acrodef{SVP}{singular value projection}
\acrodef{RV}{random variable}
\acrodef{TWF}{truncated Wirtinger flow}
\acrodef{WF}{Wirtinger flow}
\acrodef{w.r.t.}{with respect to}
\acrodef{w.l.o.g.}{without loss of generality}
\acrodef{W.l.o.g.}{Without loss of generality}

\newcommand{\gpath}{figures/}

\begin{document}

\title{Positive Semidefinite Matrix Factorization:\\ A Connection with Phase Retrieval\\ and Affine Rank Minimization}

\author{Dana~Lahat, \IEEEmembership{Member, IEEE}, Yanbin Lang, \IEEEmembership{Student Member, IEEE}, Vincent Y. F. Tan, \IEEEmembership{Senior Member, IEEE}, C\'{e}dric F\'{e}votte, \IEEEmembership{Senior Member, IEEE}%
\thanks{D.~Lahat is with the School of Electrical Engineering, Tel Aviv University, 69978 Tel Aviv, Israel, and with the Department of CSEE, University of Maryland, Baltimore County, Baltimore, MD 21250, USA. Most of D.~Lahat's work was carried out when she was with IRIT, Universit\'{e} de Toulouse, CNRS, Toulouse, France (email: Dana@Lahat.org.il).
C. F\'{e}votte is with IRIT, Universit\'{e} de Toulouse, CNRS, Toulouse, France (email: Cedric.Fevotte@irit.fr). Y. Lang and V. Y. F. Tan are with Department of Electrical and Computer Engineering, National University of Singapore, Singapore 119077 (emails: e0004795@u.nus.edu, vtan@nus.edu.sg).}%
\thanks{The work of D.~Lahat and C. F\'{e}votte has received funding from the European Research Council (ERC) under the European Union's Horizon 2020 research and innovation programme under grant agreement No.~681839 (project FACTORY). V.~Y.~F.~Tan is supported by a Singapore National Research Foundation (NRF) Fellowship (R-263-000-D02-281) and a Singapore Ministry of Education Tier 2 grant (R-263-000-C83-112).} %
}

%

\maketitle


\begin{abstract}
Positive semidefinite matrix factorization (PSDMF) expresses each entry of a nonnegative matrix as the inner product of two positive semidefinite (psd) matrices. When all these psd matrices are constrained to be diagonal, this model is equivalent to nonnegative matrix factorization. Applications include combinatorial optimization, quantum-based statistical models, and recommender systems, among others. However, despite the increasing interest in PSDMF, only a few PSDMF algorithms were proposed in the literature. 
In this work, we provide a collection of tools for PSDMF, by showing that PSDMF algorithms can be designed based on phase retrieval (PR) and affine rank minimization (ARM) algorithms. This procedure allows a shortcut in designing new PSDMF algorithms, as it allows to leverage some of the useful numerical properties of existing PR and ARM methods to the PSDMF framework. Motivated by this idea, we introduce a new family of PSDMF algorithms based on iterative hard thresholding (IHT). This family subsumes previously-proposed projected gradient PSDMF methods.
We show that there is high variability among PSDMF optimization problems that makes it beneficial to try a number of methods based on different principles to tackle difficult problems. In certain cases, our proposed methods are the only algorithms able to find a solution. In certain other cases, they converge faster.
Our results support our claim that the PSDMF framework can inherit desired numerical properties from PR and ARM algorithms, leading to more efficient PSDMF algorithms, and motivate further study of the links between these models.
  
\end{abstract}

\sloppy
%
%
\bstctlcite{IEEEexample:BSTcontrol} 
%


%
\IEEEpeerreviewmaketitle

\acresetall
\section{Introduction}
\label{sec:introduction}
\IEEEPARstart{M}{atrix} factorization is a basic tool in numerous fields such as machine learning, engineering, and optimization. 
In this paper, we address \ac{PSDMF}~\cite{Gouveia2013_Lifts,Fiorini2012_LinearVsSemidefinite}, 
a recently-proposed type of factorization of nonnegative matrices. 
\ac{PSDMF}
expresses the $(i,j){\textrm{th}}$ entry $x_{ij}$ of a nonnegative matrix $\bX\in\mathbb{R}^{I\times J}$ as an inner product of two $K\times K$ symmetric \ac{psd} matrices $\bA_i$ and $\bB_j$, indexed by $\indrange{i}{I}$, $\indrange{j}{J}$:
\begin{align}
\label{eq:psdmf_x_ij=inner_prod_A^i_B^j}
x_{ij}{}\cong{} \langle\bA_i,\bB_j\rangle
 {}={} \trace{\bA_i\bB_j}
\end{align}
where $\trace{\cdot}$ denotes the trace of a matrix, $\langle \bM,\bN\rangle{}={}\trace{\bM\transpose\bN}$ is the inner product between any two real-valued matrices $\bM$ and $\bN$ with compatible dimensions, and $\cong$ stands for equality or approximation, depending on the context. 
In \ac{PSDMF} literature, the minimal number $K$ such that a nonnegative matrix $\bX$ admits an exact \ac{PSDMF} is called the \emph{\ac{psd} rank} of $\bX$~\cite{Gouveia2013_Lifts}.
Each \ac{psd} matrix $\bA_i$ and $\bB_j$ may have a different rank, denoted as $R_{A_i}$ and $R_{B_j}$, respectively.
We shall sometimes refer to $R_{A_i}$ and $R_{B_j}$ as \emph{inner ranks}~\cite{Vandaele2018}. 
Unlike the \ac{psd} rank, the values of the inner ranks are not guaranteed to be unique, in general; see, e.g.,~\cite{Fawzi2015_PSD_Rank,Vandaele2018}.

When $\bA_i$ and $\bB_j$ are constrained to be diagonal matrices for all $\indrange{i}{I}$ and $\indrange{j}{J}$, the resulting model is equivalent to \ac{NMF} (e.g.,~\cite{Thomas1974_Nonnegative,Cohen1993,Paatero1994,Donoho2004_NMF}).
In \ac{NMF}, a nonnegative matrix $\bX\in\mathbb{R}^{I\times J}$ is modeled as a product of two nonnegative matrices: $\bX\cong\bW\bH\transpose$, where $\bW\in\mathbb{R}^{I\times K_\textrm{NMF}}$, $\bH\in\mathbb{R}^{J \times K_\textrm{NMF}}$, and $K_\textrm{NMF}\leq \min(I,J)$. 
In this context, the matrices $\bW$ and $\bH$ are sometimes referred to as \emph{factors}. 
In \ac{NMF}, the rows of each factor matrix $\bW$ and $\bH$ are in the nonnegative orthant $\mathbb{R}_+^{K_\textrm{NMF}}$, which is a closed convex cone. 
It is thus possible to express \ac{NMF} using a \ac{PSDMF} model by putting the $i{\textrm{th}}$ row of $\bW$ as the diagonal of $\bA_i$, the $j{\textrm{th}}$ row of $\bH$ as the diagonal of $\bB_j$, and setting all other entries of $\bA_i,\bB_j$ to zero; in this case, $K=K_{\textrm{NMF}}$. The converse, however, does not hold in general, because the off-diagonal entries of $\bA_i$ and $\bB_j$ may take negative values. The relationship between \ac{PSDMF}, \ac{NMF}, and other matrix decompositions is further discussed in~\cref{sec:background}.

\subsection{Motivation}
\ac{PSDMF} was proposed as an extension of a well-known result~\cite{Yannakakis1991} that links \ac{NMF} with geometry and with linear constraints in linear programming. Yannakakis'~\cite{Yannakakis1991} result is fundamental in combinatorial optimization (e.g.,~\cite{Kaibel2011}), where problems can often be written as linear programs with constraints associated with the facets of a polytope. Yannakakis' result implies that if the nonnegative rank of a slack matrix of the polytope associated with the linear constraints of the optimization problem is sufficiently small, one may find a simpler representation of the problem, with \new{fewer} constraints, in a higher dimension, and thus reduce the overall complexity of the problem.
With \ac{PSDMF}, this result extends to \acl{SDP}~\cite{Gouveia2013_Lifts}, where now the \ac{psd} rank is associated with the number of constraints in the optimization.
Applications involving \ac{PSDMF} include combinatorial optimization~\cite{Gouveia2013_Lifts,Fiorini2012_LinearVsSemidefinite,Fawzi2015_PSD_Rank}, 
quantum computing (e.g.,~\cite{vanApeldoorn2020_QuantumSDP}), quantum information theory and quantum communications~\cite{Fiorini2012_LinearVsSemidefinite,Jain2013_Quantum,Fawzi2015_PSD_Rank}, probabilistic modeling~\cite{Glasser2019_ExpressiveTN_NIPS}, and quantum-based models for recommender systems~\cite{Stark2016_RecommenderQuantum}. The relation to the quantum framework is due to the fact that quantum measurements, known as \ac{POVM}s, are represented by a set of \ac{psd} matrices whose sum is the identity matrix. 
Recently, it has been shown that \ac{PSDMF} is a special case of a more general framework of tensor networks~\cite{Glasser2019_ExpressiveTN_NIPS}.
However, despite this broad range of timely applications, a surprisingly small number of \ac{PSDMF} algorithms has been proposed in the literature, namely those in~\cite{Vandaele2018,Stark2016_RecommenderQuantum,Glasser2019_ExpressiveTN_NIPS}.

\subsection{Main Contributions}
The three main contributions of this paper are as follows. 

\begin{itemize}[wide]
\item We develop a large class of algorithms for \ac{PSDMF} by relating the problem of \ac{PSDMF} optimization to two important problems in the recent signal processing literature---\ac{ARM} (e.g.,~\cite{Fazel2002_PhD,Meka2008_Rank,Recht2010_FazelParrilo,Zheng2015_ConvergentGradient,Tu2016,Jain2010_SVP,Goldfarb2011_Convergence,Tanner2013_NIHT}) and \acl{PR} (e.g.,~\cite{Candes2013_Phaselift,Candes2015_SIRev,Candes2015_WirtingerFlow,Qiu2016_PRIME,Chen2015_TruncatedWF,Chandra2017_PhasePack}). In particular, we show that in alternating algorithms, which are the most common framework used to address \ac{PSDMF} optimization (e.g., in~\cite{Stark2016_RecommenderQuantum,Glasser2019_ExpressiveTN_NIPS,Vandaele2018,Basu2016_PSD_Factorizations}), each subproblem therein consists in approximately minimizing an objective function that is also used in \ac{ARM} or \acl{PR}. 

\item Based on this observation, we introduce a new family of \ac{PSDMF} algorithms. These algorithms are based on \emph{\ac{SVP}}~\cite{Jain2010_SVP}, sometimes referred to as \emph{\ac{IHT}}~\cite{Blumensath2009_IHT}. Our proposed \ac{SVP}-based \ac{PSDMF} algorithms subsume the \ac{PGM}~\cite{Vandaele2018} by allowing the use of inner ranks smaller than $K$. We also show that \ac{SVP} subsumes PRIME-Power~\cite{Qiu2016_PRIME}, which is a \acl{MM}-based \acl{PR} method. We further propose three variants to our basic \ac{SVP}-based \ac{PSDMF}---the first is \ac{FSVP}, which is based on Nesterov's accelerated gradient descent~\cite{Nesterov1983_AcceleratedGradient} and subsumes the \ac{FPGM}~\cite{Vandaele2018}. Our second variant is based on \ac{NIHT}~\cite{Tanner2013_NIHT,Blumensath2010_NormalizedIHT}. \ac{NIHT} was proposed as a computationally efficient version of \ac{SVP} with a specially-designed step size. The third variant is based on \ac{CGIHT}~\cite{Blanchard2015_CGIHT}, a variant of \ac{NIHT} designed to have fast asymptotic convergence rate. \ac{FPGM}, \ac{FSVP}, and \ac{CGIHT}-based \ac{PSDMF} require an additional parameter that determines the number of inner iterations within each subproblem\new{;} the user has to fine-tune \new{this parameter} to achieve sufficient acceleration.
Together with the two \ac{ABG}~\cite{Lahat2020_ICASSP_PSDMF,Lahat2020_EUSIPCO_TWF} algorithms, recently proposed by two of the authors of this paper, we provide a collection of tools for \ac{PSDMF} optimization. These methods are based on different principles and thus can assist in addressing a variety of \ac{PSDMF} problems.

\item Finally, we carry out an extensive set of numerical experiments on random and geometric datasets to compare and contrast the proposed methods with the state-of-the-art \ac{CD}~\cite{Vandaele2018}. We show that there exist cases in which \ac{NIHT} and \ac{CGIHT} dramatically outperform all other methods, being the \emph{only} algorithms able to converge towards a solution with random initialization. We also show that our proposed projection-based methods generally have a smaller per-iteration \acl{CC} than \ac{CD}\new{; t}his \new{trend} is in agreement with \new{our observation that our proposed projection-based methods} generally \new{need a} smaller CPU time to reach a target model fit error than \ac{CD}. 
We show that \ac{FSVP} and \ac{NIHT} generally achieve a desired model fit error with fewer iterations and faster than \ac{SVP}, as predicted by theory.
An advantage of \ac{NIHT} over \ac{FSVP} and \ac{CGIHT} is that it does not require an extra parameter to control the number of acceleration steps, as is the case with \ac{FSVP}  and \ac{CGIHT}. 
We exhibit some other test cases in which our \ac{ABG} algorithms succeed in decomposing the matrix faster than the competing algorithms.
Our results can serve as guidelines as to which methods might be preferred in different scenarios.
\end{itemize}

The main message of this paper is that with careful implementation, the \ac{PSDMF} framework can inherit desirable numerical properties from the multitude of \acl{PR} and \ac{ARM} methods. This then allows for the design and analysis of a host of efficient algorithms for \ac{PSDMF} starting from more basic signal processing primitives.

\subsection{Related Work}
\label{sec:related_work}
The first algorithms for \ac{PSDMF} were developed independently in~\cite{Glasser2019_ExpressiveTN_NIPS,Stark2016_RecommenderQuantum,Vandaele2018}. 
Stark's~\cite{Stark2016_RecommenderQuantum} work is motivated by the predictive power of quantum-inspired recommender systems. Thus, his algorithm uses \ac{psd} matrices normalized similarly to \ac{POVM}s. Stark~\cite{Stark2016_RecommenderQuantum} uses a standard \acl{SDP} solver to enforce the \ac{psd} constraint and to minimize a quadratic objective function, in an alternating optimization approach. The framework in~\cite{Stark2016_RecommenderQuantum} does not take into account the values of the inner ranks.

Glasser~\etal~\cite{Glasser2019_ExpressiveTN_NIPS} show that \ac{PSDMF} is a special case of a more general framework of tensor networks, in which the nonnegative matrix (or tensor) has a probabilistic interpretation. Their algorithm is based on \acl{ML} estimation of the tensor network parameters\new{,} and is implemented using a non-linear \ac{L-BFGS} algorithm. Due to their tensor network framework, the value of the inner ranks in their algorithm can be smaller than $K$ but must be the same for all \ac{psd} matrices.
The focus in~\cite{Stark2016_RecommenderQuantum,Glasser2019_ExpressiveTN_NIPS} is on demonstrating the applicability of the algorithms to the tasks of recommendation and expressivity of certain probabilistic models, respectively, and not on numerical properties of the algorithms.

Vandaele~\etal's~\cite{Vandaele2018} work is the closest to ours in the sense that they design general-purpose \ac{PSDMF} algorithms that are not tailored to a specific application, and study their numerical properties on exact \ac{PSDMF} tasks of matrices with geometric interpretation. Vandaele~\etal~\cite{Vandaele2018} propose two families of alternating \ac{PSDMF} algorithms that minimize a quadratic objective function: \ac{CD} and \ac{PGM}. \ac{PGM} is limited to the case where all inner ranks are equal to $K$, whereas \ac{CD} can handle any values of inner ranks---as is the case with the algorithms developed in this paper. Our \ac{SVP}- (resp., \mbox{\ac{FSVP}-)} based algorithm is a direct generalization of \ac{PGM} (resp., \ac{FPGM}) by allowing the inner ranks to take any value.

In these works~\cite{Glasser2019_ExpressiveTN_NIPS,Stark2016_RecommenderQuantum,Vandaele2018}, \ac{PSDMF} has not yet been connected with \acl{PR} or \ac{ARM}; this idea was first described by two of the authors of this paper in~\cite{Lahat2020_ICASSP_PSDMF}.
Based on the connection with \acl{PR} and \ac{ARM}, two of the authors of this paper have recently proposed a new family of efficient \ac{PSDMF} algorithms, based on \ac{ABG}~\cite{Lahat2020_ICASSP_PSDMF,Lahat2020_EUSIPCO_TWF}. 
The \ac{ABG} algorithm in~\cite{Lahat2020_ICASSP_PSDMF} is based on \new{\acl{WF}}~\cite{Candes2015_WirtingerFlow}, which is a gradient descent approach for \acl{PR}.
The \ac{ABG} variant in~\cite{Lahat2020_EUSIPCO_TWF} differs from \ac{CD}~\cite{Vandaele2018}, \new{from} \ac{ABG}~\cite{Lahat2020_ICASSP_PSDMF}\new{,} and from the other methods developed in this paper, in that it minimizes an objective function based on the \acl{GKLD}. The \acl{GKLD} is associated with the Poisson log-likelihood. However, as pointed out  in~\cite{Chen2017_TruncatedWF}, replacing a quadratic objective function with the \acl{GKLD} can significantly accelerate the convergence of \new{\acl{WF}} even in the absence of noise. Our preliminary results in~\cite{Lahat2020_EUSIPCO_TWF} indicate that this property can be inherited by \ac{PSDMF}, in certain cases. 
Compared with the conference papers~\cite{Lahat2020_ICASSP_PSDMF,Lahat2020_EUSIPCO_TWF}, this paper provides a more detailed discussion of the connection of \ac{PSDMF} with \acl{PR} and \ac{ARM}, as well as more extensive comparisons of the two recently-proposed variants of \ac{ABG}, not only with \ac{CD}~\cite{Vandaele2018}, but also, for the first time, with the \ac{IHT}-based methods that we introduce in this paper.
Further details about \ac{PSDMF} optimization, in the context of the methods developed in this paper, can be found in~\cref{sec:link_phase_retrieval}.

\subsection{Notations}
\label{sec:notations}
We use font types $a$ and $A$ to denote scalars. Column vectors and matrices are denoted with $\ba$ and $\bA$, respectively. Unless otherwise specified, $\ba_i$ is the $i{\textrm{th}}$ column of $\bA$, $a_{ij}$ is the $(i,j){\textrm{th}}$ entry of $\bA$, and $a_i$ is the $i{\textrm{th}}$ entry of $\ba$. $\bI$ denotes the identity matrix.
The operator $\vect{\cdot}$ reshapes a matrix into a column vector, and $\vectt{\cdot}$ denotes the transpose of $\vect{\cdot}$.
The linear map $\mathcal{A}:\mathbb{R}^{K_1\times K_2}\mapsto\mathbb{R}^I$ is determined by $I$ matrices $\bA_1,\ldots,\bA_I\in\mathbb{R}^{K_1\times K_2}$ and is given by \begin{align}
\label{eq:linear_map_def}
\mathcal{A}(\bM){}={}& \begin{bmatrix}
\langle \bA_1,\bM\rangle &
\cdots &
\langle \bA_I,\bM\rangle
\end{bmatrix}\transpose\in\mathbb{R}^{I}\,.
\end{align}
Let $\mathcal{A}\adj:\mathbb{R}^I\mapsto\mathbb{R}^{K_1\times K_2}$ denote the adjoint of $\mathcal{A}$; then, for any $\by\in\mathbb{R}^I$, we have (e.g.,~\cite{Fawzi2015_PSD_Rank})
$\mathcal{A}\adj(\by){}={}\sum_{i=1}^I y_i\bA_i$.
In this paper, we focus on \ac{PSDMF} over the real numbers $\mathbb{R}$. We do so for the sake of simplicity, but also because real-valued \ac{PSDMF} is used in most applications (e.g.,~\cite{Gouveia2013_Lifts,Fiorini2012_LinearVsSemidefinite}). However, \ac{PSDMF} over other fields has been considered as well, e.g.~\cite{Fawzi2015_PSD_Rank}. The algorithms proposed in this paper, as well as those in~\cite{Lahat2020_EUSIPCO_TWF,Lahat2020_ICASSP_PSDMF} and (F)\ac{PGM}~\cite{Vandaele2018}, work equally well over the complex numbers $\mathbb{C}$: one only has to change the transpose operation to Hermitian in the appropriate locations.
For the same reasons, and in order to simplify the transitions between the \acl{PR} and \ac{ARM} framework and \ac{PSDMF}, we shall use real-valued notations (transpose) also for the \acl{PR} and \ac{ARM} equations.

\subsection{Outline}
In~\cref{sec:background}, we provide theoretical background material about \ac{PSDMF}.
In~\cref{sec:link_phase_retrieval}, we explain how \ac{PSDMF} is related to \acl{PR} and \ac{ARM}, and how this link can be used to design new \ac{PSDMF} algorithms. Based on this link, we present in~\cref{sec:projection_based_methods} three new projection-based alternating algorithms for \ac{PSDMF}. \Cref{sec:numerical_experiments} is dedicated to numerical issues and comparisons with state of the art. In~\cref{sec:discussion}, we discuss the impact of our results from a broader perspective.

\section{Background}
\label{sec:background}
In this section, we provide theoretical preliminaries that are necessary for the exposition of our results in the next sections of this paper.
In~\cref{sec:psdmf_rank_decomposition_psd}\new{,} we \new{explain how \ac{PSDMF} is related to usual matrix factorization. \Cref{,sec:psdmf_rank_decomposition_factors} presents a factor-based representation}. In~\cref{sec:psdmf_dof}, we count the number of free variables in real-valued \ac{PSDMF}. In~\cref{sec:psdmf_zeros}, we discuss implications of the presence of zero values in the input matrix $\bX$ on the factorization.

\subsection{PSDMF as a Structured Matrix Factorization}
\label{sec:psdmf_rank_decomposition_psd}
In order to better see the link to usual matrix factorizations, let us vectorize the \ac{psd} matrices and rearrange them as follows:
\begin{subequations}
\label{eq:psdmf_A_B_defn}
\begin{align}
\mathfrak{A}{}\defn{}& \begin{bmatrix}
\vect{\bA_1} & \cdots & \vect{\bA_I}
\end{bmatrix} \; \in\mathbb{R}^{K^2\times I}\\
\mathfrak{B}{}\defn{}& \begin{bmatrix}
\vect{\bB_1} & \cdots & \vect{\bB_J}
\end{bmatrix}\; \in\mathbb{R}^{K^2\times J}\,.
\end{align}
\end{subequations}
Matrices $\mathfrak{A}$ and $\mathfrak{B}$ are \emph{structured} because their columns, upon rearrangement as $K\times K$ matrices, are in $\mathbb{S}_+^K$, the closed convex cone of $K\times K$ \ac{psd} matrices.
With this notation, \ac{PSDMF} can now be expressed as a structured matrix factorization~\cite{Fawzi2015_PSD_Rank,Gouveia2015_ApproximateCone,Basu2016_PSD_Factorizations,Vandaele2018}:
\begin{align}
\label{eq:psdmf_X=At_B}
\bX{}\cong{}& \mathfrak{A}\transpose\mathfrak{B}\,.
\end{align}
The symmetry of the \ac{psd} matrices implies that $\rank(\mathfrak{A})\leq \min(\frac{K(K+1)}{2},I)$ and $\rank(\mathfrak{B})\leq \min(\frac{K(K+1)}{2},J)$. Hence, $\rank(\bX)\leq\min(\frac{K(K+1)}{2},I,J)$~\cite{Fawzi2015_PSD_Rank}.
In~\cref{app:psdmf_sum_rank_1_terms_psd} of the \acl{SM}, we demonstrate how the \ac{PSDMF} model, in terms of \ac{psd} matrices, can be written as a sum of rank-1 terms.

\subsection{A Factor-Based Representation}
\label{sec:psdmf_rank_decomposition_factors}
In \ac{PSDMF}, the \ac{psd} matrices can be written as (e.g.,~\cite{Fawzi2015_PSD_Rank})
\begin{align}
\label{eq:psdmf_factors_def}
\bA_i\defn\bU_i\bU_i\transpose\quad\textrm{and}\quad  \bB_j\defn\bV_j\bV_j\transpose\,,
\end{align} 
where $\bU_i\in\mathbb{R}^{K\times R_{A_i}}$ and $\bV_j\in\mathbb{R}^{K\times R_{B_j}}$ are referred to as \emph{factor} matrices (factors, for short). This formulation requires knowing (or guessing) the inner ranks $R_{A_i}$ and $R_{B_j}$ in advance.
The change of variables in~\cref{eq:psdmf_factors_def} implies that
\begin{align}
\label{eq:psdmf_x_ij=inner_prod_Ui_Vj_squared}
\langle\bA_i,\bB_j\rangle{}={}&\trace{\bU_{i}\bU_i\transpose  \bV_j  \bV_j \transpose} {}={} \|\bU_{i}\transpose \bV_j \|_F^2\,.
\end{align}
A change of variables as in~\cref{eq:psdmf_factors_def} is common in \acl{SDP} (e.g.,~\cite{Burer2003}), especially when the \ac{psd} matrices have low rank. Indeed, in \ac{PSDMF} applications, the \ac{psd} matrices are often of low rank, i.e., $R_{A_i},R_{B_j}<K$ for some $i$ and/or $j$ (e.g.,~\cite{Fawzi2015_PSD_Rank}; see also~\cref{sec:psdmf_zeros}).
In~\cref{app:psdmf_sum_rank_1_terms_factors} of the \acl{SM}, we demonstrate how the \ac{PSDMF} model, in the factor-based representation, can be written as a sum of rank-1 terms.

\subsection{Degrees of Freedom}
\label{sec:psdmf_dof}
We now use the factor-based formulation to calculate the effective number of \ac{d.o.f.}~in a real-valued \ac{PSDMF} model. Due to symmetry, a \ac{psd} matrix $\bA_i$ of rank $R_{A_i}$ has $R_{A_i}K-R_{A_i}(R_{A_i}-1)/2$ free variables.
Due to the invariance of the trace operator to rotation of its variables, and given any arbitrary nonsingular $K\times K$ matrix $\bL$,
\begin{align}
\trace{\bA_i\bB_j}{}={}& \trace{\bL\transpose\bA_i\bL\cdot\bL\inv\bB_j\bL\itranspose}
\end{align}
which means that the matrices
\begin{align}
\label{eq:psdmf_inherent_ambiguity}
\bL\transpose\bA_1\bL,\ldots,\bL\transpose\bA_I\bL,\bL\inv\bB_1\bL\itranspose,\ldots,\bL\inv\bB_J\bL\itranspose
\end{align}
also form a \ac{PSDMF} of $\bX$~\cite{Fawzi2015_PSD_Rank}.
We thus have to subtract $K^2$ from the number of variables in all factors. 
Hence, \ac{PSDMF} with \ac{psd} rank $K$ has (at most) 
\begin{multline}
\label{eq:psdmf_dof}
N_{\textrm{model}}{}={}\sum_{i=1}^I \left(R_{A_i}K-\frac{R_{A_i}(R_{A_i}-1)}{2}\right)\\
 + \sum_{j=1}^J \left(R_{B_j}K-\frac{R_{B_j}(R_{B_j}-1)}{2}\right) -K^2
\end{multline} free variables that we have to learn \new{from} the $N_{\textrm{data}}{}={}IJ$ observations. 
In practice, this task is not always achievable, because the \ac{psd} structure may result in additional constraints, e.g., due to zeros, see~\cref{sec:psdmf_zeros}\new{, and because of the highly non-convex nature of the optimization problem}. 
Besides, if $\rank(\bX)>\frac{K(K+1)}{2}$, an exact \ac{PSDMF} does not exist even if $N_{\textrm{model}}\gg N_{\textrm{data}}$.
Thus, $N_{\textrm{model}}\geq N_{\textrm{data}}$ does not guarantee the existence of an exact factorization. \new{If an exact factorization exists, intuitively, the more degrees of freedom we have \ac{w.r.t.}~the number of constraints imposed by the input matrix, the easier it is to satisfy all the constraints.}
For this reason, finding (bounds on) the \ac{psd} rank of structured matrices is a major endeavor (e.g.,~\cite{Gouveia2013_Lifts,Fawzi2015_PSD_Rank,Gribling2019,Shitov2019_EDM,Glasser2019_ExpressiveTN_NIPS,Vandaele2018}). As demonstrated by~\cite{Vandaele2018}, \ac{PSDMF} algorithms can contribute to this effort by validating conjectures and finding new factorizations even in the absence of sufficient theory.

\subsection{\texorpdfstring{How the Presence of Zeros Affects the Inner Ranks or Why \ac{PSDMF} Differs from NMF in representing Nonnegative Data}{How the Presence of Zeros Affects the Inner Ranks}}
\label{sec:psdmf_zeros}
In exact \ac{NMF}, a zero observation $x_{ij}=0$ imposes zero values in the factors because the product of two nonnegative vectors can be zero only if each non-zero entry in one vector has a zero counterpart in the other vector. Hence, if a given matrix $\bX$ contains zeros (or values relatively close to zero), the factors $\bW$ \new{and} $\bH$ will be sparse (or close to sparse). In applications, this turns out to enhance uniqueness and thus yield interpretable factors (e.g.,~\cite{Anttila1995_NMF,Lee1999_NMF,Donoho2004_NMF}).

Consider now the factor-based formulation of \ac{PSDMF}, as in~\cref{eq:psdmf_factors_def}, with $\bu_r^{(i)}$ and $\bv_s^{(j)}$ the $r{\textrm{th}}$ and $s{\textrm{th}}$ columns of $\bU_i$ and $\bV_j$, respectively.
A zero value in the observations implies $x_{ij}{}={} \|\bU_{i}\transpose \bV_j \|_F^2{}={}0$ and imposes $\bu_r^{(i)\top}\bv_s^{(j)}=0$ for all $1\leq r\leq R_{A_i}$, $1\leq s\leq R_{B_j}$. 
That is, the subspace spanned by the $R_{A_i}$ columns of $\bU_i$ must be orthogonal to the subspace spanned by the $R_{B_j}$ columns of $\bV_j$. Since the dimension of this subspace is $K$, this can hold only if $R_{A_i}+R_{B_j}\leq K$ (e.g.,~\cite{Lee2013_Support},\cite[Proposition~1]{Fawzi2015_PSD_Rank}). We conclude that in \ac{PSDMF}, the effect of zeros in the input matrix $\bX$ is different than that in \ac{NMF} because for \ac{PSDMF}, zeros in the input do not, in general, result in zero values in the factors, but only affect the inner ranks. When there are several zeros in $\bX$, these orthogonality constraints must be satisfied simultaneously for all pairs of factors indexed by $(i,j)$ for which $x_{ij}=0$. These additional constraints imply that the balance of free model variables~in~\cref{sec:psdmf_dof} should be used with caution, as it provides only partial and limited information on the expressivity of the model. We shall validate this numerically in~\cref{sec:numerical_experiments_zeros_in_X}.

\section{PSDMF Optimization Based on a Link with Phase Retrieval and Affine Rank Minimization}
\label{sec:link_phase_retrieval}

This section presents in detail our concept of designing \ac{PSDMF} algorithms based on \acl{PR} and \ac{ARM} methods.
In~\cref{sec:psdmf_algorithms_in_general}, we describe the alternating optimization framework for \ac{PSDMF}, and review the relevant state of the art.
In~\cref{sec:psdmf_link_phase_retrieval}, we revisit this alternating optimization framework and explain its relation to \acl{PR} and \ac{ARM}.
In~\cref{sec:psdmf_caveats_wrt_PR_ARM}, we discuss the caveats of this approach.

\subsection{\texorpdfstring{Background: Alternating Optimization for \ac{PSDMF}}{Background: Alternating Optimization for PSDMF}}
\label{sec:psdmf_algorithms_in_general}
The \ac{psd} matrices can be estimated by minimizing the following quadratic objective function~\cite{Vandaele2018,Stark2016_RecommenderQuantum,Basu2016_PSD_Factorizations}:
\begin{subequations}
\label{eq:psdmf_objective_function}
\begin{align}
f={}&f(\{\bA_i\}_{i=1}^I,\{\bB_j\}_{j=1}^J){}=\frac{1}{2}\sum_{i=1}^I \sum_{j=1}^J (x_{ij} -\trace{\bA_i\bB_j})^2
\label{eq:psdmf_objective_function_trace_Ai_Bj}
\\
{}={}&f(\mathfrak{A},\mathfrak{B})
{}={}\frac{1}{2}\|\bX-\mathfrak{A}\transpose\mathfrak{B}\|_F^2\,.\label{eq:psdmf_objective_function_At_B}
\end{align}
\end{subequations}
The formulation in~\cref{eq:psdmf_objective_function_At_B} is equivalent to matrix factorization (with structural constraints), a problem known to be non-convex in general~(e.g.,~\cite{Vandaele2018}). However, when one matrix is fixed, the objective function \ac{w.r.t.}~the other matrix variable is convex. This observation motivated~\cite{Vandaele2018} to propose optimizing \ac{PSDMF} in a scheme that alternates between two subproblems, one to update $\mathfrak{A}$, the other to update $\mathfrak{B}$. Since the objective function in~\cref{eq:psdmf_objective_function} is symmetric in $\mathfrak{A}$ and $\mathfrak{B}$, one can use the same optimization procedure for the two subproblems.
This alternating scheme was proposed independently also in~\cite{Basu2016_PSD_Factorizations,Stark2016_RecommenderQuantum,Glasser2019_ExpressiveTN_NIPS}; however, it was not motivated by convexity arguments. This alternating scheme is outlined in~\cref{alg:psdmf_alternating_strategy}, based on~\cite[Algorithm~1]{Vandaele2018} and~\cite[Algorithm~1]{Stark2016_RecommenderQuantum}. 

\begin{algorithm}[H]
\caption{Alternating strategy for \ac{PSDMF}~\cite{Vandaele2018,Basu2016_PSD_Factorizations,Stark2016_RecommenderQuantum}}
\label{alg:psdmf_alternating_strategy}
\begin{algorithmic}[1]
\Require $\bX\in\mathbb{R}^{I\times J}_+$.
\Ensure $\bA_{1},\ldots,\bA_{I}$, $\bB_1,\ldots,\bB_J$. 
\State \textbf{Initialize} $\bA_{1},\ldots,\bA_{I}$, $\bB_1,\ldots,\bB_J$
\While{Convergence criterion not satisfied}\State $\{\bB_j\}_{j=1}^J\gets \mathtt{update\_psd}(\bX,\{\bA_i\}_{i=1}^I,\{\bB_j\}_{j=1}^J)$\label{alg:psdmf_alternating_strategy_alternate_B_j}
\State $\{\bA_i\}_{i=1}^I\gets \mathtt{update\_psd}(\bX\transpose,\{\bB_j\}_{j=1}^J,\{\bA_i\}_{i=1}^I)$\label{alg:psdmf_alternating_strategy_alternate_A_i}
\EndWhile
\end{algorithmic}
\end{algorithm}

As for implementing the subproblems in~\cref{alg:psdmf_alternating_strategy},
Vandaele~\etal~\cite{Vandaele2018} developed dedicated algorithms and showed that they outperform the use of general convex solvers. In this paper, we adopt the approach of~\cite{Vandaele2018}.
Stark~\cite{Stark2016_RecommenderQuantum} optimized each subproblem using an \acl{SDP} solver.
Motivated by their probabilistic framework,~\cite{Glasser2019_ExpressiveTN_NIPS} proposed minimizing a \acl{KLD} objective function using a non-linear \ac{L-BFGS} optimization algorithm.

One of the methods proposed by~\cite{Vandaele2018} consists in minimizing the objective function $f{}={}f(\mathfrak{A},\mathfrak{B})$ alternately \ac{w.r.t.}~the matrix variables $\mathfrak{A}$ and $\mathfrak{B}$ using gradient descent, where each update is followed by projecting the columns of $\mathfrak{A}$ or $\mathfrak{B}$ on $\mathbb{S}_+^K$ in order to guarantee the \ac{psd} structure.
This approach is termed \ac{PGM}~\cite{Vandaele2018}. 
This type of projection is mentioned also in~\cite{Basu2016_PSD_Factorizations} and implied in~\cite{Stark2016_RecommenderQuantum}.
Vandaele~\etal~\cite{Vandaele2018} proposed also a variant of \ac{PGM} based on Nesterov's accelerated gradient descent~\cite{Nesterov1983_AcceleratedGradient}, called \ac{FPGM}. \ac{PGM} and \ac{FPGM} implicitly assume that $R_{A_i}=R_{B_j}=K$ for all $\indrange{i}{I}$ and $\indrange{j}{J}$.
The methods we shall present in~\cref{sec:projection_based_methods} do not have this limitation, and coincide with \ac{PGM} and \ac{FPGM} when this special case holds.

Another optimization approach proposed in~\cite{Vandaele2018} is based on the factor-based representation in~\cref{sec:psdmf_rank_decomposition_factors}, where now the objective function in~\cref{eq:psdmf_objective_function_trace_Ai_Bj} is written as
\begin{align}
\label{eq:psdmf_objective_function_factors}
f{}={}& f(\{\bU_i\}_{i=1}^I,\{\bV_j\}_{j=1}^J){}={}\frac{1}{2}\sum_{i=1}^I\sum_{j=1}^J(x_{ij}-\|\bU_i\transpose\bV_j\|_F^2)\new{^2}\,.
\end{align}
The idea of~\cite{Vandaele2018} is to minimize~\cref{eq:psdmf_objective_function_factors} using a \ac{CD} method operating on the entries of the factors $\{\bU_i\}_{i=1}^I$ and $\{\bV_j\}_{j=1}^J$ alternately. 
Instead of working on each scalar entry of $\bU_i$ and $\bV_j$,
\ac{ABG}~\cite{Lahat2020_ICASSP_PSDMF} minimizes~\cref{eq:psdmf_objective_function_factors} in a gradient descent approach \ac{w.r.t.}~each factor matrix. 
In the \ac{ABG} algorithm in~\cite{Lahat2020_EUSIPCO_TWF}, the quadratic objective function in~\cref{eq:psdmf_objective_function_factors} is replaced with the \new{\acl{GKLD}}.
\ac{CD} and \ac{ABG} can handle any values of the inner ranks, as is the case with the methods proposed in this paper.


\subsection{\texorpdfstring{How is \ac{PSDMF} Related to Phase Retrieval and ARM?}{How is PSDMF Related to Phase Retrieval and Affine Rank Minimization?}}
\label{sec:psdmf_link_phase_retrieval}
Assume for a moment that we are given a system of quadratic equations, where
\begin{align}
\label{eq:phase_retrieval_y_i}
y_i{}\cong{}& |\langle \bu_i,\bv\rangle|^2{}={}
|\bu_i\transpose\bv|^2\quad,\quad \indrange{i}{I}
\end{align}
and $\bv\in\mathbb{R}^{K}$ is unknown. This is ``almost'' a system of linear equations, the difference being that the phase, or sign, of $\bv$, is not available. 
The problem of recovering a signal $\bv\in\mathbb{R}^{K}$ from phaseless measurements as in~\cref{eq:phase_retrieval_y_i} given \emph{sensing vectors} $\bu_i\in\mathbb{R}^{K}$, $\indrange{i}{I}$, is known as (generalized) \emph{\acl{PR}} (e.g.,~\cite{Candes2013_Phaselift,Candes2015_SIRev,Candes2015_WirtingerFlow,Qiu2016_PRIME,Chen2015_TruncatedWF,Chandra2017_PhasePack}).

Instead of addressing the unknown $\bv$ directly, it is possible to ``lift'' the quadratic measurements into linear measurements in the rank-one matrix $\bB{}={}\bv\bv\transpose$~\cite{Candes2015_SIRev}. In this case, we can write~\cref{eq:phase_retrieval_y_i} as
\begin{align}
\label{eq:phase_retrieval_lifted}
\!\!\!{}y_i\cong& |\langle \bu_i,\bv\rangle|^2
= \trace{\bv\transpose\bu_i\bu_i\transpose\bv}=
\trace{\bA_i\bB} =[\mathcal{A}(\bB)]_i\,,
\end{align}
where $\bA_i{}={}\bu_i\bu_i\transpose\in\mathbb{R}^{K\times K}$,
and $\mathcal{A}:\mathbb{R}^{K\times K}\mapsto\mathbb{R}^{I}$ is an affine transformation that maps matrices to vectors.

We point out that classical \acl{PR} often deals with complex-valued entities, whence the use of the term ``phase'' instead of sign. However, in this work, we are dealing with real-valued entities in \ac{PSDMF}, and thus we restrict ourselves to real-valued terminology.

\Cref{eq:phase_retrieval_y_i,eq:phase_retrieval_lifted} can be generalized from the vector case to an unknown matrix $\bV\in\mathbb{R}^{K\times R_B}$,
\begin{align}
\label{eq:lrmr_y_i}
y_i{}\cong{}&\|\bU_i\transpose\bV\|_F^2
{}={}\trace{\bV\transpose\bA_i\bV}
{}={} \trace{\bA_i\bB}
\,,
\end{align}
where $\bU_i\in\mathbb{R}^{K\times R_A}$ for all $i$.
The problem of recovering a low-rank matrix $\bB\in\mathbb{R}^{K\times K}$ of rank $R_B$ from a set of linear measurements as in~\cref{eq:lrmr_y_i}, not necessarily with a \ac{psd} constraint on $\bB$, is known as \ac{ARM}.

\ac{ARM} (e.g.,~\cite{Fazel2002_PhD,Meka2008_Rank,Recht2010_FazelParrilo,Zheng2015_ConvergentGradient,Tu2016,Jain2010_SVP,Goldfarb2011_Convergence,Tanner2013_NIHT}) can be stated as:
\begin{align}
\label{eq:lrmr_ARMP_in_terms_of_psdmf_min_B_j}
\min_\bB \;\rank(\bB)\quad\textrm{s.t.}\quad \mathcal{A}(\bB){}={}\by
\end{align}
where $\bB\in\mathbb{R}^{K_1\times K_2}$ is the unknown matrix (not necessarily \ac{psd}), $\by\in\mathbb{R}^{I}$ is the vector of observations, and  $\mathcal{A}:\mathbb{R}^{K_1\times K_2}\mapsto \mathbb{R}^{I}$ is a known linear mapping. \ac{ARM} underlies numerous problems in the signal processing literature, including \acl{LRMR}, matrix completion, and compressed sensing, to name a few (e.g.,~\cite{Fazel2008_CS,Recht2010_FazelParrilo,Candes2015_SIRev}).
\ac{ARM} is NP-hard and hard to approximate (e.g.,~\cite{Meka2008_Rank}). Hence, numerous relaxations and variants have been proposed in the literature, among which we mention relaxations to the equality $\mathcal{A}(\bB){}={}\by$ using a quadratic loss (e.g.,~\cite{Jain2010_SVP}), and relaxations to the rank constraint by optimizing over $\bV\in\mathbb{R}^{K\times R}$, where $\bB\defn\bV\bV\transpose$ (e.g.,~\cite{Candes2015_WirtingerFlow,Zheng2015_ConvergentGradient,Tu2016}).
With this in mind, we are ready to show how \acl{PR} and \ac{ARM} are related to \ac{PSDMF}.

\Acf{w.l.o.g.}, the quadratic objective function in~\cref{eq:psdmf_objective_function} can be written as a sum of $J$ terms:
\begin{align}
\label{eq:psdmf_objective_function_sum_f_j}
f{}={}&f(\{\bA_i\}_{i=1}^I,\{\bB_j\}_{j=1}^J){}={}
\sum_{j=1}^J f_j\,,
\end{align}
where
\begin{subequations}
\label{eq:psdmf_objective_function_fixed_j}
\begin{align}
f_j{}={}f_j(\{\bA_i\}_{i=1}^I,\bB_j){}={}&\frac{1}{2}\sum_{i=1}^I (x_{ij}-\trace{\bA_i\bB_j})^2\label{eq:psdmf_objective_function_trace_Ai_Bj_fixed_j}\\
{}={}& \frac{1}{2}\|\bx_j-\mathcal{A}(\bB_j)\|_2^2\label{eq:psdmf_objective_function_affine_per_j}\\
{}={}& \frac{1}{2}\|\bx_j-\mathfrak{A}\transpose\bb_j\|_2^2\label{eq:psdmf_objective_function_At_b_j_fixed_j}\,.
\end{align}
\end{subequations}
In~\cref{eq:psdmf_objective_function_fixed_j},
$\|\cdot\|_2$ is the Euclidean norm, $\bx_j\in\mathbb{R}^{I}$ is the $j{\textrm{th}}$ column vector of $\bX$, \new{and}
$\bb_j{}\defn{}\vect{\bB_j}$. 
The formulation in~\cref{eq:psdmf_objective_function_affine_per_j} is implicit in~\cite{Basu2016_PSD_Factorizations}.
It follows from~\cref{eq:psdmf_objective_function_fixed_j,eq:psdmf_objective_function_sum_f_j} that minimizing the objective function $f{}={}f(\{\bA_i\}_{i=1}^I,\{\bB_j\}_{j=1}^J)$ \ac{w.r.t.}~the \ac{psd} matrix $\bB_j$ is equivalent to minimizing the objective function $f_j{}={}f_j(\{\bA_i\}_{i=1}^I,\bB_j)$ in~\cref{eq:psdmf_objective_function_fixed_j} \ac{w.r.t.}~the same variable. This optimization problem can be written as
\begin{align}
\label{eq:psdmf_optimize_Bj_psd}
\min_{\bB_j}\quad \|\bx_j-\mathcal{A}(\bB_j)\|_2^2 \quad\textrm{s.t.}\quad \bB_j\in\mathbb{S}_+^K\,,
\end{align}
possibly also subject to the rank constraint
\begin{align}
\label{eq:psdmf_optimize_Bj_psd_w_rank_constraint}
\rank(\bB_j)\leq R_{B_j}\,.
\end{align}
For a specific value of $j$, the optimization problem in~\cref{eq:psdmf_optimize_Bj_psd} can be associated with the problem of estimating a matrix $\bB_j$ from the vector of observations $\bx_j$ given the linear operator $\mathcal{A}$ and the observation model~\cite{Basu2016_PSD_Factorizations}
\begin{align}
\label{eq:psdmf_lrmr_affine_approx_psd}
\bx_j{}\cong{}&\mathcal{A}(\bB_j)\,,
\end{align}
subject to additional structural constraints, e.g., \ac{psd} and low-rank, on the variable $\bB_j$.
The key point is that the optimization problems in~\cref{eq:psdmf_optimize_Bj_psd,eq:psdmf_optimize_Bj_psd_w_rank_constraint}, which we wrote down as subproblems in alternating \ac{PSDMF} optimization, can be optimized using existing methods in the literature that were developed for \acl{PR} and \ac{ARM}. The reverse holds as well: methods for optimizing \ac{PSDMF} subproblems in the alternating framework that we have just described may, essentially, be used for \acl{PR} and \ac{ARM}.

\Cref{alg:psdmf_subproblem_affine} outlines a subproblem in~\cref{alg:psdmf_alternating_strategy_alternate_B_j} in~\cref{alg:psdmf_alternating_strategy} when the update of the variable $\bB_j$ is carried out using the approach that we have just described. 
More specifically, in~\cref{alg:psdmf_subproblem_affine}, \cref{alg:psdmf_subproblem_affine_unvec_col_j_of_B} stands for the update of the variable $\bB_j$ using a \acl{PR} or \ac{ARM} method. Concrete examples will be given in~\cref{sec:projection_based_methods}.
The fact that each subproblem updates each $\bB_j$ independently of the others allows for parallelization of the computations within each subproblem.
\begin{algorithm}[H]
\caption{Subproblem to update $\{\bB_j\}_{j=1}^J$ given $\mathcal{A}$}
\label{alg:psdmf_subproblem_affine}
\begin{algorithmic}[1]
\Require $\bX\in\mathbb{R}^{I\times J}_+$, $\mathcal{A}$,  $\{\bB_j\}_{j=1}^J$.
\Ensure  $\{\bB_j\}_{j=1}^J$ (updated). 
\For{$j=1:J$}
\State $\bB_j\gets \mathtt{PR\_or\_ARM\_algorithm} (\mathcal{A}(\bB_j),\bx_j)$ \label{alg:psdmf_subproblem_affine_unvec_col_j_of_B}
\EndFor
\end{algorithmic}
\end{algorithm}

\subsection{\texorpdfstring{Alternating \ac{PSDMF} \new{Versus} \acl{PR} and \ac{ARM}}{How Alternating PSDMF Differs from PR and ARM}}
\label{sec:psdmf_caveats_wrt_PR_ARM}
Before moving on to specific numerical methods, we point out some fundamental differences between alternating \ac{PSDMF} and the \acl{PR} or \ac{ARM} methods they rely on that must be taken into account in the algorithm design process. 
Probably the most important difference is that in \acl{PR} and \ac{ARM}, the transformation $\mathcal{A}$ is given and known. Based on this fact, various strategies for initializing \acl{PR} and \ac{ARM} have been proposed. These initialization methods are often critical to guarantee convergence of the \acl{PR} and \ac{ARM} method to the desired solution, and play an important part in the analysis of the minimal number of observations required for reliable reconstruction of the desired signal (see, e.g.,~\cite{Candes2015_WirtingerFlow}). In the alternating \ac{PSDMF} framework, however, the affine operator $\mathcal{A}$ in~\cref{alg:psdmf_subproblem_affine} consists of \ac{psd} matrices that are unknowns themselves. Hence, initialization methods for \acl{PR} and \ac{ARM} that are based on knowing the \emph{true} $\mathcal{A}$ cannot be applied to \ac{PSDMF}. Consequently, methods that promise high convergence speed subject to appropriate initialization in the \acl{PR} and \ac{ARM} setting may perform poorly in the alternating \ac{PSDMF} framework.
For similar reasons, it is not clear to which extent convergence guarantees that were derived for \acl{PR} and \ac{ARM} are relevant to the alternating \ac{PSDMF} framework. 
Another issue that is of utmost importance in \acl{PR} and \ac{ARM} is finding bounds on the number of measurements that guarantee an exact reconstruction of the desired signal. However, this question does not apply naturally to \ac{PSDMF} because this analysis relies on the prerequisite that the \emph{true} sensing vectors or matrices are known.
Due to these differences, our interest in \acl{PR} and \ac{ARM} is restricted to borrowing algorithms that update the variables in each \ac{PSDMF} optimization step.

\section{\texorpdfstring{Projection-Based Algorithms for \ac{PSDMF}}{Projection-Based Algorithms for PSDMF}}
\label{sec:projection_based_methods}
In this section, we proceed from concept to practice.
The new \ac{PSDMF} algorithms that we introduce in this section are based on \ac{SVP}~\cite{Jain2010_SVP}, an \ac{ARM} method that we describe in~\cref{sec:psdmf_svp_background}.
Inspecting \ac{SVP} from the perspective of the link of \ac{PSDMF} to \acl{PR} and \ac{ARM} makes it natural to \new{extend} \ac{PGM}~\cite{Vandaele2018} to a framework that can handle any value of inner ranks. Our \new{\ac{SVP}-based alternating \ac{PSDMF} algorithm}, and its accelerated variant \ac{FSVP}, are described in~\cref{sec:psdmf_algorithm_based_on_SVP}.
In~\cref{sec:psdmf_niht}, we present another \ac{PSDMF} algorithm, based on \ac{NIHT}~\cite{Tanner2013_NIHT}. In~\cref{sec:psdmf_cgiht}, we describe our \ac{CGIHT}-based \ac{PSDMF} algorithm.
In~\cref{sec:psdmf_svp_link_mm}, we show the link between \ac{SVP} and a \acl{MM}-based algorithm for \acl{PR}~\cite{Qiu2016_PRIME}.

\subsection{Background: Singular Value Projection}
\label{sec:psdmf_svp_background}
Consider the objective function $f_j(\bB_j)$ in~\cref{eq:psdmf_objective_function_fixed_j} for a given $\mathcal{A}$ and a specific value of $j$. This function is convex in $\bB_j$. In this section, we discuss minimizing $f_j(\bB_j)$ \ac{w.r.t.}~$\bB_j$ subject to $\rank(\bB_j)\leq R_{B_j}$. The set of low-rank matrices is not convex, and thus this optimization problem is non-convex, in general, when $R_{B_j}<K$. 
This problem was formulated by~\cite{Jain2010_SVP} as a robust variant of the \ac{ARM} problem in~\cref{eq:lrmr_ARMP_in_terms_of_psdmf_min_B_j}. In~\cite{Jain2010_SVP}, the matrix $\bB_j$ was not constrained to be \ac{psd}. 
The first step in Jain~\etal's~\cite{Jain2010_SVP} approach involves gradient descent.
To simplify our notation, we omit the index $j$ and write  $\psi(\bB)$ instead of $f_j(\bB_j)$.
The gradient of $\psi(\bB)$ \ac{w.r.t.}~$\bB$ can take any of the following forms:
\begin{subequations}
\label{eq:lrmr_SVP_gradient}
\begin{align}
\nabla\psi(\bB){}={}& \mathcal{A}\adj(\mathcal{A}(\bB)-\by)
{}={} \sum_{i=1}^I \left[ (\mathcal{A}(\bB)-\by)\right]_i\bA_i\\
{}={}& \sum_{i=1}^I (\trace{\bA_i\transpose\bB}-y_{i})\bA_i\,,
\end{align}
\end{subequations}
where the conjugate map $\mathcal{A}\adj$ was defined in~\cref{sec:notations}.
Now, a gradient step based on~\cref{eq:lrmr_SVP_gradient} does not take into account the low-rank structure of $\bB_j$. Instead, the rank constraint is imposed by orthogonal projection of the updated version of $\bB$ onto the set of low-rank matrices, an operation that consists of taking the truncated \ac{SVD} of $\bB$. This procedure is termed \ac{SVP} in~\cite{Jain2010_SVP}.

The \ac{psd} variant of \ac{SVP} was addressed by~\cite{Tu2016}, who used a few iterations of \ac{SVP} to initialize another non-convex \ac{ARM} algorithm. In the \ac{psd} case, the projection of a matrix on the set of $K\times K$ \ac{psd} matrices of rank at most $R$ is denoted by $\Hardpsd{R}{\cdot}$. This projection can be computed as $\Hardpsd{R}{\bB} = \bV\bLambda_R\bV\transpose$, where  $\bLambda_R\in\mathbb{R}^{R\times R}$ is a diagonal nonnegative matrix with the $R$ largest nonnegative eigenvalues of $\bB$ on its main diagonal, and the columns of $\bV\in\mathbb{R}^{K\times R}$ are the eigenvectors of $\bB$ associated with these $R$ largest nonnegative eigenvalues.
If $\bB$ has less than $R$ positive eigenvalues, some of the diagonal values of $\bLambda_R$ will be zero, and the corresponding columns of $\bV$ can be zero vectors as well. 
The letter ``$\mathrm{H}$'' is reminiscent of the fact that $\Hardpsd{R}{\cdot}$ is a hard thresholding operator (see also~\cite{Blumensath2009_IHT}). Note that in the \ac{psd} case, we cannot use \ac{SVD} or any other method that extracts the eigenvectors by the magnitude of the $R$ leading eigenvalues because we must have access to the signs of the eigenvalues.

\ac{SVP} with \ac{psd} constraint on the unknown matrix $\bB$ is outlined in~\cref{alg:lrmr_SVP_psd}, based on~\cite[Algorithm~1]{Jain2010_SVP}.
The update rule based on projected gradient descent is given in~\cref{alg:lrmr_SVP_psd_update_and_project_step} of~\cref{alg:lrmr_SVP_psd}, where $\eta$ is the step size.
The zero initialization in~\cref{alg:lrmr_SVP_psd_initialization} of~\cref{alg:lrmr_SVP_psd} was proposed by~\cite{Jain2010_SVP}. Other initialization procedures, such as \emph{spectral initialization}~\cite{Candes2015_WirtingerFlow}: $\bB\gets\mathcal{A}\adj(\by)$, are possible.
Jain~\etal~\cite{Jain2010_SVP} prove that \ac{SVP} can converge to a desired low-rank solution if the step size $\eta$ is smaller than a certain bound that depends on geometric properties of $\mathcal{A}$ and on the rank of the desired solution.

\begin{algorithm}[H]
\caption{\ac{SVP} Algorithm~\cite{Jain2010_SVP}---the \ac{psd} case.}
\label{alg:lrmr_SVP_psd}
\begin{algorithmic}[1]
\Require $\mathcal{A}$, $\by$, $R$, $\eta$
\Ensure $\bB$ of $\rank(\bB)\leq R$. 
\State \textbf{Initialize:} $\bB\gets\bzero$ \label{alg:lrmr_SVP_psd_initialization}
\While{stopping criterion not satisfied}
\State $\bB \gets\Hardpsd{R}{\bB-\eta\mathcal{A}\adj(\mathcal{A}(\bB)-\by)}$\label{alg:lrmr_SVP_psd_update_and_project_step}
\EndWhile
\end{algorithmic}
\end{algorithm}

\subsection{\texorpdfstring{A \ac{PSDMF} Algorithm Based on \ac{SVP}}{A PSDMF Algorithm Based on SVP}}
\label{sec:psdmf_algorithm_based_on_SVP}
Our proposed \ac{SVP}-based \ac{PSDMF} algorithm is constructed by using the update step of \ac{SVP} to optimize the subproblems in~\cref{alg:psdmf_subproblem_affine}, along with the necessary adaptations to the alternating framework. \Cref{alg:psdmf_SVP} outlines one subproblem to update $\{\bB_j\}_{j=1}^J$, given $\mathcal{A}$ and $\{\bB_j\}_{j=1}^J$ from the previous subproblem (see~\cref{alg:psdmf_alternating_strategy}).

\begin{algorithm}[H]
\caption{One subproblem of \ac{SVP}-based \ac{PSDMF} to update $\{\bB_j\}_{j=1}^J$}
\label{alg:psdmf_SVP}
\begin{algorithmic}[1]
\Require $\bX$, $\mathcal{A}$, $\bB_1,\ldots, \bB_J$, $R_{B_1},\ldots, R_{B_J}$, $D$.
\Ensure $\bB_1,\ldots, \bB_J$. 
\State  $\eta\gets (\lambda_{\textrm{max}}(\mathfrak{A}\mathfrak{A}\transpose))\inv$ \label{alg:psdmf_SVP_choosing_step_size}
\For{$j=1:J$}
\For{$d=1:D$}
\State $\bB_j \gets\Hardpsd{R_{B_j}}{\bB_j-\eta \mathcal{A}\adj(\mathcal{A}(\bB_j)-\bx_j)}$\label{alg:psdmf_SVP_update_and_project_step}
\EndFor
\EndFor
\end{algorithmic}
\end{algorithm}
The optional parameter $D$ in~\cref{alg:psdmf_SVP} determines the number of inner iterations~\cite{Vandaele2018}.
In the special case that $R_{A_j}=R_{B_j}=K$ for all \new{$i$} and \new{$j$}, the \ac{SVP}-based \ac{PSDMF} described in~\cref{alg:psdmf_SVP} coincides with \ac{PGM}~\cite{Vandaele2018}.
In~\cite{Vandaele2018}, Vandaele~\etal~proposed a variant to \ac{PGM} in which the gradient step is replaced by $D$ steps of Nesterov-based accelerated gradient descent~\cite{Nesterov1983_AcceleratedGradient}.
It is thus natural to propose an accelerated variant of~\cref{alg:psdmf_SVP} that will subsume \ac{FPGM}~\cite{Vandaele2018}. In this case, we replace the loop on $D$ in~\cref{alg:psdmf_SVP} with:
\begin{algorithmic}[1]
\State $\bB_j^{\textrm{prev}}\gets \bB_j$
\For{$d=1:D$}
\State $\bY \gets \bB_j + \frac{d-2}{d+1}(\bB_j-\bB_j^{\textrm{prev}})$
\State $\bB_j^{\textrm{prev}}\gets \bB_j$
\State $\bB_j \gets\Hardpsd{R_{B_j}}{\bY-\eta \mathcal{A}\adj(\mathcal{A}(\bY)-\bx_j)}$
\EndFor
\end{algorithmic}
We call this variant \emph{\ac{FSVP}} (henceforth, we omit the suffix ``-\ac{PSDMF}'' when the context is clear).  \ac{FSVP} reduces to \ac{FPGM} when all inner ranks are equal to $K$. 
When $D=1$, \ac{FSVP} is equivalent to \ac{SVP}. Note that in a non-alternating framework, $D$ is simply the number of iterations of the accelerated gradient descent algorithm until a stopping criterion is achieved. Thus, in order to benefit from the acceleration, one has to choose a sufficiently large value of $D$. However, due to the alternating framework, $D$ should not be too large, otherwise the performance of \ac{FSVP} degrades again, as demonstrated by~\cite{Vandaele2018}. The optimal value of $D$ depends on many \new{factors}, including the values in $\bX$, the factorization parameters, and the stopping criterion. Hence, in practice, $D$ is chosen based on empirical evaluation~\cite{Vandaele2018}; see~\cref{sec:numerical_issues_D}.

As for the step size in~\cref{alg:psdmf_SVP_choosing_step_size} of~\cref{alg:psdmf_SVP}, in (F)\ac{PGM}~\cite{Vandaele2018} ($R=K$), the step size within each subproblem is fixed and equal to $\eta{}={}1/L$, where $L=\lambda_{\textrm{max}}(\mathfrak{A}\mathfrak{A}\transpose)$ is the Lipschitz constant of the gradient $\nabla \psi(\bB_j)$ [see~\cref{eq:lrmr_SVP_gradient}], $\mathfrak{A}\mathfrak{A}\transpose{}={} \sum_{i=1}^I \vect{\bA_i} \vectt{\bA_i}$, and $\lambda_{\textrm{max}}(\cdot)$ is the leading eigenvalue of its operand.
In numerical experiments, we observed that the objective function decreased monotonically also for \ac{SVP} and \ac{FSVP}, i.e., $R<K$, with the same step size $\eta{}={}(\lambda_{\textrm{max}}(\mathfrak{A}\mathfrak{A}\transpose))\inv$. This empirical observation is not obvious, because this $\eta$ is no longer guaranteed to be the optimal fixed step size when projecting on the non-convex set of low-rank matrices, as noted, e.g., in~\cite{Jain2010_SVP}.

\subsection{\texorpdfstring{A \ac{PSDMF} Algorithm Based on \ac{NIHT}}{A PSDMF Algorithm Based on NIHT}}
\label{sec:psdmf_niht}
Since it was first proposed, several improvements to the basic \ac{SVP} algorithm appeared, see, e.g.,~\cite{Jain2010_SVP,Blanchard2015_CGIHT,Xue2019_TurboAffineRank,Vandereycken2013_Riemannian} and references therein. It is thus natural to consider these variants as candidates for more efficient \ac{PSDMF} methods, and to see if they can offer numerical advantages also in the alternating \ac{PSDMF} framework. We now describe a \ac{PSDMF} algorithm based on \ac{NIHT}~\cite{Blumensath2010_NormalizedIHT,Tanner2013_NIHT}.
One subproblem to update $\bB_j$, $\indrange{j}{J}$,  based on \ac{NIHT}, is outlined in~\cref{alg:psdmf_NIHT}. 
Each update step of $\bB_j$ in~\cref{alg:psdmf_NIHT} is identical to an update step in the original \ac{NIHT} algorithm.
\begin{algorithm}[H]
\caption{One subproblem of \ac{NIHT}-based \ac{PSDMF} to update $\{\bB_j\}_{j=1}^J$}
\label{alg:psdmf_NIHT}
\begin{algorithmic}[1]
\Require $\bX\in\mathbb{R}^{I\times J}_+$, $\mathcal{A}$, $\bB_1,\ldots, \bB_J$, $R_{B_1},\ldots, R_{B_J}$.
\Ensure $\bB_1,\ldots, \bB_J$. 
\For{$j=1:J$}
\State  $\bU\gets$ $R_{B_j}$ eigenvectors of $\bB_j$ associated with the $R_{B_j}$

\quad\quad largest nonnegative eigenvalues of $\bB_j$ \label{alg:psdmf_NIHT_initialization}
\State $\bP_U\gets\bU{\bU}\transpose$
\State $\eta\gets \frac{\|\bP_U\mathcal{A}\adj(\bx_j-\mathcal{A}(\bB_j))\|_F^2}{\|\mathcal{A}(\bP_U\mathcal{A}\adj(\bx_j-\mathcal{A}(\bB_j)))\|_2^2}$ \label{alg:psdmf_NIHT_step_size}
\State $\bB_j\gets \Hardpsd{R_{B_j}}{\bB_j-\eta\mathcal{A}\adj(\mathcal{A}(\bB_j)-\bx_j)}$\label{alg:psdmf_NIHT_update_and_proj_step}
\EndFor
\end{algorithmic}
\end{algorithm}

Our \ac{NIHT}-based method in~\cref{alg:psdmf_NIHT} differs from \ac{SVP} (\cref{alg:psdmf_SVP}) in the evaluation of the step size $\eta$ for the gradient descent. Compared with \ac{SVP}, which has a fixed step size, the step size in \ac{NIHT} is adaptive. In each iteration, the current estimate of $\bB_j$ is updated along the gradient descent direction with the locally steepest descent stepsize, followed by thresholding to the manifold of rank-$R_{B_j}$ matrices~\cite{Wei2016_RiemannianRecovery,Tanner2013_NIHT}.
When $R_{B_j}<K$, $\eta$ in~\cref{alg:psdmf_NIHT_step_size} of~\cref{alg:psdmf_NIHT} depends not only on $\mathcal{A}$ but also on $\bB_j$, and may be different for different $\bB_j$. When $R_{B_j}=K$, the projection operator is the identity matrix: $\bP_U=\bU\bU\transpose=\bI$, and thus $\eta$ no longer depend\new{s} on $\bB_j$. However, for any choice of $\bP_U,\mathcal{A},\bx_j,$ and $\bB_j$, the step size of \ac{NIHT} is never smaller than that of (F)\ac{SVP}: $\eta^{\textrm{NIHT}} \geq \eta^{\textrm{(F)SVP}}=(\lambda_{\textrm{max}}(\mathfrak{A}\mathfrak{A}\transpose))\inv$. To see why this inequality always holds, 
let $\bM\neq \bzero$ be an arbitrary $K\times K$ matrix. Then,
\begin{align}
\label{eq:psdmf_niht_step_size_inequality}
\frac{\|\mathcal{A}(\bM)\|_2^2}{\|\bM\|_F^2} {}={} \frac{\|\mathfrak{A}\transpose\vect{\bM}\|_2^2}{\|\vect{\bM}\|_2^2} {}\leq{}\sqrt{\lambda_{\textrm{max}}(\mathfrak{A}\mathfrak{A}\transpose)}\new{\,,}
\end{align}
where the inequality follows from the definition of the spectral norm. Setting $\bM = \bP_U\mathcal{A}\adj(\bx_j-\mathcal{A}(\bB_j))$ in~\cref{eq:psdmf_niht_step_size_inequality}, the \acl{LHS} of~\cref{eq:psdmf_niht_step_size_inequality} is equal to $(\eta^{\textrm{NIHT}})\inv$. Hence, $\eta^{\textrm{NIHT}} \geq \eta^{\textrm{(F)SVP}}{}={}(\lambda_{\textrm{max}}(\mathfrak{A}\mathfrak{A}\transpose))\inv$. 
This inequality provides an intuitive explanation why our \ac{NIHT}-based \ac{PSDMF} algorithm can achieve the same model fit error with \new{fewer} iterations than our \ac{SVP}-based \ac{PSDMF} that has step size $(\lambda_{\textrm{max}}(\mathfrak{A}\mathfrak{A}\transpose))\inv$, as we shall demonstrate in~\cref{sec:numerical_experiments}. Even more interesting is our observation that
in general, our \ac{NIHT}-based method also outperforms \ac{FSVP} in terms of number of iterations required to achieve the same model fit error, as we show in~\cref{sec:numerical_experiments}. The latter fact is significant because \ac{NIHT} does not require to estimate or adjust an additional acceleration parameter $D$ as is the case with \ac{FPGM} and \ac{FSVP} in order to achieve its best performance. Another noteworthy property of \ac{NIHT} is that its step size does not guarantee a monotonous decrease of the objective function~\cite{Tanner2013_NIHT}, in contrast to \ac{ABG}~\cite{Lahat2020_EUSIPCO_TWF,Lahat2020_ICASSP_PSDMF} and \ac{CD}~\cite{Vandaele2018}. As we shall demonstrate in~\cref{sec:numerical_experiments_geometrical_submatrix_CORn}, this property may explain why in certain cases, our \ac{NIHT}-based algorithm is the only method able to properly minimize the objective function.

\subsection{\texorpdfstring{A PSDMF Algorithm based on CGIHT}{A PSDMF Algorithm based on CGIHT}}
\label{sec:psdmf_cgiht}

\ac{CGIHT}~\cite{Blanchard2015_CGIHT} was proposed as an improvement to \ac{NIHT}~\cite{Tanner2013_NIHT} by combining the fast asymptotic convergence rate of the conjugate gradient method with the low per-iteration complexity of hard thresholding methods.
One subproblem to update $\bB_j$, $\indrange{j}{J}$, based on \ac{CGIHT}~\cite{Blanchard2015_CGIHT}, is outlined in~\cref{alg:psdmf_CGIHT}.

\begin{algorithm}[H]
\caption{One subproblem of \ac{CGIHT}-based \ac{PSDMF} to update $\{\bB_j\}_{j=1}^J$}
\label{alg:psdmf_CGIHT}
\begin{algorithmic}[1]

\Require $\bX\in\mathbb{R}^{I\times J}_+$, $\mathcal{A}$, $\bB_1,\ldots, \bB_J$, $R_{B_1},\ldots, R_{B_J}$, $D$.
\Ensure $\bB_1,\ldots, \bB_J$. 
\For{$j=1:J$}
\State $\bQ=0$, $d\gets 0$, 
\For{$d=1:D$}

\State $\bU\gets$  $R_{B_j}$ eigenvectors of $\bB_j$ associated with  

\quad the  $R_{B_j}$ largest nonnegative eigenvalues of $\bB_j$,

\State $\bG\gets \mathcal{A}\adj(\by-\mathcal{A}(\bB_j))$

\State $\bP_U\gets\bU{\bU}\Her$

\If{$d=1$}
\State $\beta\gets 0$
\Else
\State $\beta\gets \frac{\langle \mathcal{A}(\bP_U\bG), \ \mathcal{A}(\bP_U\bQ) \rangle}{\|\mathcal{A} \bP_U\bQ\|_2^2}$\label{alg:psdmf_CGIHT_beta}

\EndIf

\State $\bQ\gets \bG + \beta\bQ$\label{alg:psdmf_CGIHT_search_direction}

\State $\eta \gets \frac{\langle \bP_U\bG , \bP_U\bQ \rangle}{\|\mathcal{A}(\bP_U\bQ)\|_2^2}$
\State $\bB_j\gets \Hardpsd{R}{{\bB_j-\eta \bQ}}$
\EndFor
\EndFor
\end{algorithmic}
\end{algorithm}

Each update step of $\bB_j$ in~\cref{alg:psdmf_CGIHT} is identical to an update step in the original \ac{CGIHT} algorithm. If the orthogonalization weight $\beta$ is zero, \ac{CGIHT} becomes equivalent to \ac{NIHT}. For $\beta$ to be different from zero, we need at least one inner iteration, $D\geq 2$, in~\cref{alg:psdmf_CGIHT}.
Matrix $\bQ$ in~\cref{alg:psdmf_CGIHT_search_direction} defines the search direction, and $\eta $ the step size.

Similarly to other \acl{PR} and \ac{ARM} methods, the stability and recovery guarantees of \ac{NIHT} and \ac{CGIHT} depend on satisfying conditions on the restricted isometry constants of the sensing operators. However, these conditions are generally not satisfied within the alternating \ac{PSDMF} optimization framework. As our numerical experiments show, the error evolution trajectories of \ac{CGIHT} are often irregular and erratic. Nevertheless, as we shall show in~\cref{sec:numerical_experiments}, such a behaviour can in fact turn out useful.
In order to improve the stability of our algorithm, the following rules were implemented: if $\beta$, $\eta $, or the norm of the gradient, become excessively large, we set them to zero.

\subsection{\texorpdfstring{A Link Between SVP, MM, and PRIME-Power}{A Link Between SVP, MM, and PRIME-Power}}
\label{sec:psdmf_svp_link_mm}
In~\cite[Sec.~III.D]{Qiu2016_PRIME}, Qiu~\etal~proposed an algorithm for \acl{PR}, called PRIME-Power, whose derivation is based purely on \acl{MM} considerations. 
The objective function minimized by PRIME-Power is equivalent to~\cref{eq:psdmf_objective_function_affine_per_j} for a specific value of $j$. 
We establish a new connection between PRIME-Power and gradient descent.
More specifically, we now show that PRIME-Power is equivalent to \ac{SVP}~\cite{Jain2010_SVP} in the \ac{psd} case (\cref{alg:lrmr_SVP_psd}) for $R=1$, and that the same \acl{MM} procedure in~\cite{Qiu2016_PRIME} leads to \ac{SVP} (in the \ac{psd} case) for \emph{any} $R\leq K$.
Specifically, in~\cite{Qiu2016_PRIME}, Qiu~\etal~use \acl{MM} considerations to construct a tight majorizer to~\cref{eq:psdmf_objective_function_affine_per_j} that we denote $h$. 
Let $\bC$ denote the previous value of $\bB_j$, and $\bB_j$ is the variable to update.
By construction of $h(\bB_j\mid \bC)$ as a majorization function of $f_j(\bB_j)$ at the point $\bC$, $h(\bB_j\mid \bC)$ satisfies~\cite{Qiu2016_PRIME}:
\begin{subequations}
\label{eq:psdmf_MM_h}
\begin{align}
h(\bB_j\mid \bC){}\geq{}&f_j(\bB_j)\quad \textrm{for all}\quad \bB_j\in\mathbb{R}^{K\times K}\\
h(\bC\mid \bC){}={}& f_j(\bC)\,.
\end{align}
\end{subequations}
A function $h(\bB_j\mid \bC)$ that majorizes the objective $f_j(\bB_j)$ in~\cref{eq:psdmf_objective_function_affine_per_j} is
 given in~\cite[Eq.~(28)]{Qiu2016_PRIME}:
\begin{multline}
\label{eq:phase_retrieval_MM_majorizer_PRIME_quadratic_low_rank}
h(\bB_j\mid \bC){}\defn{} \gamma \trace{\bB_j\bB_j} + 2\sum_{i=1}^I\trace{\bB_j\bA_i}\trace{\bC\bA_i} \\ - 2\gamma \trace{\bB_j\bC} -\sum_{i=1}^I 2x_{ij}\trace{\bA_i\bB_j}\,,
\end{multline}
where $\gamma\geq \lambda_{\textrm{max}}(\mathfrak{A}\mathfrak{A}\transpose)$.
\Cref{eq:phase_retrieval_MM_majorizer_PRIME_quadratic_low_rank} and the bound on $\gamma$ were derived in~\cite{Qiu2016_PRIME} using the \ac{psd} matrices $\bB$ and $\bA_i$, without making any assumption about their ranks.
Our key observation is that the value of $\bB_j$ satisfying $\frac{\partial h(\bB_j\mid \bC)}{\partial \bB_j}=0$ is:
\begin{align}
\label{eq:phase_retrieval_MM_Bopt_equal_to_W_in_PRIME}
\bB_j^{\textrm{opt}}{}={}\bC + \frac{1}{\gamma }\sum_{i=1}^I(y_i-\trace{\bA_i\bC})\bA_i\,,
\end{align}
which, using~\cref{eq:lrmr_SVP_gradient}, leads directly to the update step of \ac{SVP} in~\cref{alg:lrmr_SVP_psd}, where the rank constraint of PRIME-Power is imposed by the projection operator $\Hardpsd{R}{\cdot}$ with $R=1$. Hence, the \acl{MM} update step of PRIME-Power is equivalent to the gradient-based update step of \ac{SVP} in the $R=1$ \ac{psd} case.
Note also that the step size $\frac{1}{\gamma}\leq (\lambda_{\textrm{max}}(\mathfrak{A}\mathfrak{A}\transpose))\inv$ in~\cref{eq:phase_retrieval_MM_Bopt_equal_to_W_in_PRIME}, which was obtained in~\cite{Qiu2016_PRIME} from \acl{MM} considerations, is the same step size arising from the Lipschitz constant of the gradient, see our discussion in~\cref{sec:psdmf_algorithm_based_on_SVP}. 
Next, given that the derivation of $h$ and $\gamma$ in~\cite{Qiu2016_PRIME} did not rely on the rank of the \ac{psd} matrices $\bA_i$, we conclude that \ac{SVP}~\cite{Jain2010_SVP} (in the \ac{psd} case) is equivalent to a straightforward extension of PRIME-Power to the recovery of full-column-rank $K\times R$ matrices, instead of $K\times 1$ vectors, from their phaseless measurements. The fact that \ac{SVP}~\cite{Jain2010_SVP} coincides with a \acl{MM}-based method (in the \ac{psd} case) can serve as a reminder that \ac{SVP}-based \ac{PSDMF} methods can equally be regarded as being derived based on \acl{MM} considerations. We mention that the fact that the gradient update step in \ac{SVP} is related to \acl{MM} (regardless of the link to the PRIME-Power method) can be deduced directly by noting that the quadratic objective function in~\cref{eq:psdmf_objective_function} is strongly convex in $\mathfrak{A}$ (when $\mathfrak{B}$ is fixed) with a Lipschitz continuous gradient.


\section{Numerical Experiments}
\label{sec:numerical_experiments}
In this section, we exhibit the potential of algorithms for \ac{PSDMF} based on \acl{PR} and \ac{ARM} optimization methods. We also illustrate numerically some of the properties of \ac{PSDMF} that we discussed theoretically in~\cref{sec:background}.
We consider the projection-based \ac{PSDMF} algorithms introduced in~\cref{sec:projection_based_methods}: \ac{SVP}, \ac{FSVP}, \ac{NIHT} and \ac{CGIHT}. We consider also the \ac{ABG} methods with their two types of objective functions: quadratic~\cite{Lahat2020_ICASSP_PSDMF} and \acl{GKLD}~\cite{Lahat2020_EUSIPCO_TWF}. The latter is denoted \ac{ABG}-P in our plots, where `P' stands for Poisson log-likelihood, which is the likelihood function associated with \new{the} \acl{GKLD}.
Among the methods that were not designed based on \acl{PR} or \ac{ARM} principles, we focus on \ac{CD}~\cite{Vandaele2018} as the main competing method. We do not compare with the algorithms in~\cite{Glasser2019_ExpressiveTN_NIPS,Stark2016_RecommenderQuantum} because each of them has some restriction on the model or on the type of data addressed, as discussed in~\cref{sec:related_work}. 
We consider both cyclic and greedy (also known as \ac{GS}) variants of \ac{CD}~\cite{Vandaele2018}, where we set the ``greediness'' coefficient to $0.5$, as in~\cite{Vandaele2018}.
We implement \ac{ABG} as in~\cite{Lahat2020_ICASSP_PSDMF,Lahat2020_EUSIPCO_TWF}.
The backtracking line search parameters of \ac{ABG} are set to $\alpha = 0.1$ and $\beta = 0.35$. 
These values were chosen after verifying they provided satisfying performance in our experiments.

\subsection{Numerical Issues}
\label{sec:numerical_issues}

\subsubsection{Initialization}
\label{sec:numerical_issues_initialization}
We initialize the algorithms with factors whose entries are drawn independently from the standard normal distribution $\mathcal{N}(0,1)$.
We normalize the input matrix to $\|\bX\|_F=1$.
We then scale one set of initial factors: $\bU_i\gets \sqrt{\lambda_*}\bU_i$, $\indrange{i}{I}$, where $\lambda_*{}={}\argmin_\lambda\|\bX-\lambda\mathfrak{A}\transpose\mathfrak{B}\|_F^2$. This scaling procedure, suggested in~\cite{Vandaele2018}, along with normalizing the input matrix, turned out to improve the convergence properties in our experiments.
As explained in~\cref{sec:psdmf_caveats_wrt_PR_ARM}, strategies that are useful for \acl{PR} and \ac{ARM}, such as spectral initialization, cannot be applied to \ac{PSDMF}.

\subsubsection{Figure of Merit}
The \ac{RMFE} is defined as $\frac{\|\bX-\widehat{\bX}\|_F}{\|\bX\|_F}$, where $\widehat{\bX}\defn\widehat{\mathfrak{A}}\transpose\widehat{\mathfrak{B}}$ denotes the approximation of $\bX$ based on the approximated model parameters
$\widehat{\mathfrak{A}}$ and $\widehat{\mathfrak{B}}$ when a stopping criterion is achieved.
As a stopping criterion, we use a tolerance on the \ac{RMFE}: $\frac{\|\bX-\widehat{\bX}\|_F}{\|\bX\|_F}\leq \TolRMFE$. We also use a tolerance on the relative change in the \ac{QMFE}: $\frac{|f^{\ell+1}-f^{\ell}|}{f^1} {}<{}\TolFun$,
where the
\ac{QMFE} is defined as $f^{\ell}\defn\frac{1}{2}\|\bX-{\mathfrak{A}^{\ell}}\transpose\mathfrak{B}^{\ell}\|_F^2$, and $\mathfrak{A}^{\ell},\mathfrak{B}^{\ell}$ are the estimates of $\mathfrak{A},\mathfrak{B}$ at iteration index $\ell$. By default, $\TolFun{}={}0{}={}\TolRMFE$.

\subsubsection{\texorpdfstring{Choosing $D$}{Choosing D}}
\label{sec:numerical_issues_D}
In our experiments, we report on the number of \emph{overall iterations} $\ell\times D$, where $\ell$ is the (outer) iteration index.
Together with the \acl{CC} in~\cref{table:psdmf_complexity,sec:psdmf_complexity}, the number of overall iterations give\new{s} an idea about the amount of computation needed to achieve a stopping criterion, regardless of a specific implementation or programming platform.

We set $D=1$ for \ac{SVP}, \ac{NIHT}, \ac{ABG}, and \ac{ABG}-P. For methods that rely on acceleration\new{---}\ac{FSVP} and \ac{CGIHT}\new{:} in some cases, we made preliminary tests on candidate values of $D$. However, it is not convenient nor practical to run a preliminary test of the optimal value of $D$ whenever we wish to use an algorithms. Therefore, in the remainder of cases, we chose $D_{\textrm{FSVP}}$ and $D_{\textrm{CGIHT}}$ arbitrarily, with values similar to those that turned out useful in other experiments.
In each experiment, we specify the values of $D_{\textrm{FSVP}}$ and $D_{\textrm{CGIHT}}$ that we use.

\subsubsection{CPU Time}
\label{sec:numerical_issues_CPU_time}
In our plots showing error evolution versus CPU time, all methods are coded in Matlab R2019a and run on a MacBook Pro with a 2.8GHz Intel Core i7 processor and 16 GB memory.

\subsubsection{Code and Implementation}
We use the code in~\cite{VandaelePSDMFcode} to generate the geometric data matrices and run the \ac{CD} algorithms, which are written in c and thus run faster (Matlab does not handle loops efficiently~\cite{Vandaele2018}), in our \ac{MC} trials. For runtime comparison, however, we implemented the \ac{CD} algorithms of~\cite{Vandaele2018} in Matlab.
\new{Code for the algorithms proposed in this paper is available at~\url{https://www.dana.lahat.org.il/psdmf.html}.}

\subsubsection{Complexity}
\label{sec:psdmf_complexity}
In~\cref{table:psdmf_complexity}, we compare the per-subproblem \acl{CC} of the different methods in updating $\{\bB_j\}_{j=1}^J$. Deducing the results for the other set of variables is straightforward. The purpose is to show how each algorithm scales in each of the dimensions of the problem.
The cost of the projection-based methods is dominated by: (i) computing the Lipschitz constant for the step size in \ac{SVP} and \ac{FSVP}, which involves an \ac{SVD} of $\mathfrak{A}\in\mathbb{R}^{K^2\times I}$ and thus costs $\mathcal{O}(\min(K^4I,K^2I^2))$, (ii) computing the gradient $\mathcal{A}\adj(\bx_j-\mathcal{A}(\bB_j))$, which costs $\mathcal{O}(IK^2)$ per $j$, (iii) projecting on $\mathbb{S}_+^K$, which requires an \acl{EVD} of $\bB_j$ for each $j$, and costs $\mathcal{O}(K^3)$, (iv) projecting $\bB_j$ on the set of rank-$R$ matrices, which costs $\mathcal{O}(K^2R)$. In \ac{NIHT}, computing the matrix $\bU$ when $R<K$ requires an additional \acl{EVD} of $\mathcal{O}(K^3)$, and computing the adjoint operator costs $\mathcal{O}(IK^2)$, per $j$.
Since in \ac{PSDMF} we always have $K\leq I$ and $R\leq K$, the cost of the \acl{EVD} and the projection on rank-$R$ \ac{psd} matrices are dominated by $\mathcal{O}(IK^2)$.
Within each inner iteration, \ac{CGIHT} requires more computations---in fact, approximately twice---compared to \ac{NIHT}; however, these computations are of the same nature as those for \ac{NIHT} and thus they scale the same in the dimensions of the problem.
The \acl{CC} of one subproblem in \ac{CD} is given in~\cite{Vandaele2018}. However, in contrast to~\cite{Vandaele2018}, here we do not omit the term $\mathcal{O}(IJK^2)$ because it might dominate $\mathcal{O}(IK^4R)$ when $I>K^3$.
The \acl{CC} of \ac{ABG} is dominated by $\mathcal{O}(IJK^2)$~\cite{Lahat2020_ICASSP_PSDMF}.  Computing the denominator in the gradient in \ac{ABG}-P has the same \acl{CC} as the computation of the gradient of \ac{ABG} and thus the overall \acl{CC} does not change.

\begin{table}[!t]
\renewcommand{\arraystretch}{1.3}
\caption{Computational complexity per subproblem: updating $\{\bB_j\}_{j=1}^J$}
\label{table:psdmf_complexity}
\centering
\begin{tabular}{cc}
\hline
Method & Computational Complexity\\
\hline\hline
\ac{SVP} \& \ac{FSVP} & $\mathcal{O}(\min(IK^4,I^2 K^2)+DIJK^2)$\\
\hline
\ac{NIHT} \& \ac{CGIHT} & $\mathcal{O}(DIJK^2)$ \\
\hline
\ac{CD} & $\mathcal{O}(IJK^2 + IK^4 R)$ \\
\hline
\ac{ABG} \& \ac{ABG}-P & $\mathcal{O}(IJK^2)$ \\
\hline
\end{tabular}
\end{table}

\subsection{\texorpdfstring{Performance Comparison: Euclidean Distance Matrices}{Performance Comparison: EDM}}
\label{sec:numerical_experiments_EDM}
Consider an $I\times I$ matrix $\bD_I$ whose $(i,j)$th entry is equal to $d_{ij}{}={}(\alpha_i-\alpha_j)^2$, where $\alpha_i$ are real numbers and $\alpha_i\neq\alpha_j$ for any $\indrange{i,j}{I}$. Thus, $d_{ii}{}={}0$ for all $i$. The \ac{psd} rank of $\bD_I$ is equal to 2 and its inner ranks are all equal to 1, because $\bD_I$ admits a \ac{PSDMF} with factors $\bu_i{}\defn{}\begin{bmatrix}
\alpha_i & 1
\end{bmatrix}\new{\transpose}$ and $\bv_j{}\defn{}\begin{bmatrix}
1 & -\alpha_j
\end{bmatrix}\new{\transpose}$ for all $i,j$~\cite{Gouveia2013_Lifts}.
The usual matrix rank of $\bD_I$ is 3 for $I\geq 3$, whereas its nonnegative rank becomes arbitrarily large as $I\rightarrow\infty$~\cite{Beasley2009,Shitov2019_EDM}. This matrix is known as \ac{EDM}.

In this experiment, we compare the ability of the algorithms to factorize $\bD_I$. We draw the values of $\alpha_i$ independently from the standard uniform distribution: $\alpha_i\sim\mathcal{U}[0,1]$. We set $I{}={}100$, and run 100 \ac{MC} trials, each with a new initialization and a new draw of $\alpha_i$ for all $i$. We fit $\bD_I$  to a \ac{PSDMF} model with $K=2$ and $R_A\defn{}R_{A_i}=1$ for all $i$, $R_B{}\defn{}R_{B_j}=1$ for all $j$. 
\new{In this case, $N_{\textrm{data}}=10^4 {}>{}N_{\textrm{model}}=396$.}
Our stopping criterion is $\TolFun{}={}10^{-15}$.

\begin{figure}[!t]
\centering
{\includegraphics[width=0.58\columnwidth]{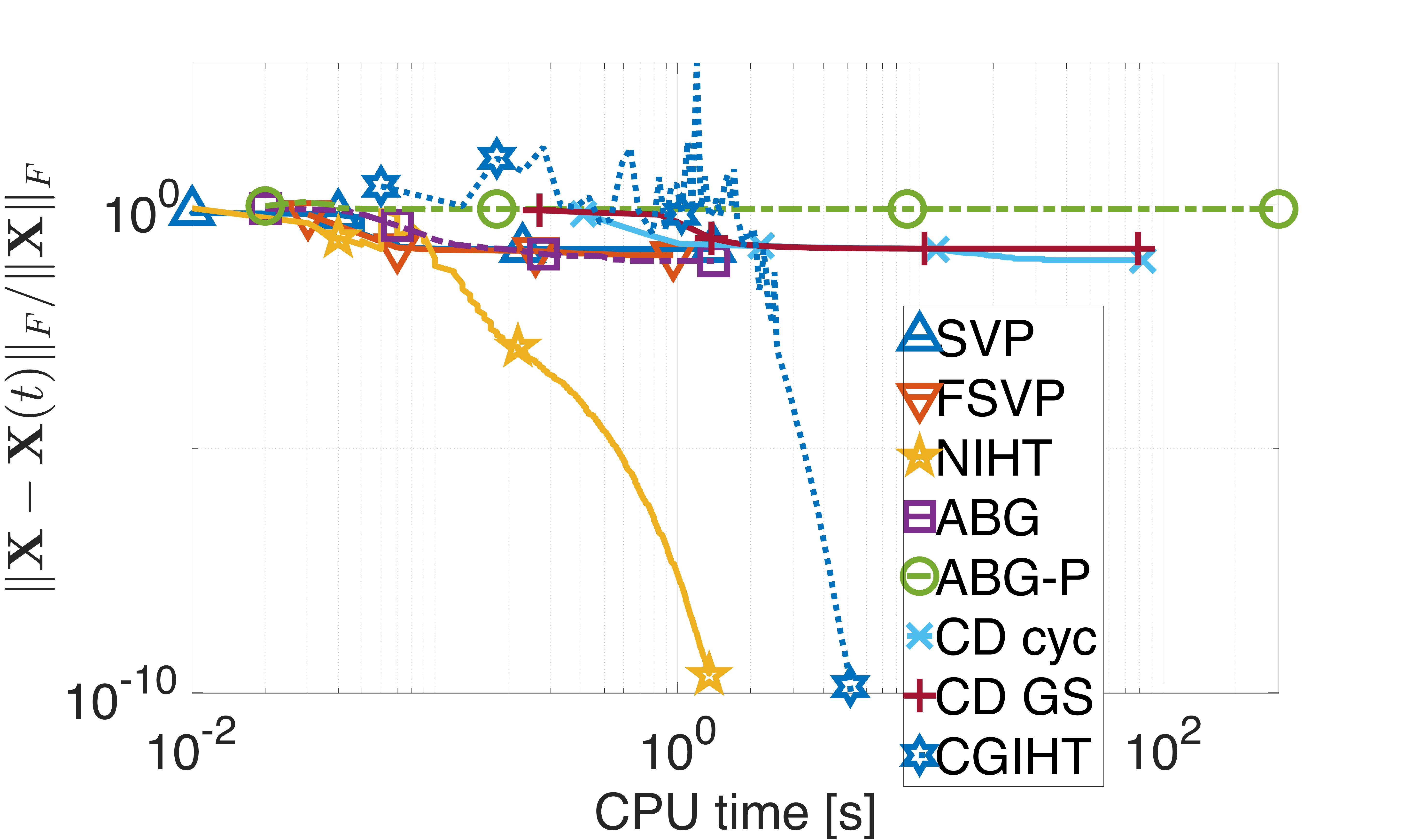}\label{fig:psdmf_EDM_100x100_Trajectory_CPU}}
\caption{Evolution of the \acl{RMFE} in the \ac{PSDMF} of a $100\times 100$ \acl{EDM} with $K=2$, $R_A=1=R_B$.}
\label{fig:psdmf_RandomEDM_100x100_K2_ra1_rb1}
\end{figure}

In this setting, we obtain two distinct and clearly-separated clusters of results: one consists of trials that did not decrease the \ac{RMFE} below $\approx 10^{-2}$, which is a rather large value. The other cluster consists of the successful trials, defined as achieving \ac{RMFE}$\leq 10^{-4}$.
For \ac{CGIHT} and \ac{FSVP}, we tested several candidate values of $D$ and found that $D_{\textrm{FSVP}}=19$ and $D_{\textrm{CGIHT}}=14$ were associated with higher rates of successful factorizations that other values of $D$ that we tried. Increasing $D$ for other methods did not provide any substantial improvement in the rate of success, as expected.
\Cref{fig:psdmf_RandomEDM_100x100_K2_ra1_rb1} exemplifies the trajectory of each algorithm in one trial and demonstrates the separation into the two categories of ``success'' and ``failure''. \Cref{table:psdmf_RandomEDM_100x100_MC100_K2_ra1_rb1} shows the number of successful factorizations per algorithm.
\begin{table}[!t]
\caption{Number of successful factorizations of $100\times 100$ Euclidean distance matrices in 100 \ac{MC} trials.}
\centering
\tabcolsep=0.11cm 
\begin{tabular}{|c|c|c|c|c|c|c|c|}
\hline
 SVP & FSVP & NIHT & ABG& ABG-P & CD cyc & CD GS & CGIHT\\
\hline\hline
  1  &   2 &   37  &   0   &  0  &   1  &   9 &   91\\
\hline
\end{tabular}
\label{table:psdmf_RandomEDM_100x100_MC100_K2_ra1_rb1}
\end{table}

A possible explanation of the dominance of \ac{NIHT} and its accelerated variant \ac{CGIHT} can be seen from the irregular and non-monotonous error trajectories in~\cref{fig:psdmf_RandomEDM_100x100_K2_ra1_rb1}. Their trajectories drop sharply (in the log-log scale) at a certain point. The other algorithms decrease the objective function monotonically until they reach a plateau. This is in agreement with the fact that \ac{NIHT} and \ac{CGIHT} are the only methods among the ones we compare that do not have a guarantee for monotone decrease of the objective function. We postulate that this property allows \ac{NIHT} and \ac{CGIHT} to escape certain local stationary points that the other methods get trapped in.

\subsection{Performance Comparison: Geometric Data}
\label{sec:numerical_experiments_geometrical}
In this section, we compare our algorithms on slack matrices, which are matrices associated with the geometry of polytopes. Finding the \ac{psd} rank of such matrices, and sometimes also the inner ranks, is one of the central applications of \ac{PSDMF} (e.g.,~\cite{Gouveia2013_Lifts,Fawzi2015_PSD_Rank,Vandaele2018}). In these applications, one is interested in exact \acs{PSDMF} or at least in a good approximation thereof~\cite{Gouveia2015_ApproximateCone}. As demonstrated by~\cite{Vandaele2018}, even the factorization of small slack matrices can be numerically challenging. The matrices considered in this section were studied also in~\cite{Vandaele2018}.

\subsubsection{Submatrix of the slack matrix of the correlation polytope}
\label{sec:numerical_experiments_geometrical_submatrix_CORn}
Consider a $2^n\times 2^n$ binary matrix $\bM_n$ whose rows and columns are indexed by vectors $\bc,\bd\in\{0,1\}^n$ such that 
$\bM_n(\bc,\bd){}={}(1-\bc\transpose\bd)^2$. The \ac{psd} rank of $\bM_n$ is equal to $n+1$ and its inner ranks are all equal to 1, because $\bM_n(\bc,\bd)$ admits a \ac{PSDMF} with factors $\bu\defn\begin{bmatrix}
1 &
-\bc\transpose
\end{bmatrix}\transpose$ and $\bv\defn\begin{bmatrix}
1 & \bd\transpose
\end{bmatrix}\transpose$~\cite{Fiorini2012_LinearVsSemidefinite}. This matrix is known as a submatrix of the slack matrix of the correlation polytope~\cite{Fiorini2012_LinearVsSemidefinite}.
As an example, $\bM_2= \left[\begin{smallmatrix}
     1  &   1  &   1 &    1\\
     1   &  0  &   1 &    0\\
     1   &  1  &   0 &    0\\
     1  &   0  &   0 &    1
\end{smallmatrix}\right]$.

In this experiment, we compare the ability of the algorithms to factorize $\bM_n$.
For each value of $n=2,\ldots, 7$, we run 100 \ac{MC} trials, each with a new initialization. We fit $\bM_n$ to a \ac{PSDMF} model with $K=n+1$ and $R_A\defn R_{A_i}=1$ for all $i$, $R_B\defn R_{B_j}=1$ for all $j$. Our stopping criterion is $\TolFun=10^{-12}$.

In this setting, similarly to the \ac{EDM} example in~\cref{sec:numerical_experiments_EDM}, for each value of $n$, we obtain two distinct and clearly-separated clusters of results: one cluster consists of a significant number of trials that did not decrease the \ac{RMFE} below $\approx 10^{-2}$, which is a relatively large value. The other cluster consists of the remaining trials, which managed to decrease the error to $\approx 10^{-4}$ or less. For the trials in the latter cluster, using a smaller $\TolFun$ can further decrease the \ac{RMFE} by several orders of magnitude, whereas for trials in the former cluster, their error remains on the same scale.
For each algorithm, we define as ``success'' any trial with a sufficiently small \ac{RMFE} to place it in the second cluster.

We set $D_{\textrm{FSVP}}{}={}14$ for all $n$ for acceleration.  We did not observe, in our preliminary tests, any particular influence of the value of $D_{\textrm{FSVP}}$ on the rate of success.
However, for $n=2,3,4$, we set $D_{\textrm{CGIHT}}{}={}14, 110, 9$, respectively, as we observed that these values allow to increase the rate of successful factorizations significantly. For $n=5,6,7$ we did not observe any influence of $D_{\textrm{CGIHT}}$ on the rate of success and thus for these three matrices, we set $D_{\textrm{CGIHT}}=1$. Since \ac{CGIHT} with $D_{\textrm{CGIHT}}=1$ is equivalent to \ac{NIHT}, we do not show \ac{CGIHT} in the plots for $\bM_5$, $\bM_6$, and $\bM_7$. Note that here, $D_{\textrm{CGIHT}}$ is not used in its original role as an acceleration parameter but to control the rate of success.

\begin{figure}[!t]
\centering
\subfloat[]{\includegraphics[width=0.48\columnwidth]{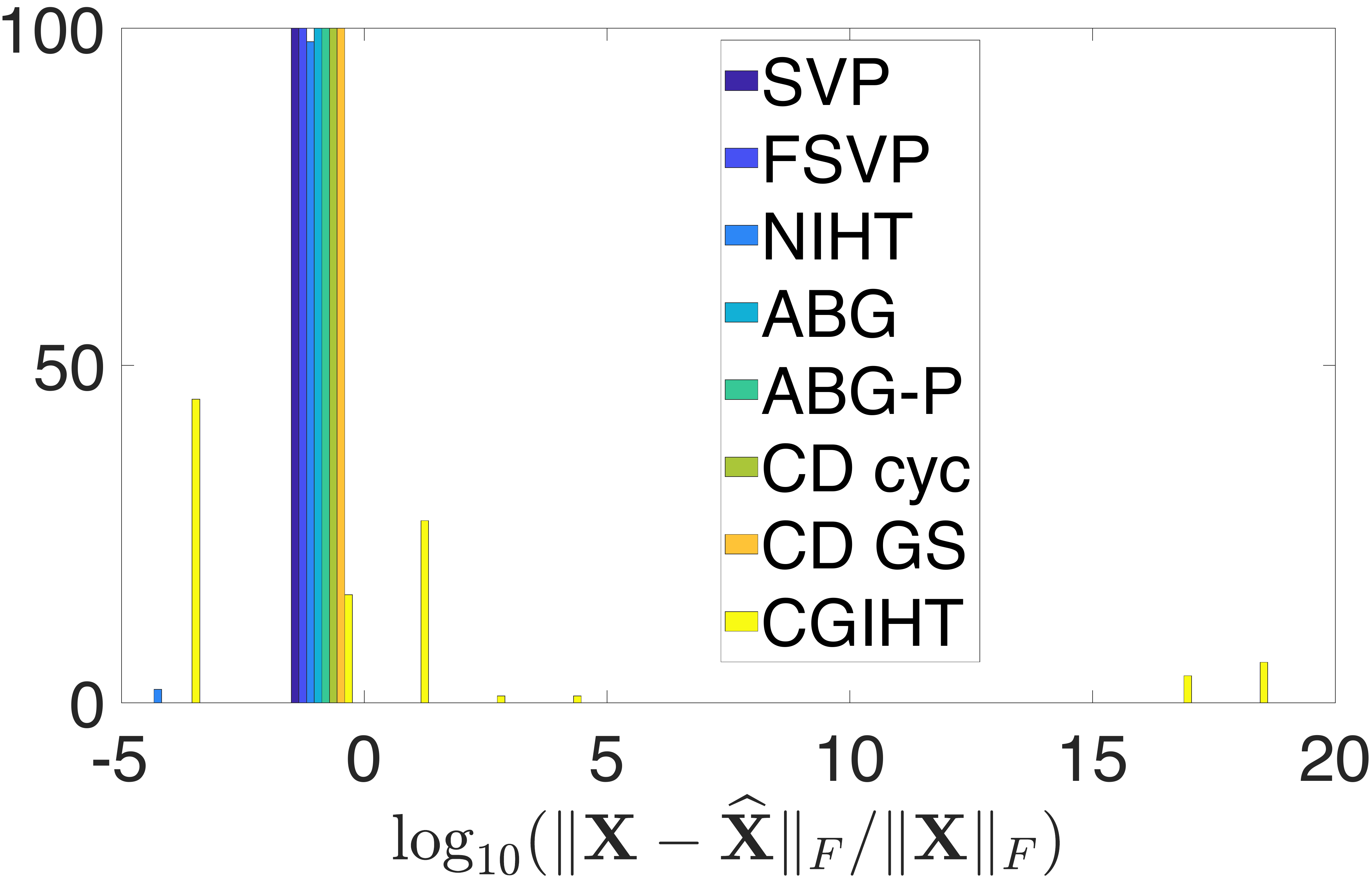}\label{fig:psdmf_Correl4_MC100_K5_ra1_rb1_TolFun_1e12_TolFval_0_MaxCPUTime_120_MaxIter_1e6_Hist_LogErr}}
\hfil
\subfloat[]{\includegraphics[width=0.48\columnwidth]{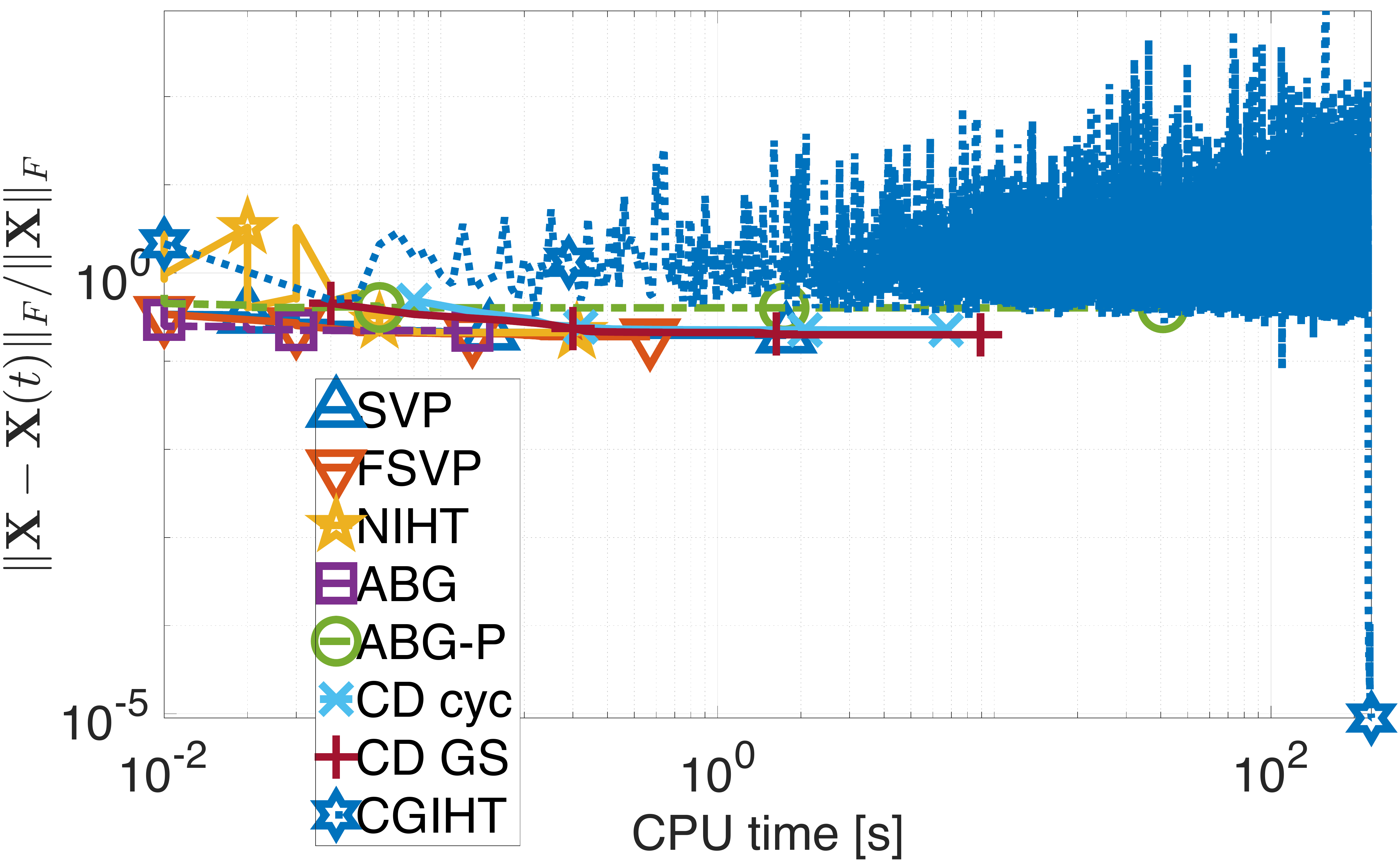}\label{fig:psdmf_Correl4_K5_ra1_rb1_TolFun_1e14_TolFval_0_MaxCPUTime_780_MaxIter_1e6_Trajectory_LogCPU}}
\\
\subfloat[]{\includegraphics[width=0.48\columnwidth]{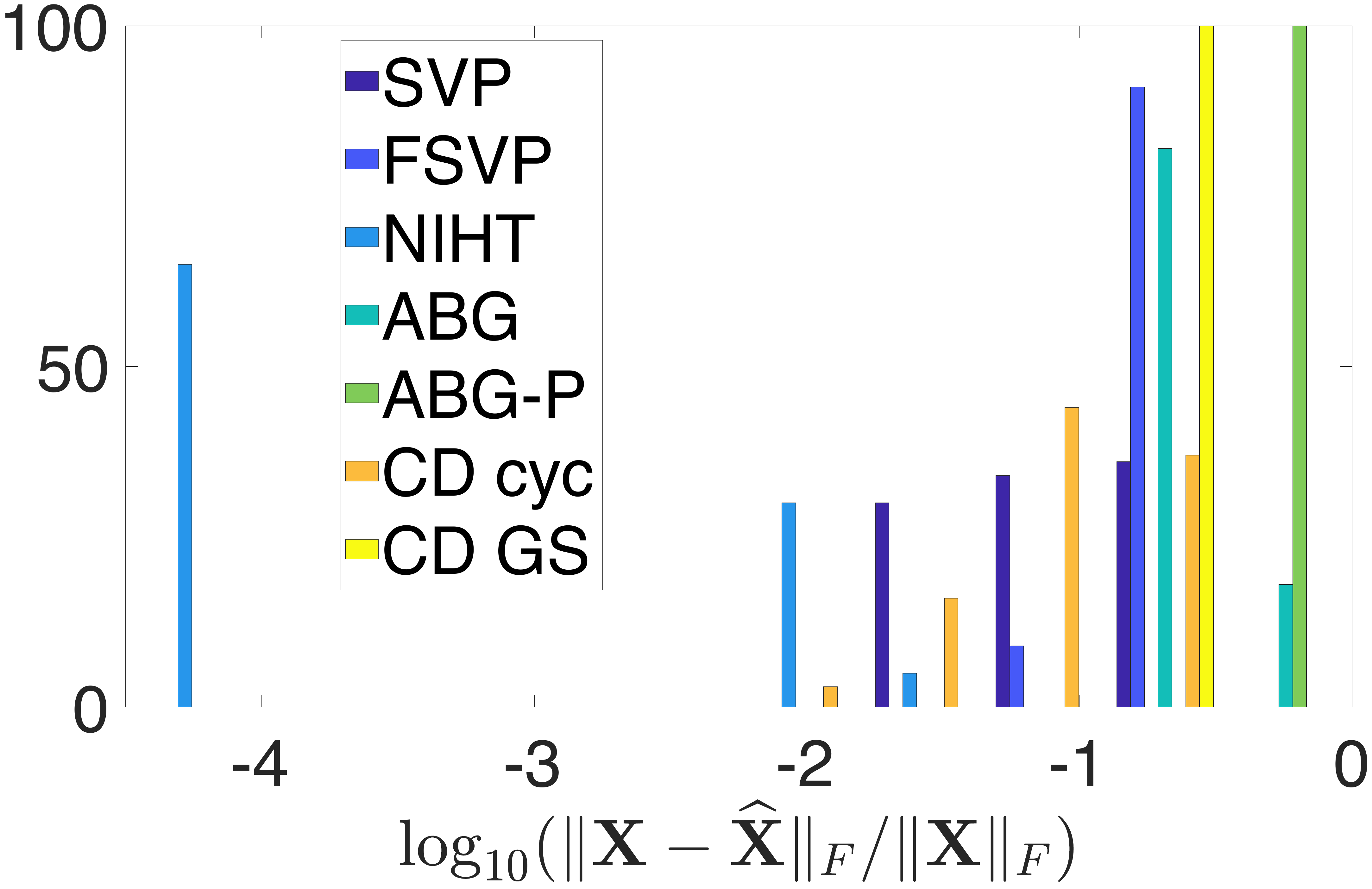}\label{fig:psdmf_Correl7_MC100_K8_ra1_rb1_TolFun_1e12_TolFval_0_MaxCPUTime_720_Hist_LogErr}}
\hfil
\subfloat[]{\includegraphics[width=0.48\columnwidth]{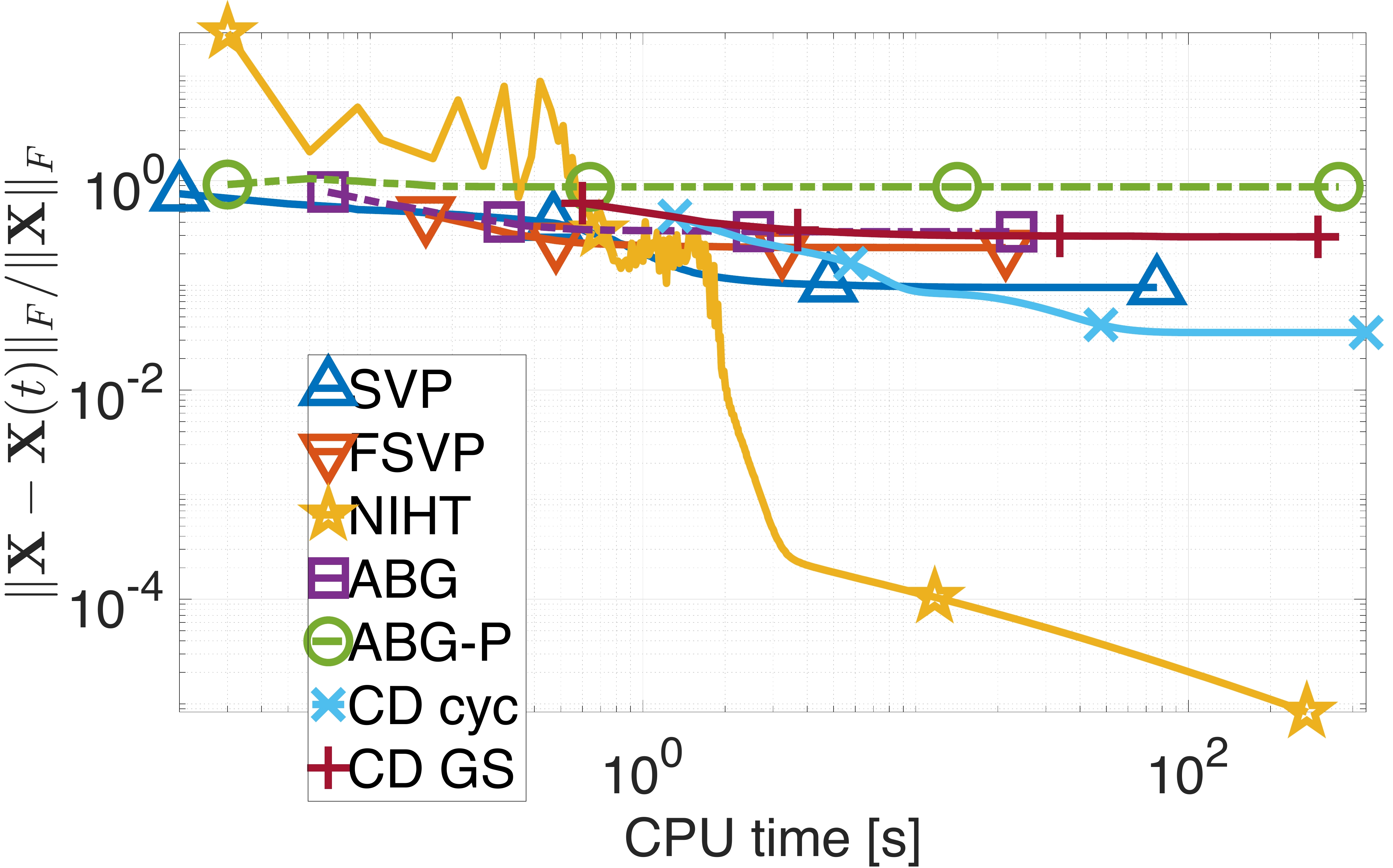}\label{fig:psdmf_Correl7_K8_ra1_rb1_TolFun_1e14_TolFval_0_MaxCPUTime_780_MaxIter_1e6_Trajectory_LogCPU}}
\caption{\ac{PSDMF} of $\bM_4$ (top) and $\bM_7$ (bottom). (a) and (c) Histogram of final error in 100 \ac{MC} trials, (b) and (d) error evolution of one trial}
\label{fig:psdmf_correl7}
\end{figure}

\Cref{fig:psdmf_correl7} exemplifies the results for $\bM_4\in\mathbb{R}^{16\times 16}$ and $\bM_7\in\mathbb{R}^{128\times 128}$. 
\Cref{fig:psdmf_Correl4_MC100_K5_ra1_rb1_TolFun_1e12_TolFval_0_MaxCPUTime_120_MaxIter_1e6_Hist_LogErr,fig:psdmf_Correl7_MC100_K8_ra1_rb1_TolFun_1e12_TolFval_0_MaxCPUTime_720_Hist_LogErr} show the histogram (with 10 bins) of the final error.
\Cref{fig:psdmf_Correl4_K5_ra1_rb1_TolFun_1e14_TolFval_0_MaxCPUTime_780_MaxIter_1e6_Trajectory_LogCPU,fig:psdmf_Correl7_K8_ra1_rb1_TolFun_1e14_TolFval_0_MaxCPUTime_780_MaxIter_1e6_Trajectory_LogCPU} exemplify the trajectory of each algorithm in one trial, where here the stopping criterion is $\TolFun=10^{-13}$.
Indeed, for $\bM_7$, \ac{NIHT} is the only algorithm that succeeded in properly decreasing the objective function, in 65 of 100 \ac{MC} trials, whereas \ac{CGIHT} dominates for $\bM_4$, as shown in~\cref{table:psdmf_correl_success}.

\Cref{table:psdmf_correl_success} summarizes the number of successful factorizations per algorithm for each value of $n$.
\Cref{table:psdmf_correl_success} shows that the ability of each algorithm to factorize $\bM_n$ varies greatly with $n$. While all methods factorize $\bM_2\in\mathbb{R}^{4\times 4}$ with 38\%--96\% success rate, 
none of \ac{ABG}'s two variants succeed in factorizing $\bM_3\in\mathbb{R}^{8\times 8}$, and the remaining algorithms have only 1\% success rate---except for \ac{CGIHT} with 55\% success. At $n=4$, our proposed \ac{NIHT}-based \ac{PSDMF} algorithm has modest success of 2\% whereas its \ac{CGIHT} variant succeeds in 45\% of the attempts.
For $n=5,6,7$, \ac{NIHT} is the only one to factorize \new{$\bM_n$}, with success rates rising from 30\% in $n=5$ to 65\%  with $n=7$. While the drop in the general success rate as $n$ increases may be explained by the non-convex nature of this problem  and the smaller number of free variables versus constraints (for $n=2$, $N_{\textrm{data}}=16 > N_{\textrm{model}}=15$, and the ratio $N_{\textrm{model}}/N_{\textrm{data}}$ only decreases with $n$), the remarkable success rate of \ac{NIHT} and \ac{CGIHT} necessitates another set of arguments.  These results resemble those in~\cref{sec:numerical_experiments_EDM} for linear \ac{EDM} with random distances.
A possible explanation may arise from~\cref{fig:psdmf_Correl4_K5_ra1_rb1_TolFun_1e14_TolFval_0_MaxCPUTime_780_MaxIter_1e6_Trajectory_LogCPU}, which shows that the error trajectory of \ac{NIHT} starts very irregular and non-monotonous, until it drops sharply (in the log-log scale). \ac{CGIHT} inherits the erratic behaviour from \ac{NIHT}, and its trajectory seems to become more erratic as $D_{\textrm{CGIHT}}$ increases. All other algorithms exhibit a monotonously decreasing trajectory that reaches a plateau. As mentioned in~\cref{sec:psdmf_niht,sec:psdmf_cgiht}, \ac{NIHT} and \ac{CGIHT} are the only methods among the ones we compare that do not have a guarantee for monotonous decrease of the objective function. It is possible that this property allows them to escape certain local stationary points that the other methods get trapped in.

\begin{table}[!t]
\caption{Number of successful factorizations of $\bM_n$ in 100 \ac{MC} trials. For $n=5,6,7$, the best results for \ac{CGIHT} were obtained when it was equivalent to \ac{NIHT}. }
\label{table:psdmf_correl_success}
\centering
\tabcolsep=0.11cm 
\begin{tabular}{|c||c|c|c|c|c|c|c|c|}
\hline
$n$ & SVP & \ac{FSVP} & \ac{NIHT} &  ABG & ABG-P & \ac{CD} cyc & \ac{CD} GS & \ac{CGIHT}\\
\hline\hline
2   & 38 & 43 & 45 & 43 & 51  & 63 & 75  & 96\\
\hline
3   & 1   & 1    &  1   &   0  &   0   &   1    &   0 & 55\\
\hline
4   & 0 & 0 & 2  &  0 &   0 & 0 & 0 & 45\\
\hline
5   & 0   &  0  &  30 &    0  &   0 &    0  &   0 & \scriptsize  NIHT\\
\hline
6   &  0  &   0  &  61   &  0  &   0 &    0  &   0 & \scriptsize  NIHT\\
\hline
7   &  0   &  0  &  65  &   0  &   0 &    0 &    0 & \scriptsize  NIHT\\
 \hline
\end{tabular}
\end{table}

\subsubsection{\texorpdfstring{Slack matrix of a regular $n$-gon}{Slack matrix of a regular n-gon}}

We consider a \emph{slack matrix of a regular $n$-gon}. An $n$-gon is a polygon with $n$ sides. Regular slack matrices, denoted $S_n$, are determined up to scaling and transposition, and have size $n\times n$. The usual matrix rank of $S_n$ is 3 for all $n$.
For most $n$-gons, neither the \ac{psd} rank nor the values of the inner ranks are known.

In this experiment, we factorize $S_{32}\in\mathbb{R}^{32\times 32}$, visualized in~\cref{fig:psdmf_S32_visualized}.
The \ac{psd} rank of $S_{32}$ is not known; we shall use $K=\new{1+\lceil\log_2(n)\rceil}=6$ based on~\cite[Conjecture~1]{Vandaele2018}. There is no conjecture about the values of the inner ranks, but due to the presence of zeros in $S_{32}$, it is likely that $R_{A_i}+R_{B_j}\leq K$ for any pair of $(i,j)$. We set $R_A=3$ and $R_B=3$. In this case, $N_{\textrm{data}}=1024 {}>{} N_{\textrm{model}}=924$. This is a numerically challenging setting even in the absence of zeros.
We set $D_{\textrm{FSVP}}=216$ and $D_{\textrm{CGIHT}}=21$.
We run 30 \ac{MC} trials, where at each trial we use a new initialization. Our stopping criterion is $\TolRMFE{}={} 0.0011$. The CPU time allotted for each trial is 600[s].
\Cref{fig:psdmf_S32_MC30_K6_ra3_rb3_TolFun_1e16_TolFval_6p05e7_MaxCPUTime_600_MaxIter_1e7_LogAllIter,fig:psdmf_S32_K6_ra3_rb3_TolFun_0_TolFval_0_MaxCPUTime_1200_MaxIter_1e6_Trajectory_LogCPU} summarize our results. 
The boxplots in~\cref{fig:psdmf_S32_MC30_K6_ra3_rb3_TolFun_1e16_TolFval_6p05e7_MaxCPUTime_600_MaxIter_1e7_LogAllIter} show the overall number of iterations $\ell\times D$ used by each algorithm to achieve the stopping criteria.
\Cref{fig:psdmf_S32_K6_ra3_rb3_TolFun_0_TolFval_0_MaxCPUTime_1200_MaxIter_1e6_Trajectory_LogCPU} exemplifies the error evolution of each algorithm in one randomly-chosen run, now stopping after 20 minutes. We add that none of the trials of \ac{ABG}-P achieved the designated \ac{RMFE} in the allotted time; this is consistent with its behaviour in~\cref{fig:psdmf_S32_K6_ra3_rb3_TolFun_0_TolFval_0_MaxCPUTime_1200_MaxIter_1e6_Trajectory_LogCPU}.

We observe in~\cref{fig:psdmf_S32_MC30_K6_ra3_rb3_TolFun_1e16_TolFval_6p05e7_MaxCPUTime_600_MaxIter_1e7_LogAllIter} is that our choice of $D_{\textrm{FSVP}}$ and $D_{\textrm{CGIHT}}$ was good: the number of overall iterations of \ac{FSVP} is the smallest on average among all methods, and the number of iterations of \ac{CGIHT} is \new{often} smaller than its non-acceleration counterpart, \ac{NIHT}. However, as we see in~\cref{fig:psdmf_S32_K6_ra3_rb3_TolFun_0_TolFval_0_MaxCPUTime_1200_MaxIter_1e6_Trajectory_LogCPU}, after a sufficiently large number of iterations, all methods eventually reach a point in which the decrease of the objective function is very slow.  One possible explanation to our results is that the inner ranks were not chosen correctly: however, we observed the same trend also with other choices of inner ranks, e.g., $R_A=4$ and $R_B=2$, and larger. The difficulty to factorize $S_n$ as $n$ increases is in agreement with the results in~\cite{Vandaele2018}.

\begin{figure}[!t]
\centering
\subfloat[]{\includegraphics[width=0.48\columnwidth]{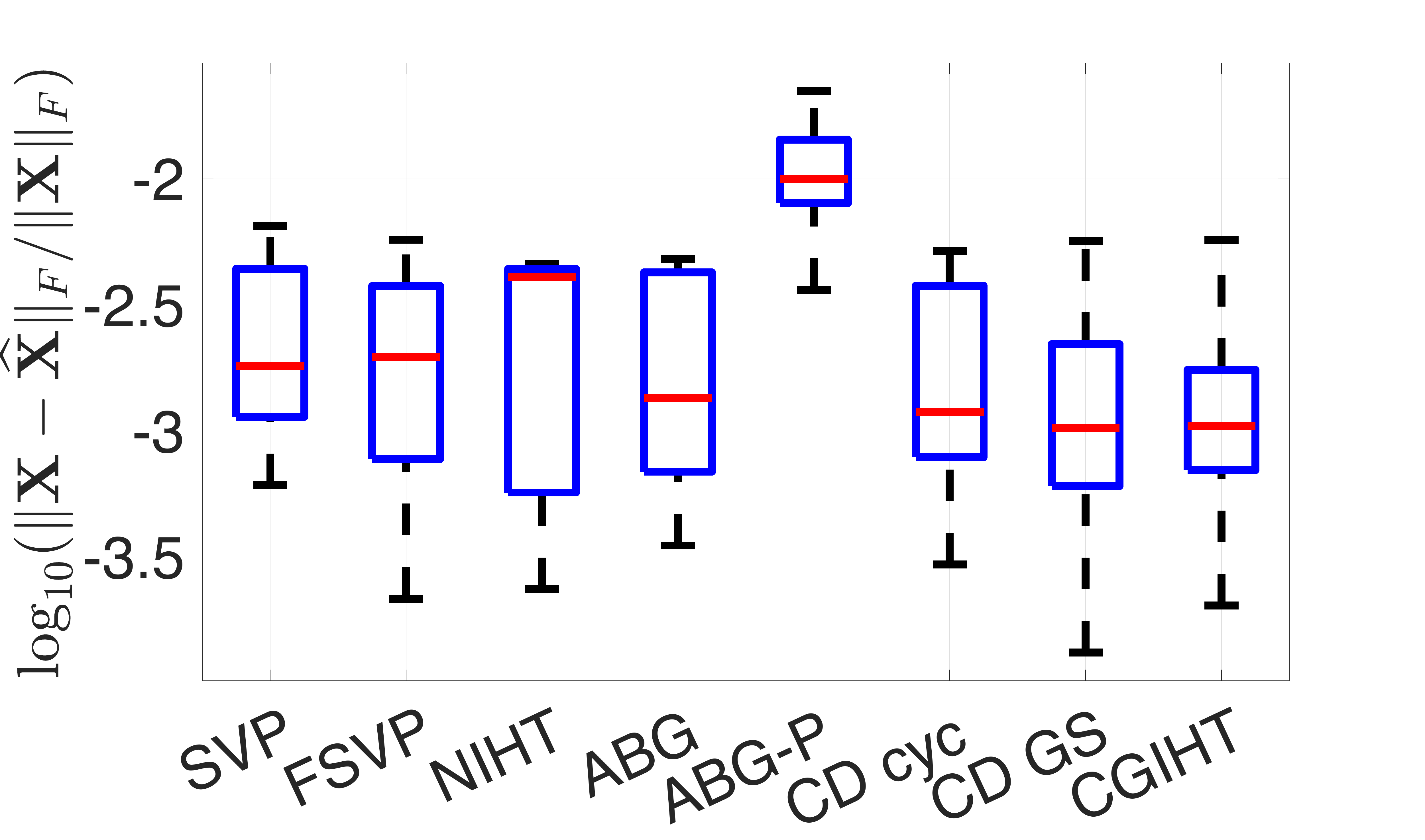}\label{fig:psdmf_S12_MC20_K5_ra3_rb1_TolFun_0_TolFval_0_MaxCPUTime_120_MaxIter_1e6_LogErr}}
\hfil
\subfloat[]{\includegraphics[width=0.48\columnwidth]{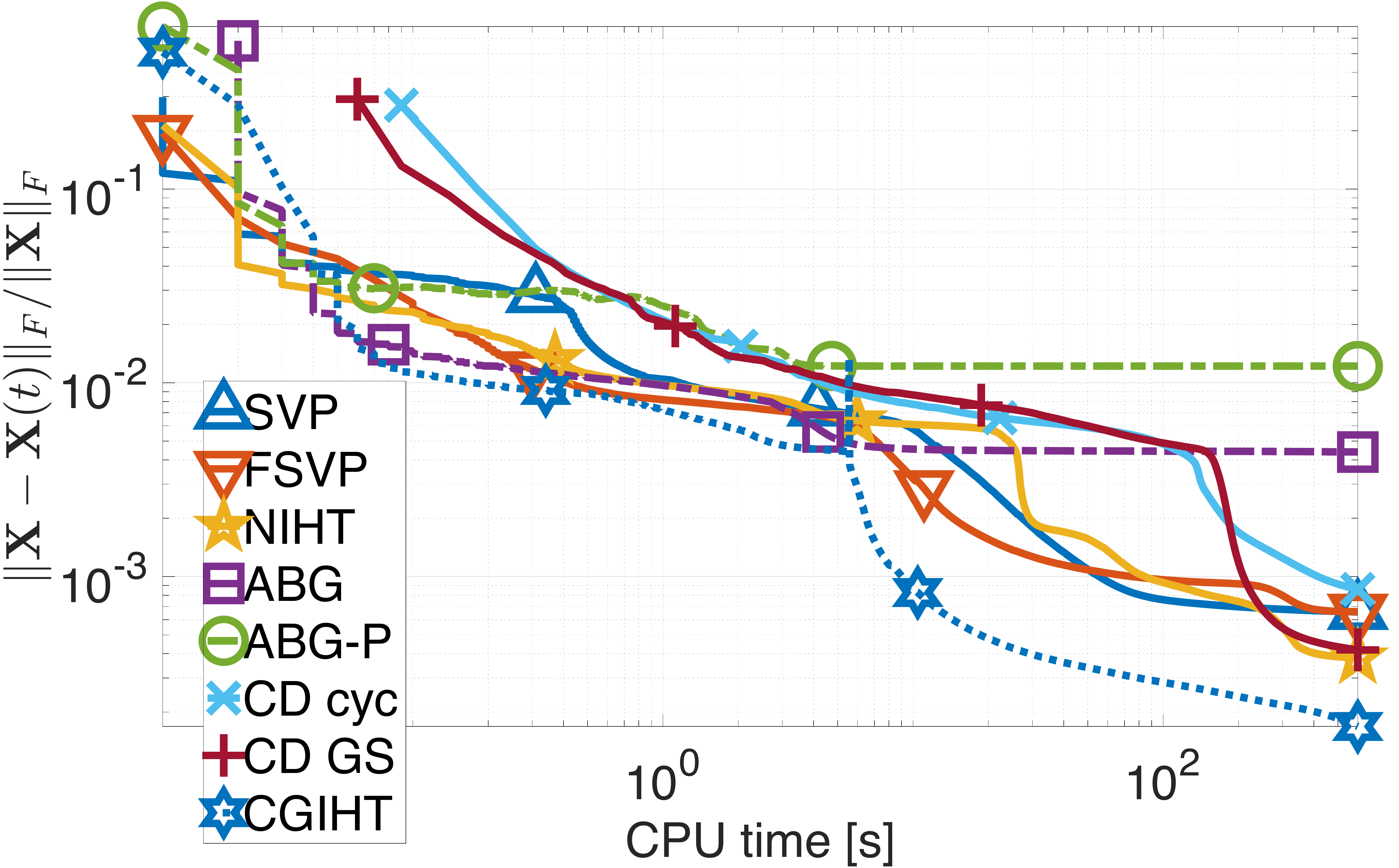}\label{fig:psdmf_S12_K5_ra3_rb1_TolFun_0_TolFval_0_MaxCPUTime_600_MaxIter_1e7_Trajectory_LogCPU}}
\caption{\ac{PSDMF} of (a) $S_{12}$ with $K=5$, $R_A=3$, $R_B=1$. (a) Final error in 20 \ac{MC} trials. (b) Error evolution.}
\label{fig:psdmf_S12}
\end{figure}

For $S_{12}\in\mathbb{R}^{12\times 12}$, the \ac{psd} rank is conjectured to be $K=5$~\cite{Vandaele2018}. The inner ranks are not known. The purpose of the following experiment is to provide evidence that the inner ranks can be equal to $R_A=3$ and $R_B=1$. We tested 20 \ac{MC} trials with stopping criteria CPU time 120[s] (24[s] for the \ac{CD} algorithms running in c). We set $D_{\textrm{FSVP}} = 10$ and $D_{\textrm{CGIHT}} = 3$. 
\Cref{fig:psdmf_S12} shows our results. \Cref{fig:psdmf_S12_MC20_K5_ra3_rb1_TolFun_0_TolFval_0_MaxCPUTime_120_MaxIter_1e6_LogErr} shows the final error when the stopping criterion is achieved, in 20 \ac{MC} trials. \Cref{fig:psdmf_S12_K5_ra3_rb1_TolFun_0_TolFval_0_MaxCPUTime_600_MaxIter_1e7_Trajectory_LogCPU} shows the error evolution of one such trial, stopping at 600[s] CPU time.
In this experiment, \ac{ABG}-P did not manage to decrease the objective function properly in all trials, whereas this happened to \ac{ABG} only occasionally. The other methods decreased the error reasonably. We note the fast decrease of the error of \ac{CGIHT} in~\cref{fig:psdmf_S12_K5_ra3_rb1_TolFun_0_TolFval_0_MaxCPUTime_600_MaxIter_1e7_Trajectory_LogCPU}, which means that the acceleration works properly in this case. These results provide evidence that this choice of inner ranks may indeed lead to exact factorization; however, experiments with \new{longer} run-time are required to determine this. \new{We mention that also for $S_{11}$ (not shown), we observed that a small reconstruction error can be achieved with $K=5$, $R_A=3$ and $R_B=1$.}

\begin{figure}[!t]
\centering
\includegraphics[width=0.6\columnwidth]{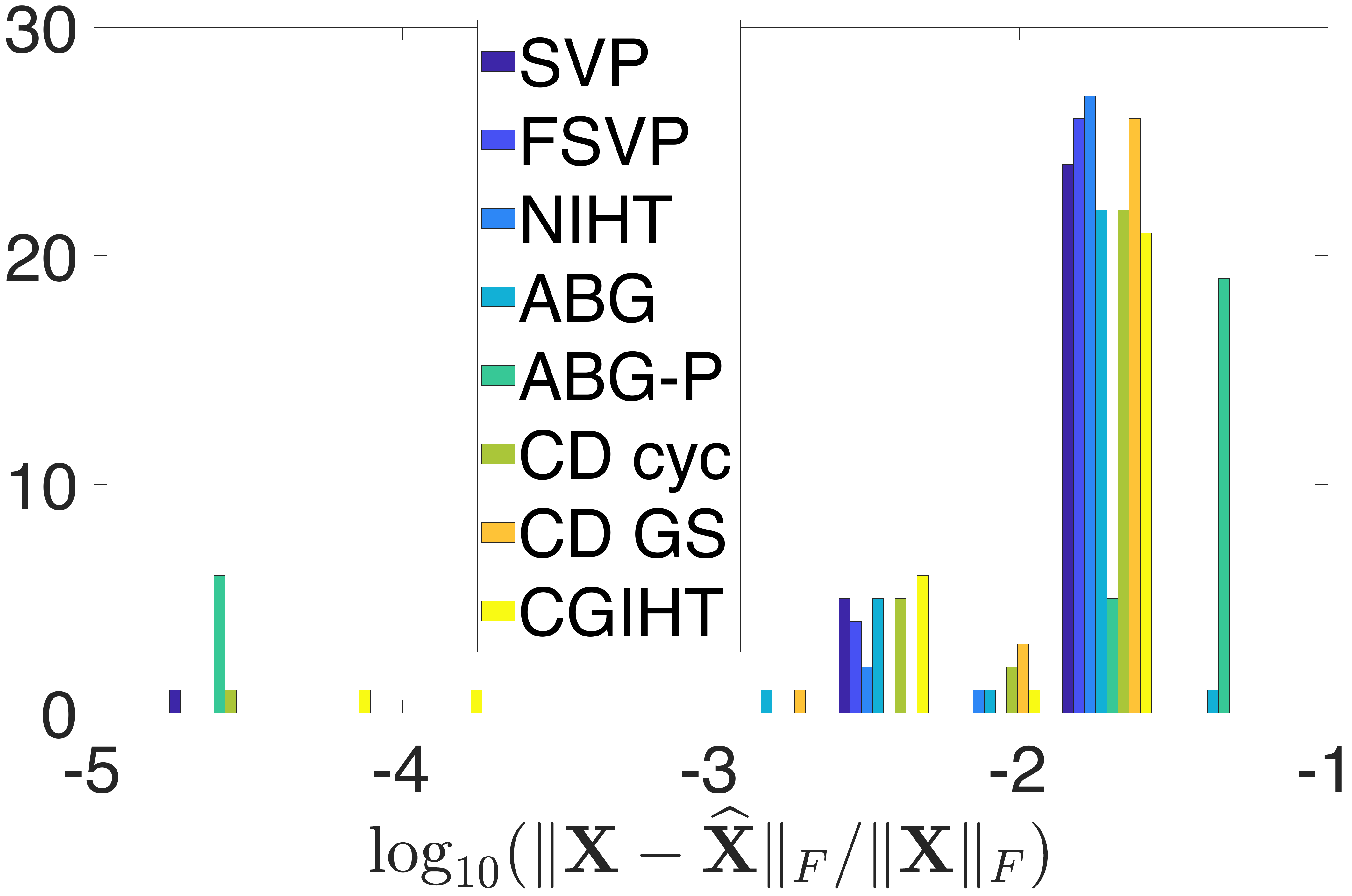}
\caption{\ac{PSDMF} of $S_8$ with $K=4$, $R_A=2$, $R_B=1$.  Histogram of final error in 30 \ac{MC} trials.}
\label{fig:psdmf_S8}
\end{figure}

\Cref{fig:psdmf_S8} shows results for factorizing $S_8\in\mathbb{R}^{8\times 8}$ with $K=4$, $R_A=2$, $R_B=1$. These inner ranks yield an exact factorization~\cite{Vandaele2018}. We set $D_{\textrm{FSVP}} = 3$ and $D_{\textrm{CGIHT}}= 9$. The stopping criteria are $\TolFun= 10^{-14}$ and $\TolRMFE= 1.9\cdot 10^{-5}$.
The histogram (with 10 bins) in~\cref{fig:psdmf_S8} shows the error when the stopping criterion is achieved. The successful trials, with \ac{RMFE}${}\leq{} 10^{-4}$, are:
6 successful trials for \ac{ABG}-P, 2 for \ac{CGIHT}, and 1 for \ac{SVP} and \ac{CD} cyclic each, out of 30 \ac{MC} trials.

\subsubsection{Subset Matrices}
\begin{figure}[!t]
\centering
\subfloat[]{\includegraphics[width=0.35\columnwidth]{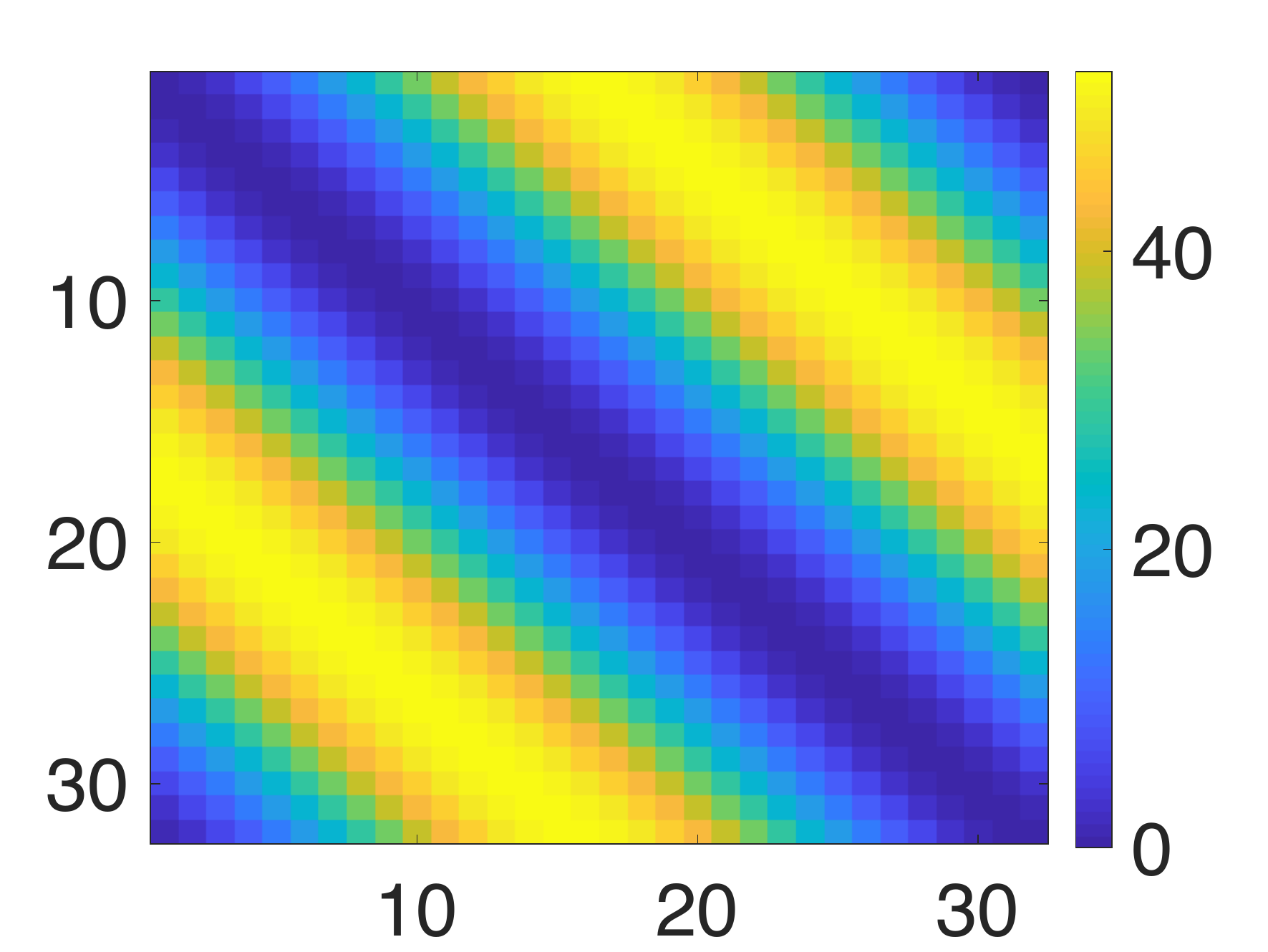}\label{fig:psdmf_S32_visualized}}
\hfil
\subfloat[]{\includegraphics[width=0.35\columnwidth]{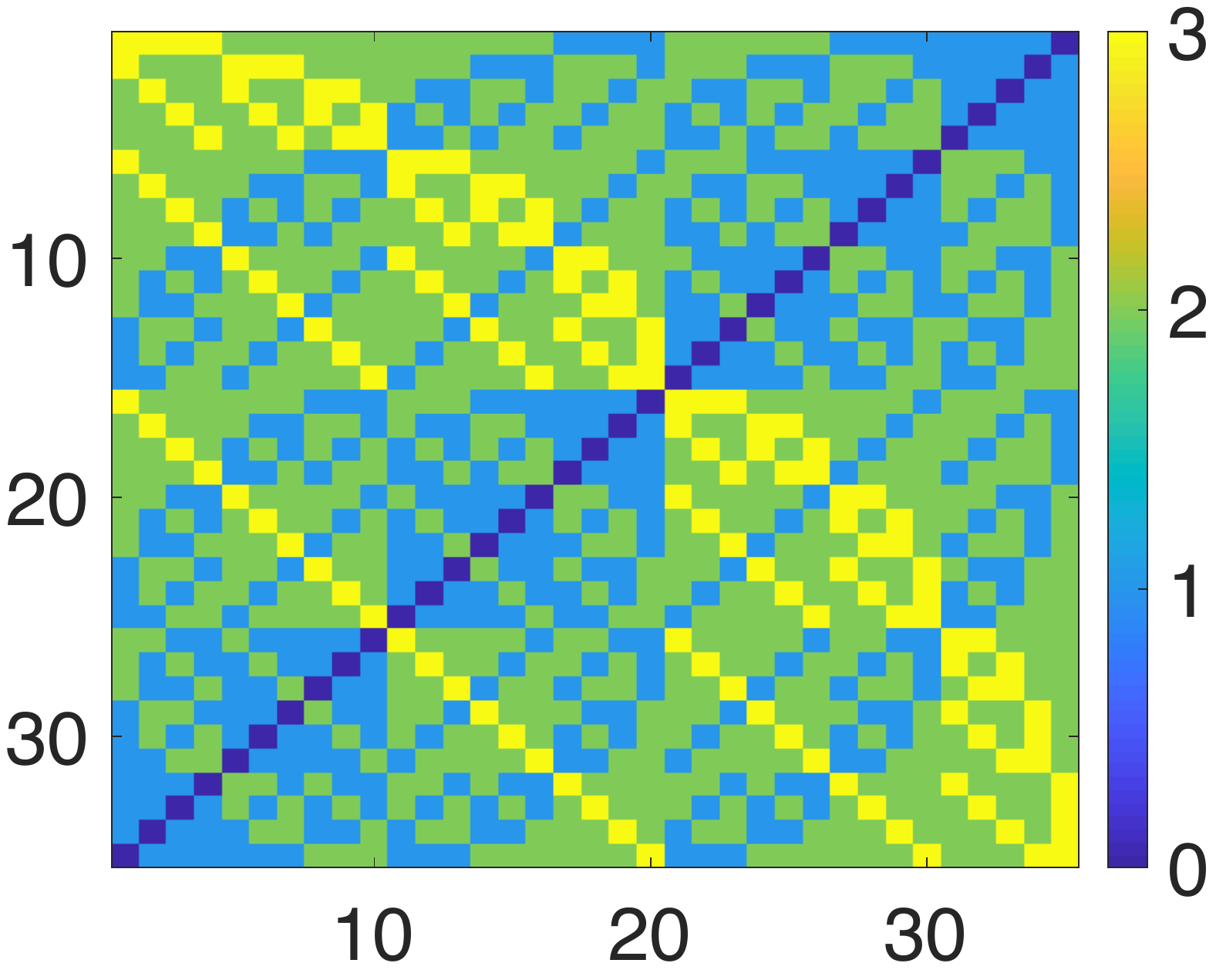}\label{fig:psdmf_P7_visualized}}
\\ \vspace{-1ex}
\subfloat[]{\includegraphics[width=0.48\columnwidth]{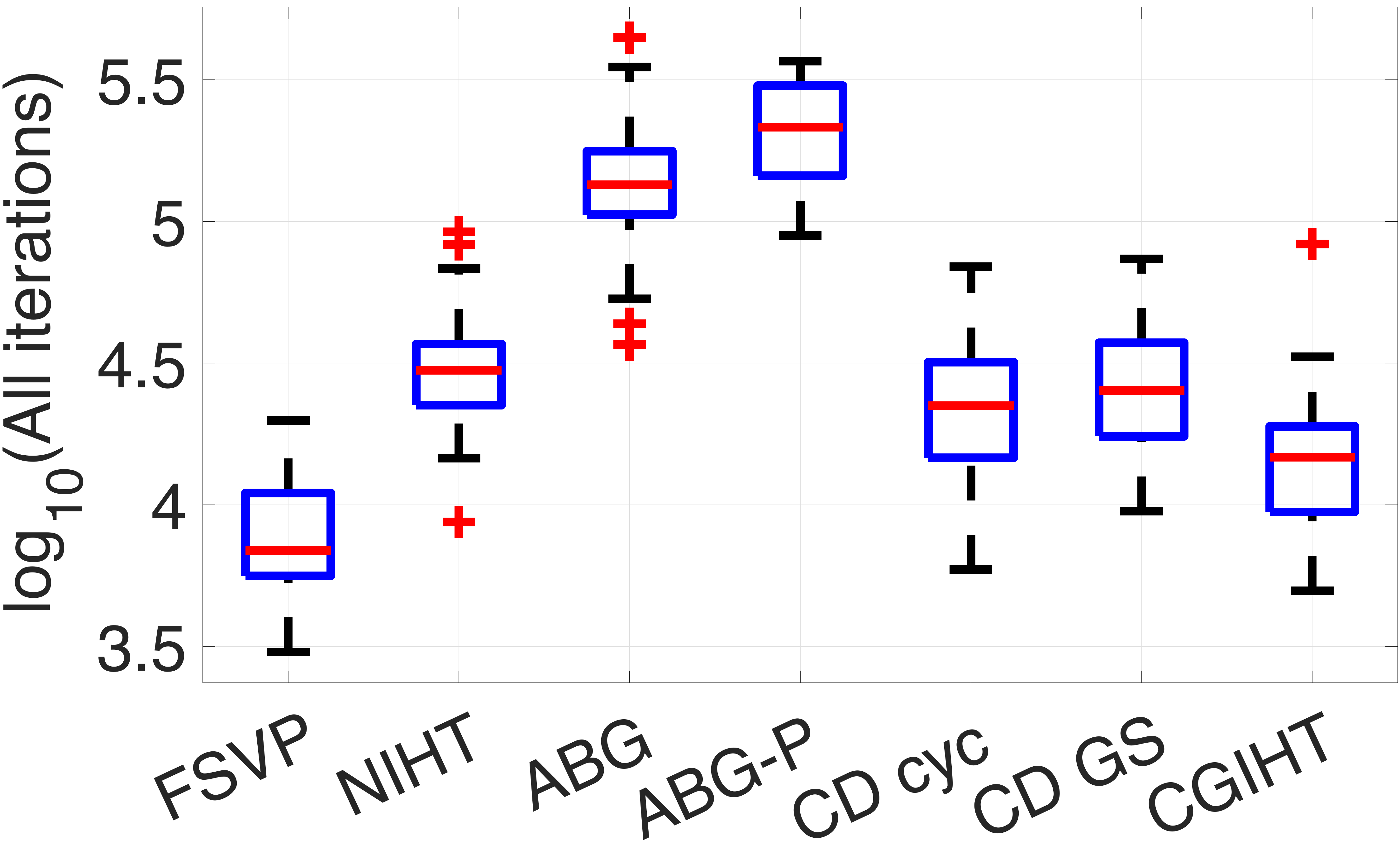}\label{fig:psdmf_S32_MC30_K6_ra3_rb3_TolFun_1e16_TolFval_6p05e7_MaxCPUTime_600_MaxIter_1e7_LogAllIter}}
\hfil
\subfloat[]{\includegraphics[width=0.48\columnwidth]{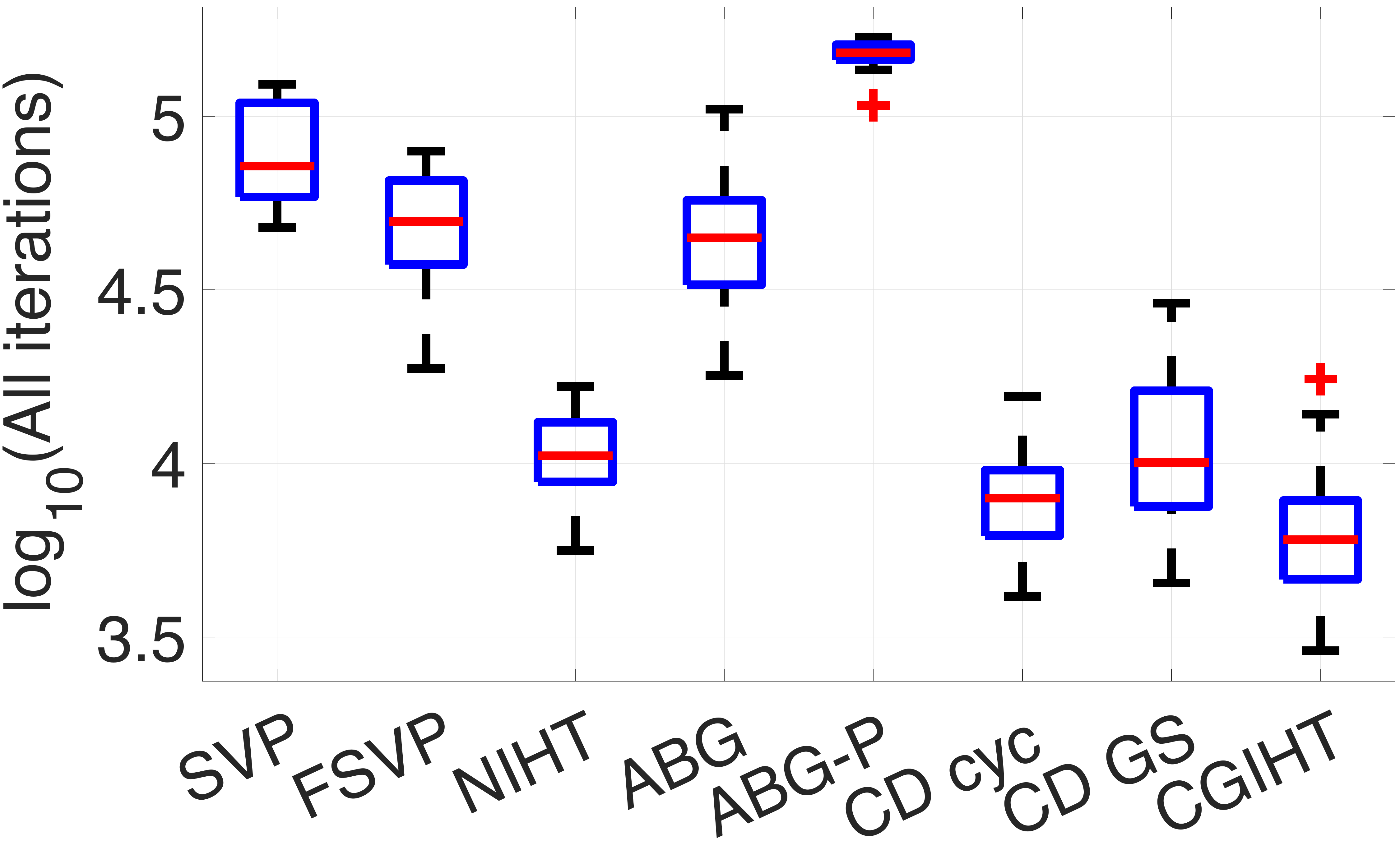}\label{fig:psdmf_Subset7_MC30_K6_ra3_rb3_TolFval_5e7_MaxCPUTime_900_MaxIter_1e6_LogAllIter}}
\\  \vspace{-1ex}
\subfloat[]{\includegraphics[width=0.48\columnwidth]{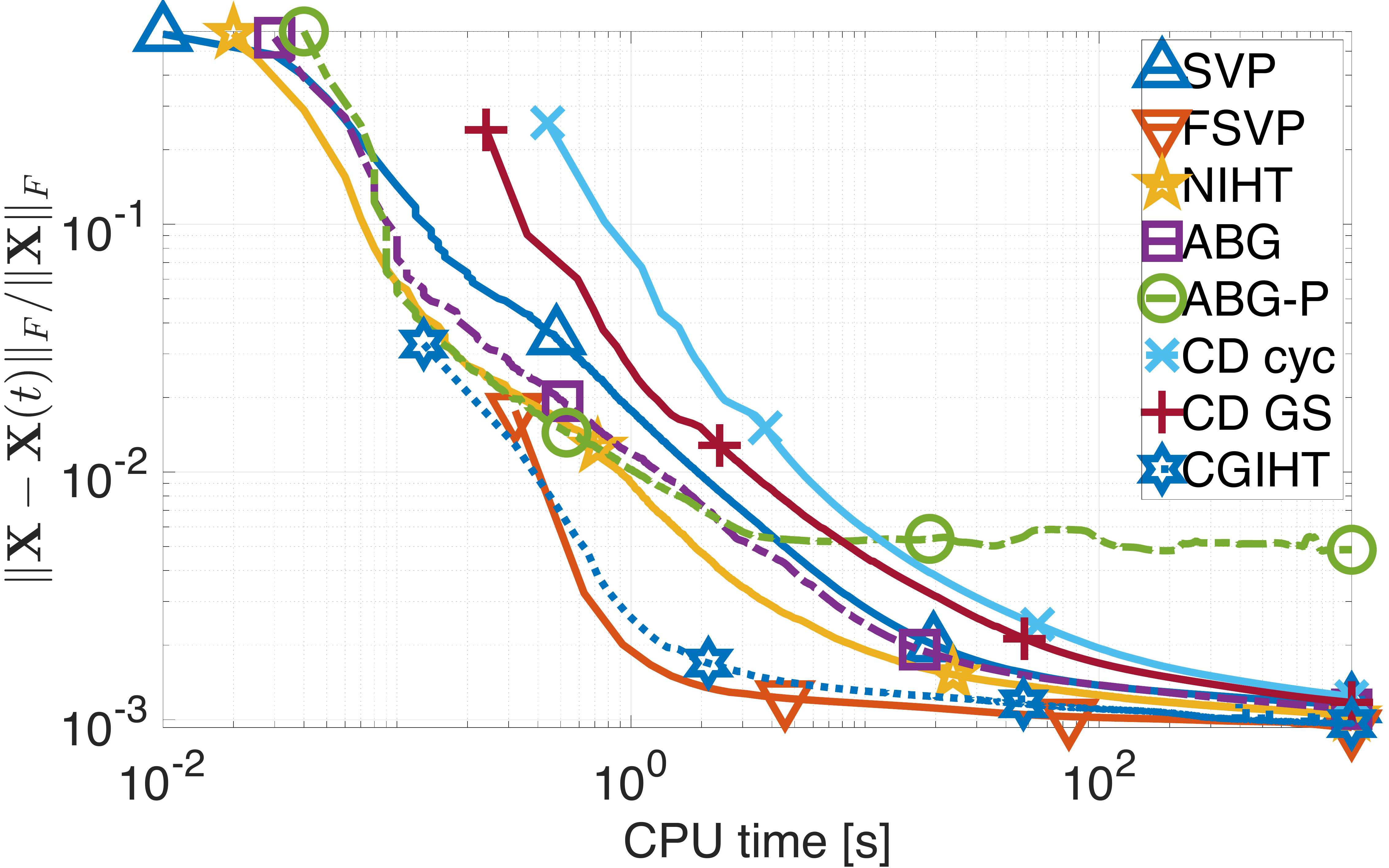}\label{fig:psdmf_S32_K6_ra3_rb3_TolFun_0_TolFval_0_MaxCPUTime_1200_MaxIter_1e6_Trajectory_LogCPU}}
\hfil
\subfloat[]{\includegraphics[width=0.48\columnwidth]{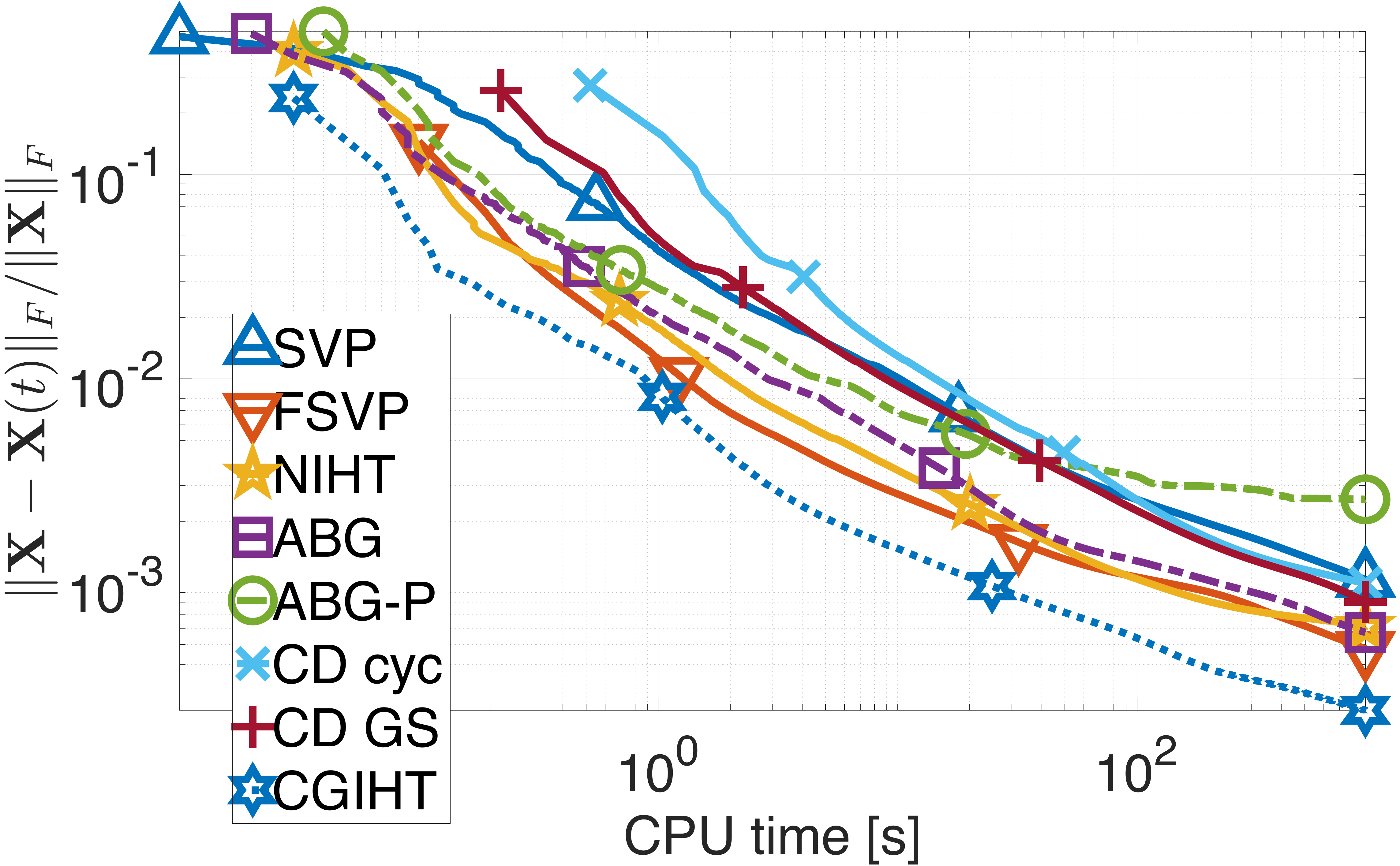}\label{fig:psdmf_Subset35_P7_K6_ra3_rb3_MaxCPUTime_900_MaxIter_Trajectory_LogAllIter}}
\caption{PSDMF of (left) $S_{32}$ and (right) $P_7$. (a) and (b) Visualization of $S_{32}$ and $P_7$, respectively. (c) and (d) Number of overall iterations in 30 \ac{MC} trials. (e) and (f) Error evolution in one trial.}
\label{fig:psdmf_S32_and_P7}
\end{figure}

\Cref{fig:psdmf_P7_visualized,fig:psdmf_Subset7_MC30_K6_ra3_rb3_TolFval_5e7_MaxCPUTime_900_MaxIter_1e6_LogAllIter,fig:psdmf_Subset35_P7_K6_ra3_rb3_MaxCPUTime_900_MaxIter_Trajectory_LogAllIter} show our results for another (generalized) slack matrix of interest, the \emph{subset matrix} $P_n$, described, e.g., in~\cite[Sec.~4.2]{Vandaele2018},~\cite[Problem~2--3]{Fawzi2015_PSD_Rank}. Here, we consider $P_7\in\mathbb{R}^{35\times 35}$, visualized in~\cref{fig:psdmf_P7_visualized}. We factorize $P_7$ with $K=6$, which is conjectured to be its \ac{psd} rank (e.g.,~\cite[Sec.~8.1]{Fawzi2015_PSD_Rank}). There are no conjectures about its inner ranks. However, based on the effect of zeros discussed in~\cref{sec:psdmf_zeros}, and the symmetry of $P_n$, we choose $R_A=3=R_B$ such that $R_A+R_B=K$. With these ranks, $N_{\textrm{model}}=1014 < N_{\textrm{data}}=1225$, which is a challenging setting due to the smaller number of model parameters versus number of constraints and the non-convexity of the objective function. We used $D_{\textrm{FSVP}}=24$ and $D_{\textrm{CGIHT}}=\new{3}$. \Cref{fig:psdmf_Subset7_MC30_K6_ra3_rb3_TolFval_5e7_MaxCPUTime_900_MaxIter_1e6_LogAllIter} shows our results for \ac{MC}${}={}30$ trials with different initializations. Our stopping criteria are $\TolRMFE= 10^{-3}$ and 900[s] in CPU time.
\Cref{fig:psdmf_Subset35_P7_K6_ra3_rb3_MaxCPUTime_900_MaxIter_Trajectory_LogAllIter} shows the error evolution for these settings, where each algorithm was stopped after 15 minutes of CPU time.

In this experiment, only once the \ac{ABG}-P algorithm managed to decrease the objective function properly. It is likely that \ac{ABG}-P converges differently due to the different objective function, which changes the optimization landscape.
In terms of the overall number of iterations in~\cref{fig:psdmf_Subset7_MC30_K6_ra3_rb3_TolFval_5e7_MaxCPUTime_900_MaxIter_1e6_LogAllIter}, we observe small average values for the \ac{CD} methods, \ac{NIHT} \new{and \ac{CGIHT}, the latter with the smallest average}. However, we should keep in mind that the update steps of \ac{ABG} are generally lighter, and that the computational complexity per update step of \ac{CD} is generally larger. This (at least partly) explains why, consistently in all our numerical experiments, \ac{CD} converged to the same target error with the highest CPU time, as shown also in~\cref{fig:psdmf_S32_K6_ra3_rb3_TolFun_0_TolFval_0_MaxCPUTime_1200_MaxIter_1e6_Trajectory_LogCPU,fig:psdmf_Subset35_P7_K6_ra3_rb3_MaxCPUTime_900_MaxIter_Trajectory_LogAllIter}.
The relations among the proposed projection-based methods are also in agreement with the theory: \ac{FSVP} has on average fewer overall iterations than \ac{SVP}, whereas \ac{NIHT} has fewer iterations \new{than both}, which means the property of \ac{NIHT} as a more numerically efficient method than \ac{SVP} can be inherited by the \ac{PSDMF} framework. \new{We also see that with a good choice of $D_{\textrm{CGIHT}}$, \ac{CGIHT} can indeed perform more efficiently than \ac{NIHT}.}
The CPU time differences between \ac{NIHT} and \ac{FSVP} are generally not so pronounced. \new{However, it is clear that \ac{CGIHT} is faster.}
\ac{ABG} is a relatively lightweight method in terms of \acl{CC} and number of operations per iteration. Hence, although its number of iterations can be higher than that of \ac{SVP}, its speed of convergence is generally closer to that of \ac{NIHT} and \ac{FSVP}.

\begin{figure}[!t]
\centering
\subfloat[]{\includegraphics[width=0.48\columnwidth]{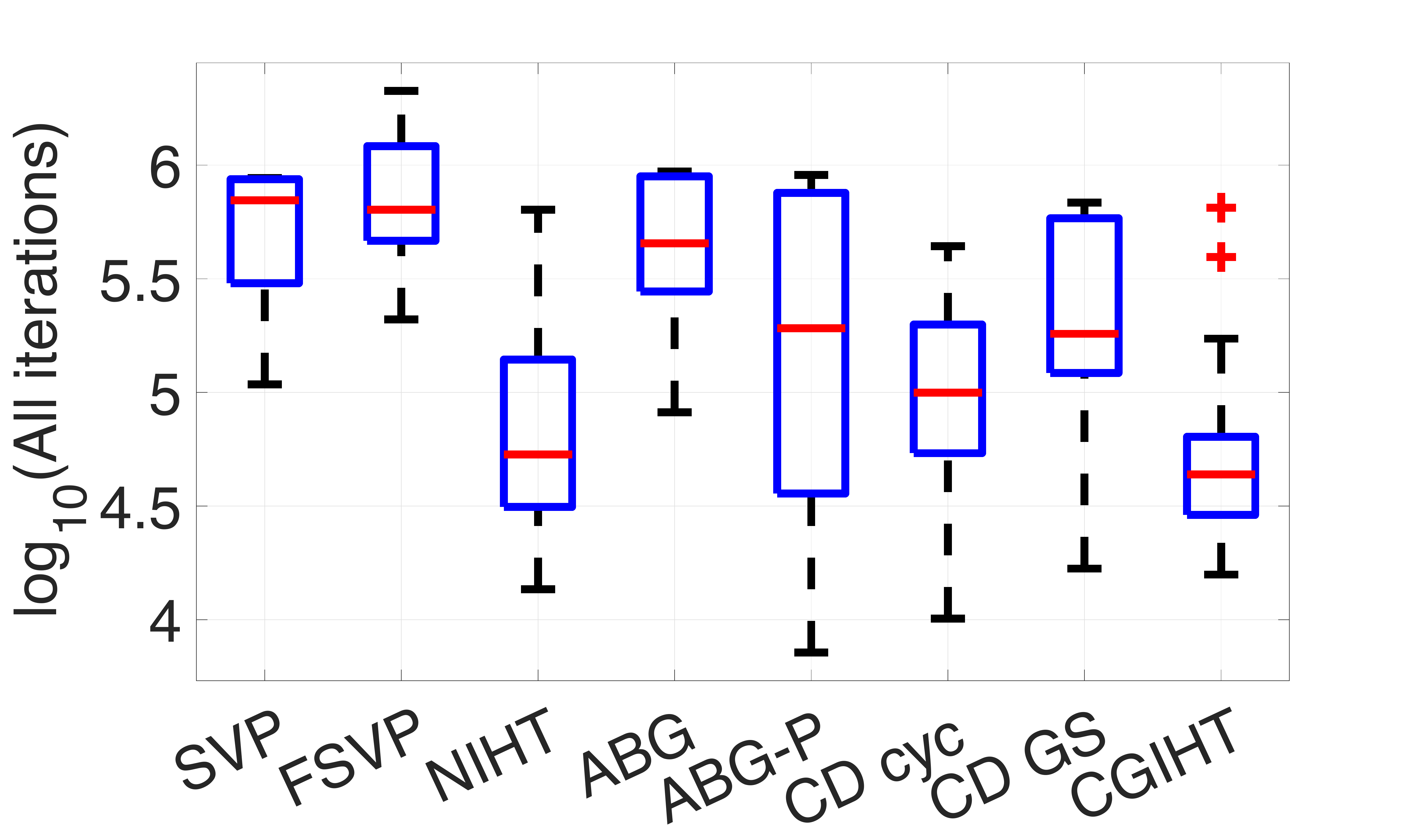}\label{fig:psdmf_Subset10_P5_MC30_K4_ra2_rb2_TolFun_0_TolFval_5e9_MaxCPUTime_900_MaxIter_1e7_LogIter}}
\hfil
\subfloat[]{\includegraphics[width=0.48\columnwidth]{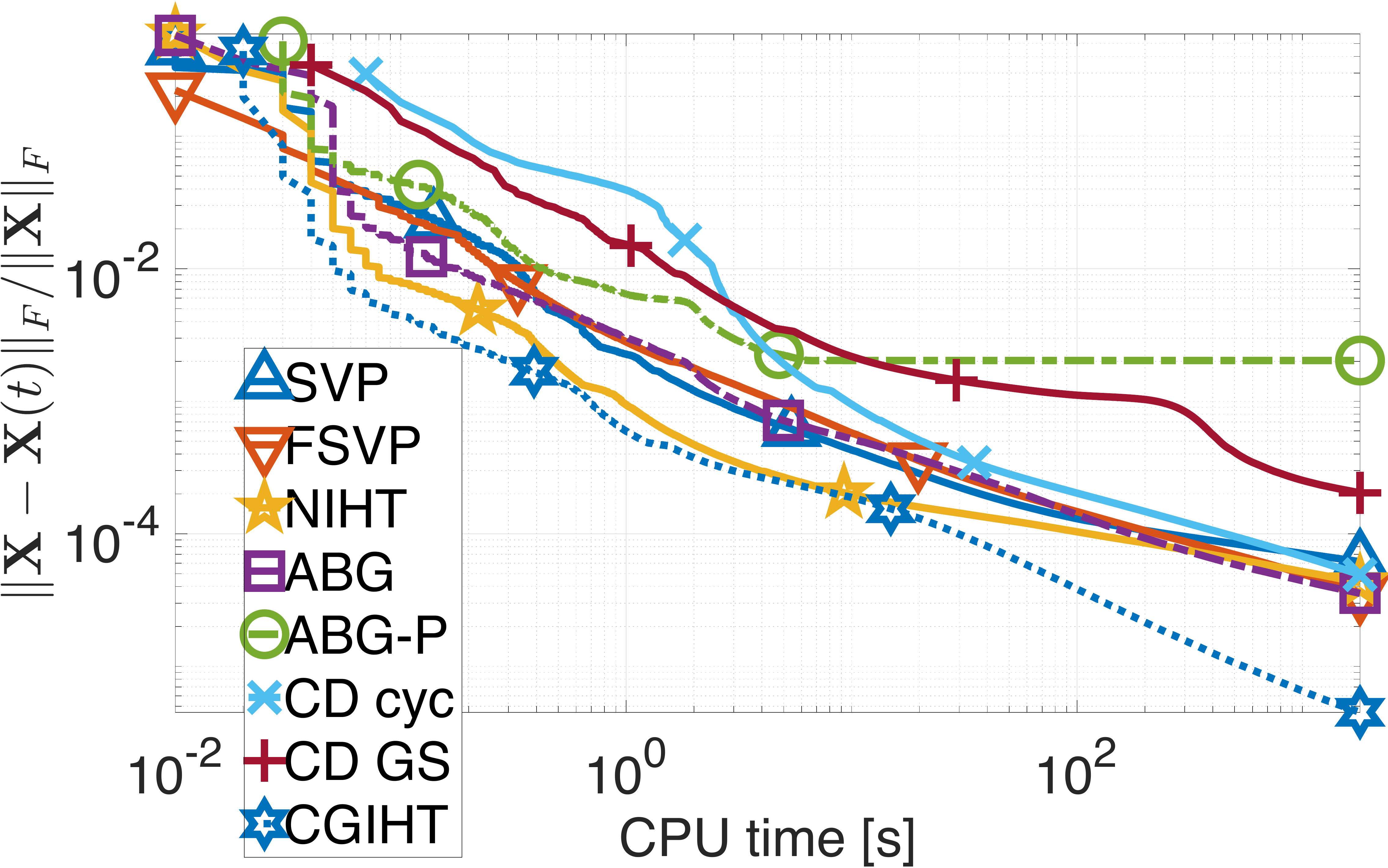}\label{fig:psdmf_Subset10_P5_K4_ra2_rb2_MaxCPUTime_1800_Trajectory_LogCPU}}
\caption{PSDMF of $P_5$. (a) Number of overall iterations in 30 \ac{MC} trials. (b) Error evolution.}
\label{fig:psdmf_P5}
\end{figure}

In~\cref{fig:psdmf_P5}, we show results for $P_5\in\mathbb{R}^{10\times 10}$. In this experiment, our stopping criterion is $\TolRMFE=10^{-4}$. We use $D_{\textrm{FSVP}}=\new{9}$ and $D_{\textrm{CGIHT}}=\new{2}$. \new{We set $K=4$ as conjectured by~\cite{Vandaele2018}. We choose $R_A=2=R_B$ due to the symmetry of $P_5$ and to satisfy $R_A+R_B\leq K$.} In~\cref{fig:psdmf_Subset10_P5_MC30_K4_ra2_rb2_TolFun_0_TolFval_5e9_MaxCPUTime_900_MaxIter_1e7_LogIter}, we observe that the number of overall iterations to achieve the \ac{RMFE} is smallest\new{, on average,} for \new{\ac{CGIHT}}. We remark that \ac{ABG}-P failed to decrease its objective function properly in \new{5} out of the 30 \ac{MC} trials. In this example, we observe that our choice of $D_{\textrm{FSVP}}$ \new{did not provide much improvement compared with its} non-accelerated counterpart. \new{These results provide supporting evidence that $P_5$ may have an exact factorization with these ranks.}
\new{For $P_6$ (not shown) we achieved, in certain trials, and after a sufficient number of iterations, small error with $K=5$ and $R_A=2=R_B$. Combined with our results for $P_7$ in~\cref{fig:psdmf_Subset35_P7_K6_ra3_rb3_MaxCPUTime_900_MaxIter_Trajectory_LogAllIter}, we conjecture that for $P_n$, $K=n-1$ and $R_A=R_B=\lfloor K/2\rfloor$ provide exact factorization.}

\subsection{Performance Comparison: Factorization of Dense Random Matrices}
\label{sec:numerical_experiments_random_matrix}

In this numerical experiment, our input is a $20\times 20$ matrix whose entries are drawn independently from the standard uniform distribution $\mathcal{U}[0,1]$. When normalized such that $\|\bX\|_1\defn\sum_{i,j}x_{ij}=1$, $\bX$ can be interpreted as a \ac{PMF} of two discrete \acl{RV}s, where one \acl{RV} takes $I$ values and the other $J$ values, such that each row or column of $\bX$ sums up to the marginal probability of each value given the other \acl{RV}.
This setup was addressed in~\cite{Glasser2019_ExpressiveTN_NIPS}, in the context of expressive power of \ac{PSDMF} (and its higher-order tensor network generalizations) in probabilistic modeling.

\begin{figure}[!t]
\centering
\subfloat[]{\includegraphics[width=0.48\columnwidth]{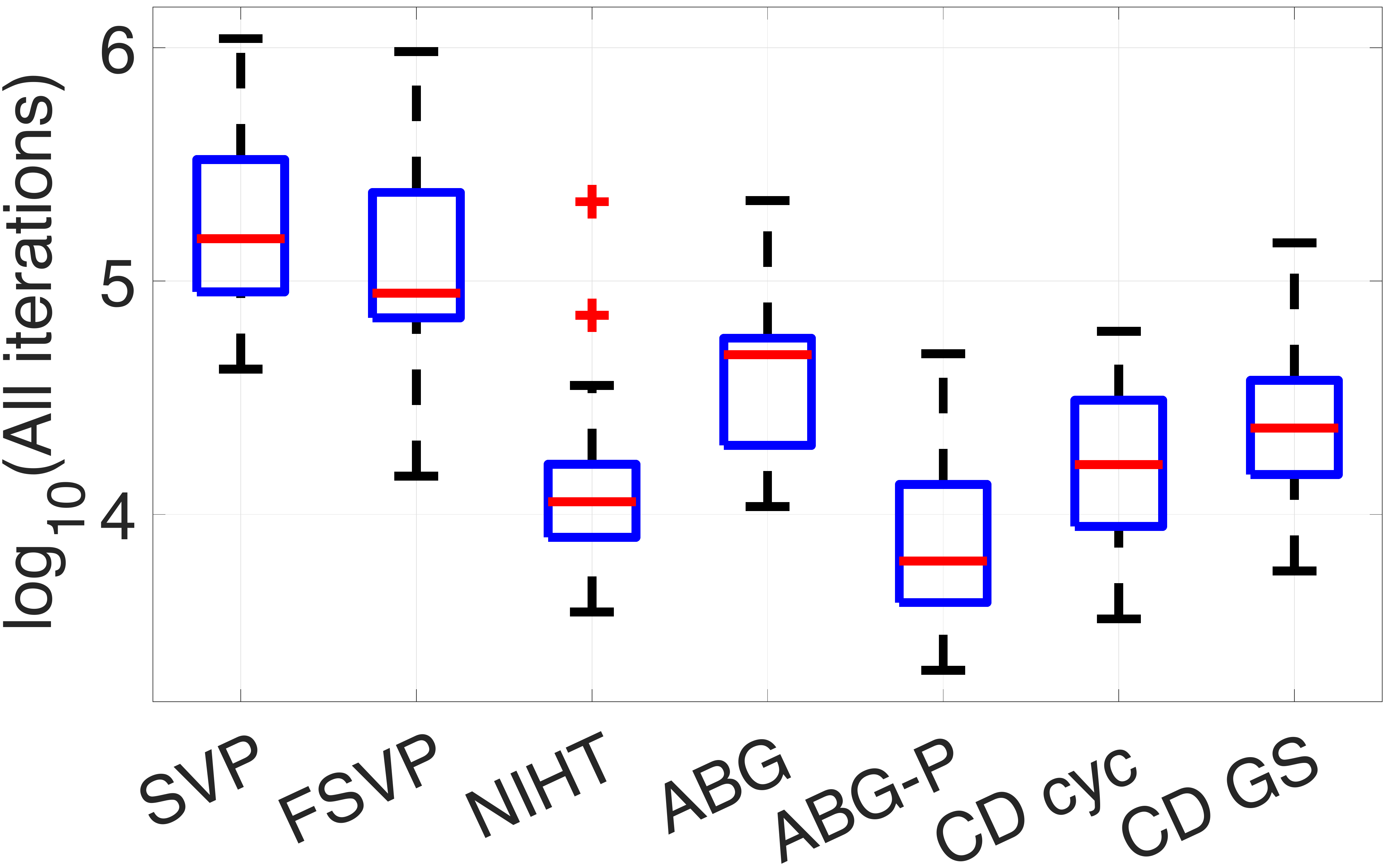}\label{fig:psdmf_20x20_MC30_K7_ra2_rb2_D_4_14_1_1_TolFval1e7_Xrand_doubleRand_Ginit_LogAllIter}}
\hfil
\subfloat[]{\includegraphics[width=0.48\columnwidth]{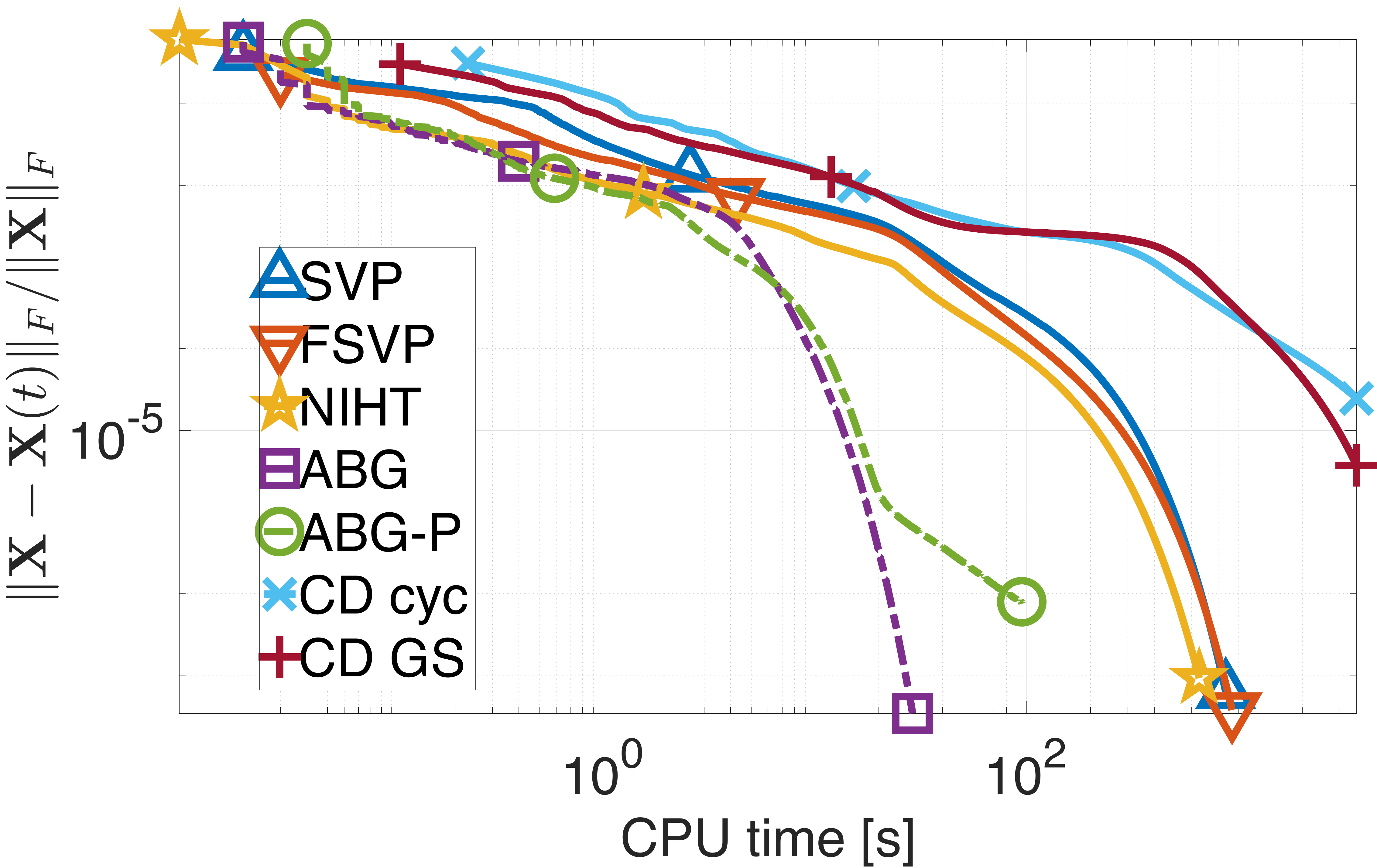}\label{fig:psdmf_20x20_K7_ra2_rb2_D_4_14_1_1_Ginit_Trajectory_CPU}}
\caption{PSDMF of a $20\times 20$ random matrix. (a) Number of overall iterations in 30 \ac{MC} trials. (b) Error evolution in one run.}
\label{fig:psdmf_20x20_K7_ra2_rb2}
\end{figure}

We fit $\bX$ to a \ac{PSDMF} model with $K=7$ and inner ranks all equal to 2, as in~\cite[Sec.~6.1]{Glasser2019_ExpressiveTN_NIPS}. In this setting, there are more free model variables than constraints: $N_{\textrm{data}}=400<N_{\textrm{model}}=471$ for $K=7$.
We run 30 \ac{MC} trials. In each trial, we generate a new random matrix $\bX$ and a new initialization. 
Our stopping criterion is $\TolRMFE= 10^{-4}$.
\new{We use $D_{\textrm{FSVP}}=14$. We did not find any value of $D_{\textrm{CGIHT}}$ with which \ac{CGIHT} outperformed \ac{NIHT}, in this scenario.}
\Cref{fig:psdmf_20x20_MC30_K7_ra2_rb2_D_4_14_1_1_TolFval1e7_Xrand_doubleRand_Ginit_LogAllIter} shows the number of overall iterations $\ell\times D$.
The error evolution as a function of CPU time is shown in~\cref{fig:psdmf_20x20_K7_ra2_rb2_D_4_14_1_1_Ginit_Trajectory_CPU}, for a randomly-chosen run, stopping at $\TolFun{}={}10^{-20}$.

In agreement with the results in~\cite{Glasser2019_ExpressiveTN_NIPS}, and our predictions from the model parameters, all trials achieved the target \ac{RMFE}. In terms of number of iterations, \ac{NIHT} has the smallest number of iterations among the projection-based methods, whereas \ac{SVP} has the largest, as expected from the theory. In agreement with the preceding experiments on geometric data and our \acl{CC} results, we observe that in general, \ac{CD} methods take the longest CPU time to reach the same \ac{RMFE} despite their relatively small number of iterations.
\Cref{fig:psdmf_20x20_K7_ra2_rb2} shows that in certain cases, \ac{ABG} and \ac{ABG}-P converges particularly fast in this setting.
\Cref{fig:psdmf_20x20_K7_ra2_rb2_D_4_14_1_1_Ginit_Trajectory_CPU} shows a case in which the two \ac{ABG} methods were the fastest to reach the designated error. Among the projection-based methods, \ac{NIHT} is the fastest, in agreement with its smaller number of overall iterations. This random matrix setting differs from the geometric data in~\cref{sec:numerical_experiments_geometrical} by (i) not containing zero entries in $\bX$ and (ii) having an excess of free model parameters versus number of constraints, which might explain the difference in convergence behavior compared with those in~\cref{fig:psdmf_S32_and_P7}. 
We mention that for $K=6$ and $R_A=2=R_B$, none of our methods managed to decrease the \ac{RMFE} below a rather large value, in agreement with~\cite[Sec.~6.1]{Glasser2019_ExpressiveTN_NIPS}, and although $N_{\textrm{data}}=400 < N_{\textrm{model}}=404$.

\subsection{Performance versus Level of Sparsity}
\label{sec:numerical_experiments_zeros_in_X}
In~\cref{sec:psdmf_zeros}, we showed how the presence of zeros in the input matrix $\bX$ imposes orthogonality constraints on the factor matrices.
In this experiment, we numerically illustrate the consequences of these constraints. The following example demonstrates that even a small percentage of zeros can result in \new{a} difference between a model that fits the data well relatively easy and another in which is it difficult to find a fit that is approximately exact. Indeed, this is so even if there is an excess of free variables compared with the number of entries in $\bX$.
We generate a $10\times 10$ matrix $\bX$ whose $(i,j){\textrm{th}}$ entry is drawn independently from the standard uniform distribution, i.e., $x_{ij}\sim\mathcal{U}[0,1]$. This matrix $\bX$ is fixed throughout the experiment except for the number and locations of the zeros we add to it that vary. 
We determine the number of zeros in $\bX$ using the parameter $0\leq p\leq 1$, where $p$ is the ratio of zeros and $p=0$ means no zeros in $\bX$.
The number of zeros in $\bX$ is $p I J$. We choose their $pIJ$ locations randomly, independently and uniformly, with the following caveat: if a row or column of $\bX$ contains only zeros, we draw new locations for all the $p I J$ zeros. We repeat this procedure if necessary. The reason is that if the $i{\textrm{th}}$ row of $\bX$ contains only zeros, it is impossible to obtain any constraints on $\bA_i$ and $\bU_i$, and similarly for $j$, and thus they can be chosen completely arbitrarily--- without needing any optimization. For each value of $p$ and at each \ac{MC} trial we draw new locations for the zeros. 
In this experiment, $p = 0, \frac{1}{I}, \frac{2}{I},\ldots,  \frac{8}{I}$. 
We fit $\bX$ to a \ac{PSDMF} model with $K=5$, $R_A\defn R_{A_i}=3$ for all $i$, $R_B\defn R_{B_j}=1$ for all $j$. 
This setup has $N_{\textrm{model}}=145$ free variables in the model that we try to fit to the $N_{\textrm{data}}=100$ observations (see~\cref{sec:psdmf_dof}). We also note that with probability 1, $\rank(\bX)=10$, which is smaller than $K(K+1)/2=15$, which is the maximal rank of an arbitrary matrix modeled with \ac{psd} rank $K=5$ (see~\cref{sec:psdmf_rank_decomposition_psd,sec:psdmf_rank_decomposition_factors}). We remind that if $\rank(\bX)>K(K+1)/2$, it is impossible to find an exact \ac{PSDMF} for $\bX$ with this value of $K$. These facts hint that fitting this model to $\bX$ should be easy---before we introduce the orthogonality constraints.
We run 20 \ac{MC} trials for each value of $p$, each with a new initialization. 
The stopping criterion is $\TolFun=10^{-13}$.

\begin{figure}[!t]
\centering
\subfloat[]{\includegraphics[width=0.48\columnwidth]{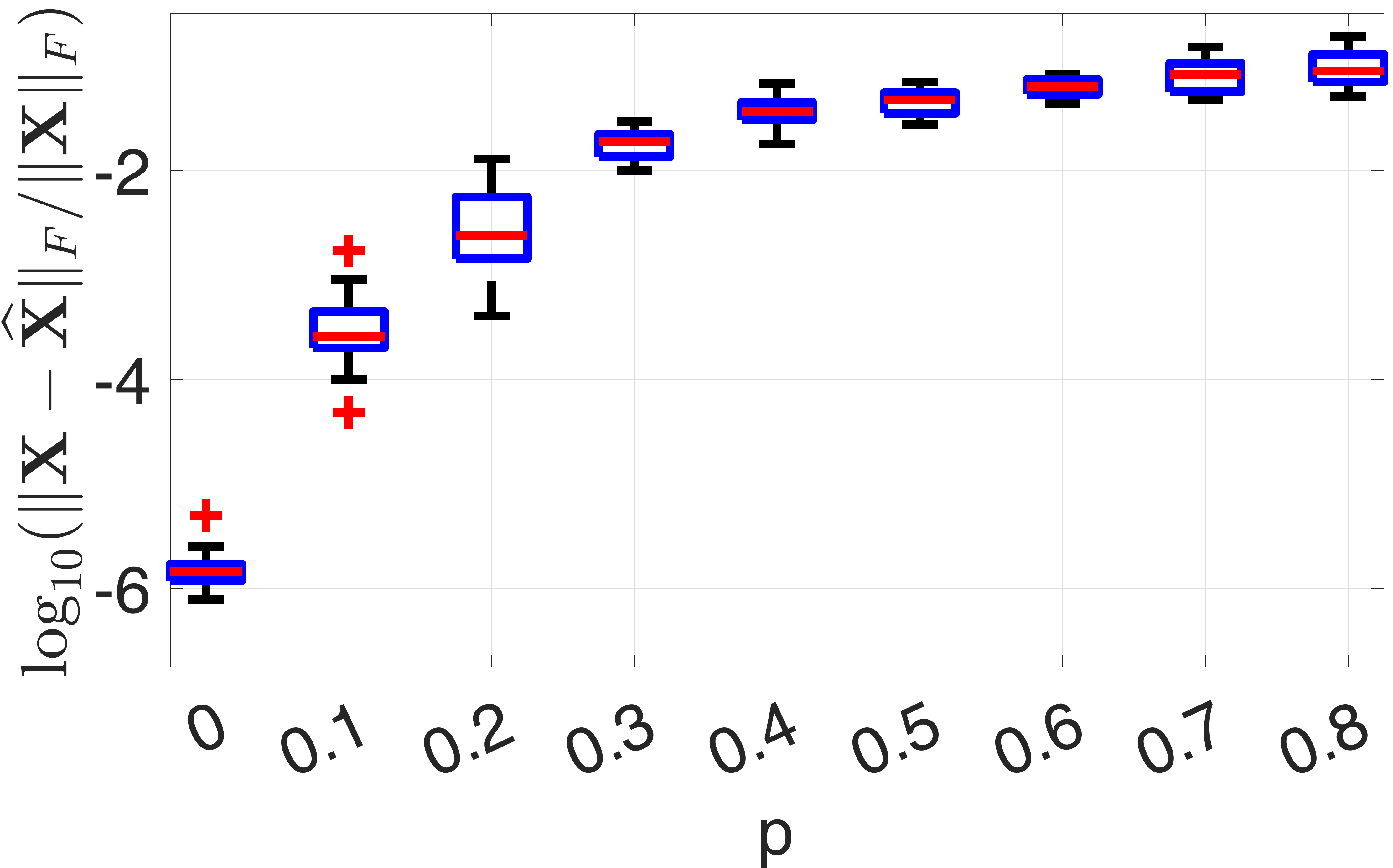}\label{fig:psdmf_SparseX_10x10_Xrand_MC20_K5_ra3_rb1_NIHT_LogErr}}
\hfil
\subfloat[]{\includegraphics[width=0.48\columnwidth]{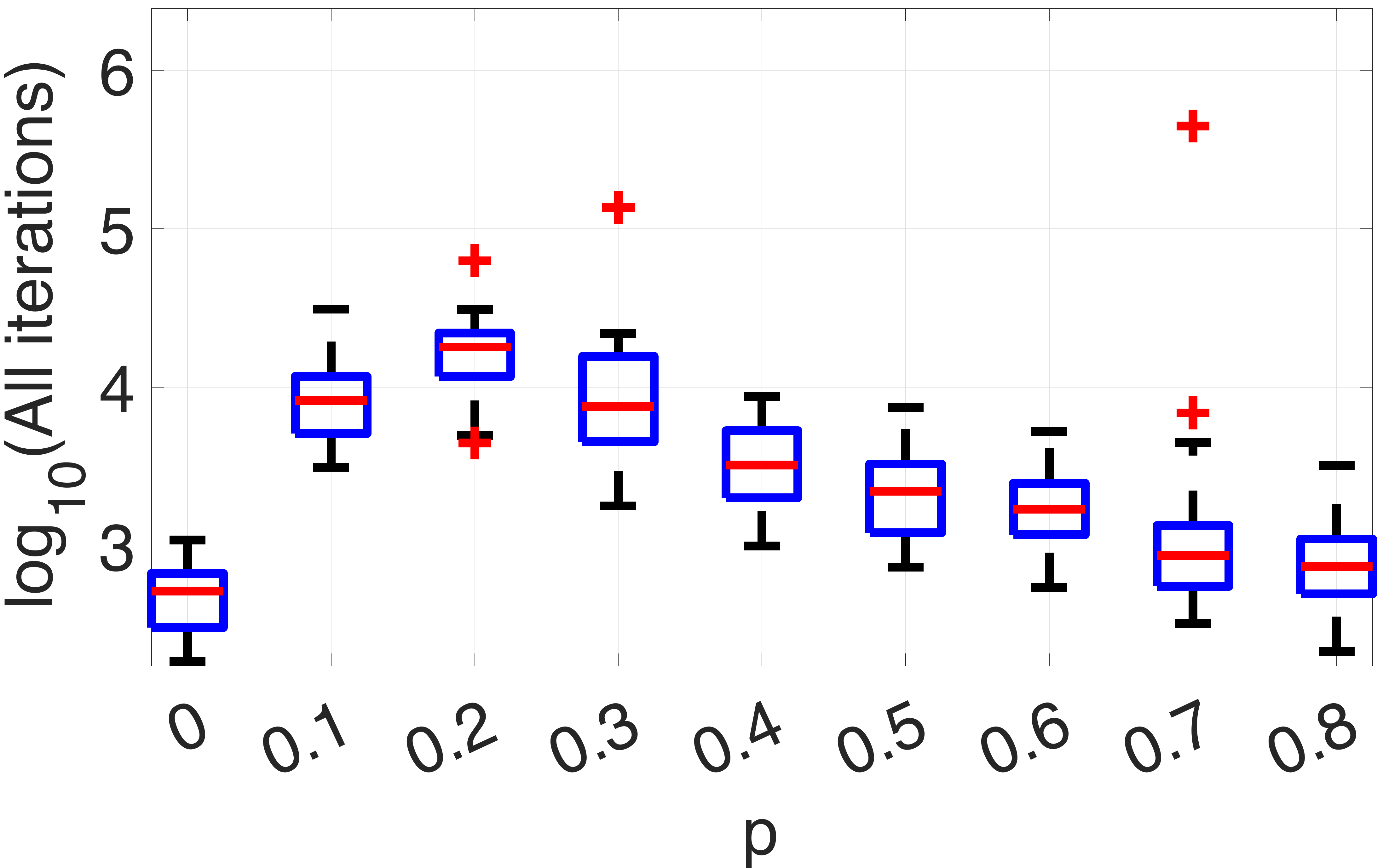}\label{fig:psdmf_SparseX_10x10_Xrand_MC20_K5_ra3_rb1_NIHT_LogAllIter}}
\caption{Influence of the proportion of zeros $p$ in a $10\times 10$ random matrix on the (a) model fit error, and (b) number of iterations.}
\label{fig:psdmf_10x10_MC20_K5_ra3_rb1_Xrand_SparseX}
\end{figure}

\Cref{fig:psdmf_10x10_MC20_K5_ra3_rb1_Xrand_SparseX} shows the results obtained using our \ac{NIHT}-based \ac{PSDMF} algorithm. Similar trends were observed with all other methods; we defer the remaining plots to~\cref{app:numerical_experiments_zeros_in_X} in the \acl{SM}.
The error in~\cref{fig:psdmf_10x10_MC20_K5_ra3_rb1_Xrand_SparseX} is calculated \ac{w.r.t.}~the value of $\bX$ at each iteration, including the zeros. 
\Cref{fig:psdmf_SparseX_10x10_Xrand_MC20_K5_ra3_rb1_NIHT_LogErr,fig:psdmf_SparseX_10x10_Xrand_MC20_K5_ra3_rb1_NIHT_LogAllIter} show the \ac{RMFE} and number of iterations, respectively, when the stopping criterion is achieved.
In the dense ($p=0$) case, the \ac{RMFE} is $\approx 10^{-6}$, which means the algorithm achieved a very good model fit, in accordance with the surplus of free model variables versus number of observations, and ``excess of rank''. This case also needed, in most \ac{MC} trials, the smallest number of iterations to achieve the stopping criterion.
\Cref{fig:psdmf_SparseX_10x10_Xrand_MC20_K5_ra3_rb1_NIHT_LogErr} shows that even with as few as $10\%$ of zeros, the algorithm does not manage to reduce the model fit error below a certain value, and this value increases with the proportion of zeros. This observation is in accordance with the theory.
The number of iterations in~\cref{fig:psdmf_SparseX_10x10_Xrand_MC20_K5_ra3_rb1_NIHT_LogAllIter} can be regarded as reflecting the ``effort'' the algorithm makes to fit the free model variables to the orthogonality constraints: with up to $20\%$ of zeros, the algorithm manages to satisfy the orthogonality constraints to a certain extent, due to the excess of free model variables over the number of observations. But, as the number of zeros increases, this task fails faster because there are too many orthogonality constraints to satisfy.


\section{Discussion}
\label{sec:discussion}
The main contribution of this paper is a connection between the new problem of \ac{PSDMF} and some canonical primitives in the recent signal processing literature such as \acl{PR} and \ac{ARM}. Based on this connection, we showed that families of \ac{PSDMF} algorithms can be readily derived from their \acl{PR} and \ac{ARM} counterparts. 
Extensive numerical experiments compared and contrasted the performance of the proposed algorithms on benchmark datasets, showing that our proposed methods can outperform state-of-the-art algorithms~\cite{Vandaele2018} in terms of their convergence rates, ability to avoid local stationary points, and computational complexities, in various cases.

From a practical point of view, we presented a collection of algorithms for \ac{PSDMF}. We showed that there is high variability among \ac{PSDMF} problems such that the same algorithm can behave differently on data of similar nature, for example, matrices generated by the same model. Our results show that there is no single algorithm that can achieve satisfying results on all data. 
Our fast method for prototyping new algorithms showed successful in that different algorithms that we designed based on different \acl{PR} methods achieved remarkable success on different matrices.
Therefore, we advise trying a number of algorithms from different families when addressing difficult \ac{PSDMF} problems.

As a practical advice which algorithms to choose and which algorithms are more likely to be successful in \ac{PSDMF}:
Methods of particularly high success rate on difficult problems, in our experiments, are \ac{NIHT} and \ac{ABG}-P: these algorithms distinguish themselves from the others by not having guarantees for a monotone decrease of the objective function (\ac{NIHT}) and a non-quadratic objective function (\ac{ABG}-P). It is possible that the fact that these methods do not ``play by the rules'' allows them to overcome local stationary points that other methods get stuck in more often, in certain cases, while this same property also explains why in other cases (other matrices or other initializations for the same matrix) these methods perform less satisfactorily than their competitors. 
As for \ac{CGIHT}, it has the disadvantage of needing a parameter $D$ that has to be fine-tuned. However, as we demonstrated, the effort sometimes pays as there exist scenarios in which \ac{CGIHT} can achieve high rates of success. 
Regarding \ac{FSVP}, we have shown that there are cases in which the acceleration is significant. However, one has to make the balance between the effort in computing an optimal $D$ and using a sub-optimal one, or another method.

These observations, and connection between \ac{PSDMF} and other signal processing primitives (\acl{PR} and \ac{ARM}), have, however, their limitations, as \ac{PSDMF} is a more challenging problem than \acl{PR} and \ac{ARM}, in certain respects, as detailed in~\cref{sec:psdmf_caveats_wrt_PR_ARM}. In particular, \ac{PSDMF} is highly non-convex, a challenge encountered in most, if not all, matrix factorization problems. This results in the problem of finding a good initialization for \ac{PSDMF} algorithms. For \acl{PR} and \ac{ARM}, good initialization methods---such as spectral initialization---have been found based on knowing the true ``dictionary'' $\mathcal{A}$. The analogue of the dictionary is unknown for \ac{PSDMF} problems. Indeed, we have discovered that methods that are fast and effective for \acl{PR} and \ac{ARM} are not always useful in the \ac{PSDMF} framework, as shown in some of our numerical experiments.

Nevertheless, revealing the connection between \ac{PSDMF}, \acl{PR} and \ac{ARM} opens the door to new algorithms and, more importantly, analyses. Like \ac{NMF}, alternating methods are almost always used to solve \ac{PSDMF} problems. Hence, we envision that some of the techniques used to analyze \ac{NMF} may also be of utility. 

Another potential impact of this work is in applications---especially those related to nonnegative and phaseless data. 
Our enrichment of the numerical tools available for \ac{PSDMF} optimization motivates considering these tools in existing and new application.
One natural extension of our work is applying our algorithms in quantum-based analysis of signal processing problems: for example, recommender systems~\cite{Stark2016_RecommenderQuantum}
or probabilistic models~\cite{Glasser2019_ExpressiveTN_NIPS}. Some of these applications may require additional constraints, e.g., normalization as in \ac{POVM}s. Another natural extension of our work is extending our algorithms from matrices to tensors, similarly to the algorithms for the tensor networks proposed by~\cite{Glasser2019_ExpressiveTN_NIPS}.
We leave these issues for future work. 


\balance

\clearpage

{\large Supplemental Material for ``Positive Semidefinite Matrix Factorization: A Connection with Phase Retrieval and Affine Rank Minimization''}, {Dana~Lahat, Yanbin Lang, Vincent Y. F. Tan,  C\'{e}dric F\'{e}votte.}

\section{PSDMF As A Sum of Rank-1 Terms}
\label{app:psdmf_sum_rank_1_terms}

In this section, we demonstrate how \ac{PSDMF} can be written as a sum of rank-1 terms.

\subsection{PSDMF As A Sum of Rank-1 Terms: The Positive Semidefinite Case}
\label{app:psdmf_sum_rank_1_terms_psd}

Next, we rearrange the rows of $\mathfrak{A}$ and $\mathfrak{B}$ such that their first $K$ rows contain the entries on the diagonals of the \ac{psd} matrices, which are nonnegative numbers. 
To do so, let $a_{kl}^{(i)}$ and $b_{kl}^{(j)}$ denote the $(k,l){\textrm{th}}$ entry of $\bA_i$ and $\bB_j$, resp., for $\indrange{k,l}{K}$, and collect the terms indexed by $l=k$ in the vectors:
\begin{align}
\!\!\ba_{kk}{}={}& \begin{bmatrix}
a_{kk}^{(1)} & \cdots & a_{kk}^{(I)}
\end{bmatrix}\transpose,\;
\bb_{kk}{}={} \begin{bmatrix}
b_{kk}^{(1)} & \cdots & b_{kk}^{(J)}
\end{bmatrix}\transpose.
\end{align}
Let $\mathfrak{A}_+\in\mathbb{R}_+^{K\times I}$ and $\mathfrak{B}_+\in\mathbb{R}_+^{K\times J}$ be the nonnegative matrices whose $k{\textrm{th}}$ rows are $\ba_{kk}\transpose$ and $\bb_{kk}\transpose$, resp..
Similarly, for $l<k$, define the vectors
\begin{align*}
\ba_{kl}{}\defn{}\sqrt{2}\begin{bmatrix}
a_{kl}^{(1)} & \cdots & a_{kl}^{(I)}
\end{bmatrix}\transpose
,\;
\bb_{kl}{}\defn{} \sqrt{2}\begin{bmatrix}
b_{kl}^{(1)} & \cdots & b_{kl}^{(J)}
\end{bmatrix}\transpose
\end{align*}	
and let $\mathfrak{A}_{\pm}\in\mathbb{R}^{((K-1)K/2)\times I}$ and $\mathfrak{B}_{\pm}\in\mathbb{R}^{((K-1)K/2)\times J}$ denote the matrices whose rows are $\ba_{kl}\transpose$ and $\bb_{kl}\transpose$, respectively, for $l<k$. The subscript ${\pm}$ reminds that these matrices generally contain both negative and positive values. However, they are structured in the sense that they must satisfy that $\bA_i$ and $\bB_j$ remain \ac{psd}.
Using the notations that we have just introduced, we can express $\bX$ as:
\begin{subequations}
\label{eq:psdmf_as_nmf_plus_structured_psd}
\begin{align}
\bX{}\cong{}& \mathfrak{A}\transpose\mathfrak{B}{}={} \begin{bmatrix}
\mathfrak{A}_+\transpose & \mathfrak{A}_{\pm}\transpose
\end{bmatrix}
\begin{bmatrix}
{\mathfrak{B}_+} \\ {\mathfrak{B}_{\pm}}
\end{bmatrix}\\
{}={}&
\mathfrak{A}_+{\mathfrak{B}_+}\transpose + \mathfrak{A}_{\pm}{\mathfrak{B}_{\pm}}\transpose\\
{}={}& \underbrace{\ba_{11}\bb_{11}\transpose+\cdots+\ba_{KK}\bb_{KK}\transpose}_{\textrm{sum of } K \textrm{ nonnegative rank-1 terms}}\label{eq:psdmf_as_nmf_plus_structured_psd_nonegative}\\
&{}+{} \underbrace{\ba_{12}\bb_{12}\transpose+\cdots+\ba_{(K-1),K}\bb_{(K-1),K}}_{\textrm{sum of } K(K-1)/2 \textrm{ rank-1 terms}}\\
{}={}& \underbrace{\bX_{\textrm{``NMF''}} + \bX_{\textrm{``structured''}}}_{\textrm{nonnegative}}\,.
\end{align}
\end{subequations}
\Cref{eq:psdmf_as_nmf_plus_structured_psd} demonstrates that a \ac{PSDMF} of a matrix $\bX$ with \ac{psd} rank $K$ can always be expressed as a sum of $K$ nonnegative rank-1 terms and up to $K(K-1)/2$ rank-1 terms that are allowed to take negative values.
Thus, if a given matrix $\bX$ can be expressed using a \ac{PSDMF} model with a certain value $K$, the usual matrix rank of $\bX$ is at most $K(K+1)/2$.

\subsection{PSDMF As A Sum of Rank-1 Terms: The Factor-Based Case}
\label{app:psdmf_sum_rank_1_terms_factors}
In analogy to the \ac{psd}-based representation in~\cref{eq:psdmf_as_nmf_plus_structured_psd}, we can write $\bX$ as a sum of rank-1 terms using the factor-based representation.
In order to have a simpler formulation, we temporarily assume that $R_{A_i}=R_A$ for all $i$ and $R_{B_j}=R_B$ for all $j$. We obtain:
\begin{align}
\label{eq:psdmf_as_nmf_plus_structured_factors}
\bX=&
\underbrace{\sum_{k=l}\balpha_{kk}\bbeta_{kk}\transpose}_{\mathclap{\textrm{sum of } K \textrm{ nonnegative rank-1 terms}}} + \overbrace{\sum_{k=1}^K \sum_{l\neq k} \balpha_{kl}\bbeta_{kl}\transpose}^{\mathclap{\textrm{sum of }K(K-1)/2 \textrm{ rank-1 terms}}}
= \underbrace{\bX_{\textrm{``NMF''}} + \bX_{\textrm{``structured''}}}_{\textrm{nonnegative}}
\end{align}
where
\begin{subequations}
\begin{align}
\balpha_{kl}{}\defn{}&  \sum_{r=1}^{R_A} \bu_{kr}\had\bu_{lr}\,\in\mathbb{R}^{I}\\
\bbeta_{kl}{}\defn{}& \sum_{s=1}^{R_B}\bv_{ks}\had\bv_{ls}\;\in\mathbb{R}^{J}\,,
\end{align}
\end{subequations}
$\had$ denotes the Hadamard (elementwise) product,
and
\begin{subequations}
\begin{align}
\bu_{kr}{}={}& \begin{bmatrix}
u_{kr}^{(1)}&
\cdots&
u_{kr}^{(i)}&
\cdots&
u_{kr}^{(I)}
\end{bmatrix}\transpose\;\in\mathbb{R}^{I}\\
\bv_{ks}{}={}& \begin{bmatrix}
v_{ks}^{(I)}&
\cdots &
v_{ks}^{(j)}&
\cdots &
v_{ks}^{(J)}
\end{bmatrix}\transpose\;\in\mathbb{R}^{J}\,,
\end{align}
\end{subequations} 
where $u_{kr}^{(i)}$, $\indrange{r}{R_A}$, and $v_{ks}^{(j)}$, $\indrange{s}{R_B}$, denote the $(k,r){\textrm{th}}$ and $(k,s){\textrm{th}}$ entry of $\bU_i$ and $\bV_j$, respectively.
We note that $\balpha_{kk}$ and $\bbeta_{kk}$ consist only of nonnegative values.

Similarly to~\cref{eq:psdmf_as_nmf_plus_structured_psd}, the model in~\cref{eq:psdmf_as_nmf_plus_structured_factors} shows that when $\bX$ can be expressed as a sum of $K$ or more nonnegative rank-1 terms only, the factorization is equivalent to \ac{NMF}, and there is no need for the \ac{PSDMF} framework. \Cref{eq:psdmf_as_nmf_plus_structured_psd,eq:psdmf_as_nmf_plus_structured_factors} provide further evidence that the usual matrix rank of $\bX$ is at most $K(K+1)/2$. 
In fact, except for very special cases (see, e.g.,~\cite{Fawzi2015_PSD_Rank}), the usual matrix rank of $\bX$ with \ac{psd} rank $K$ is $K(K+1)/2$.

\section{Performance versus Level of Sparsity--Additional Results}
\label{app:numerical_experiments_zeros_in_X}
\begin{figure*}[!ht]
\centering
\subfloat[SVP]{\includegraphics[width=0.48\columnwidth]{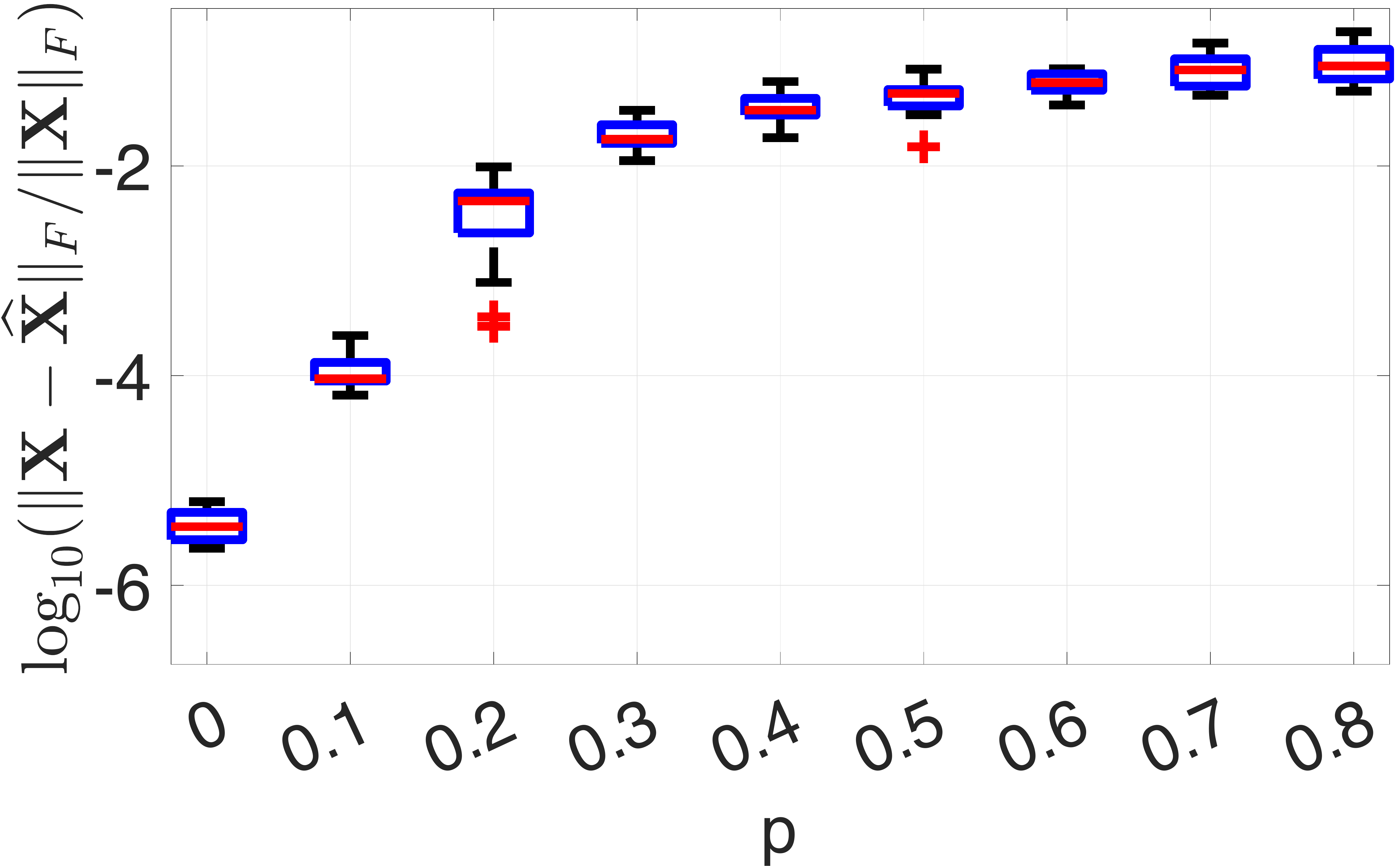}\label{fig:psdmf_SparseX_10x10_Xrand_MC20_K5_ra3_rb1_TolFun_1e13_TolFval_0_MaxCPUTime_600_MaxIter_1e6_SVP_LogErr}}
\hfil
\subfloat[FSVP]{\includegraphics[width=0.48\columnwidth]{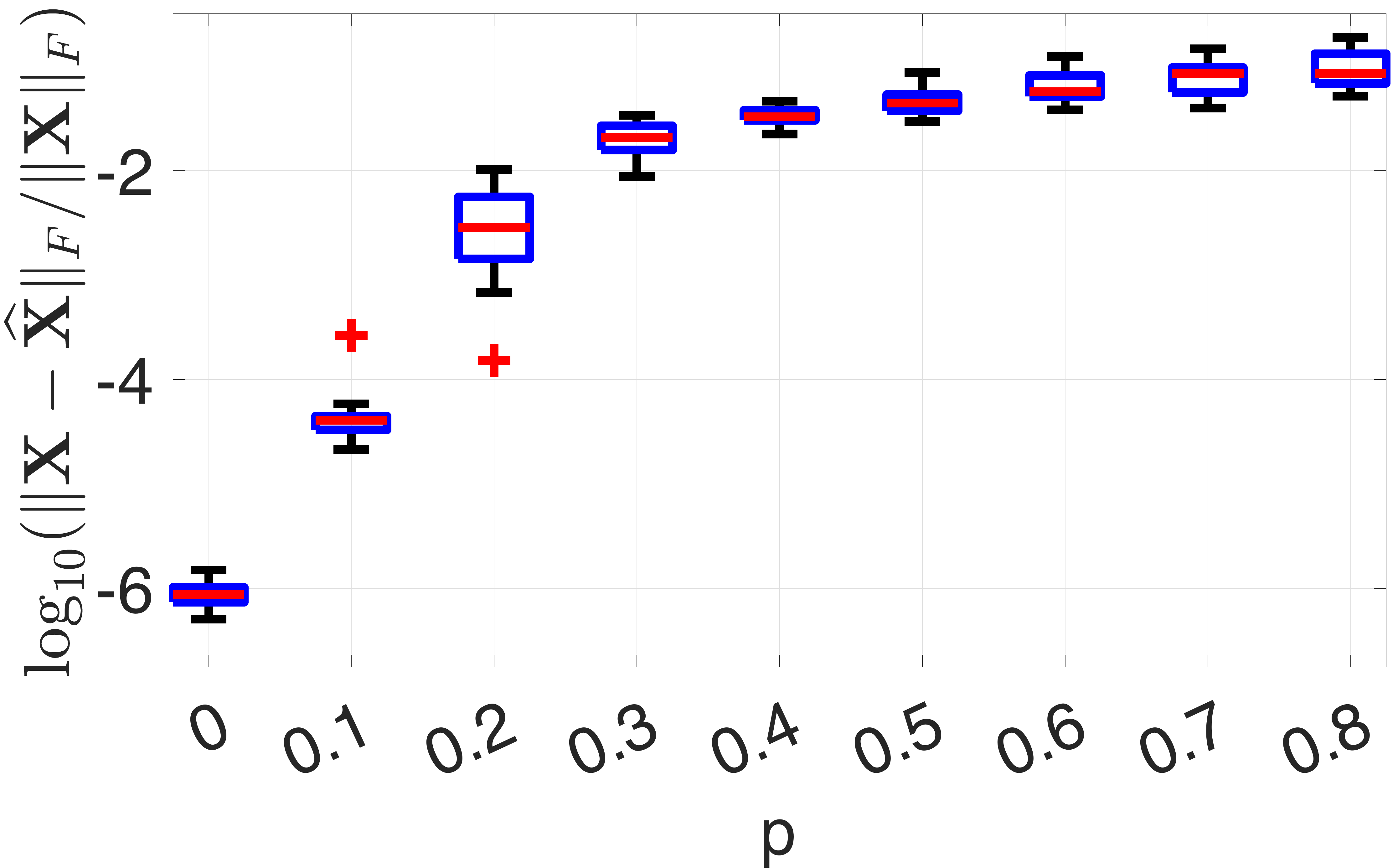}\label{fig:psdmf_SparseX_10x10_Xrand_MC20_K5_ra3_rb1_TolFun_1e13_TolFval_0_MaxCPUTime_600_MaxIter_1e6_FSVP_LogErr}}
\hfil
\subfloat[NIHT]{\includegraphics[width=0.48\columnwidth]{\gpath psdmf_SparseX_10x10_Xrand_MC20_K5_ra3_rb1_TolFun_1e13_TolFval_0_MaxCPUTime_600_MaxIter_1e6_NIHT_LogErr}\label{fig:psdmf_SparseX_10x10_Xrand_MC20_K5_ra3_rb1_TolFun_1e13_TolFval_0_MaxCPUTime_600_MaxIter_1e6_NIHT_LogErr}}
\hfil
\subfloat[ABG]{\includegraphics[width=0.48\columnwidth]{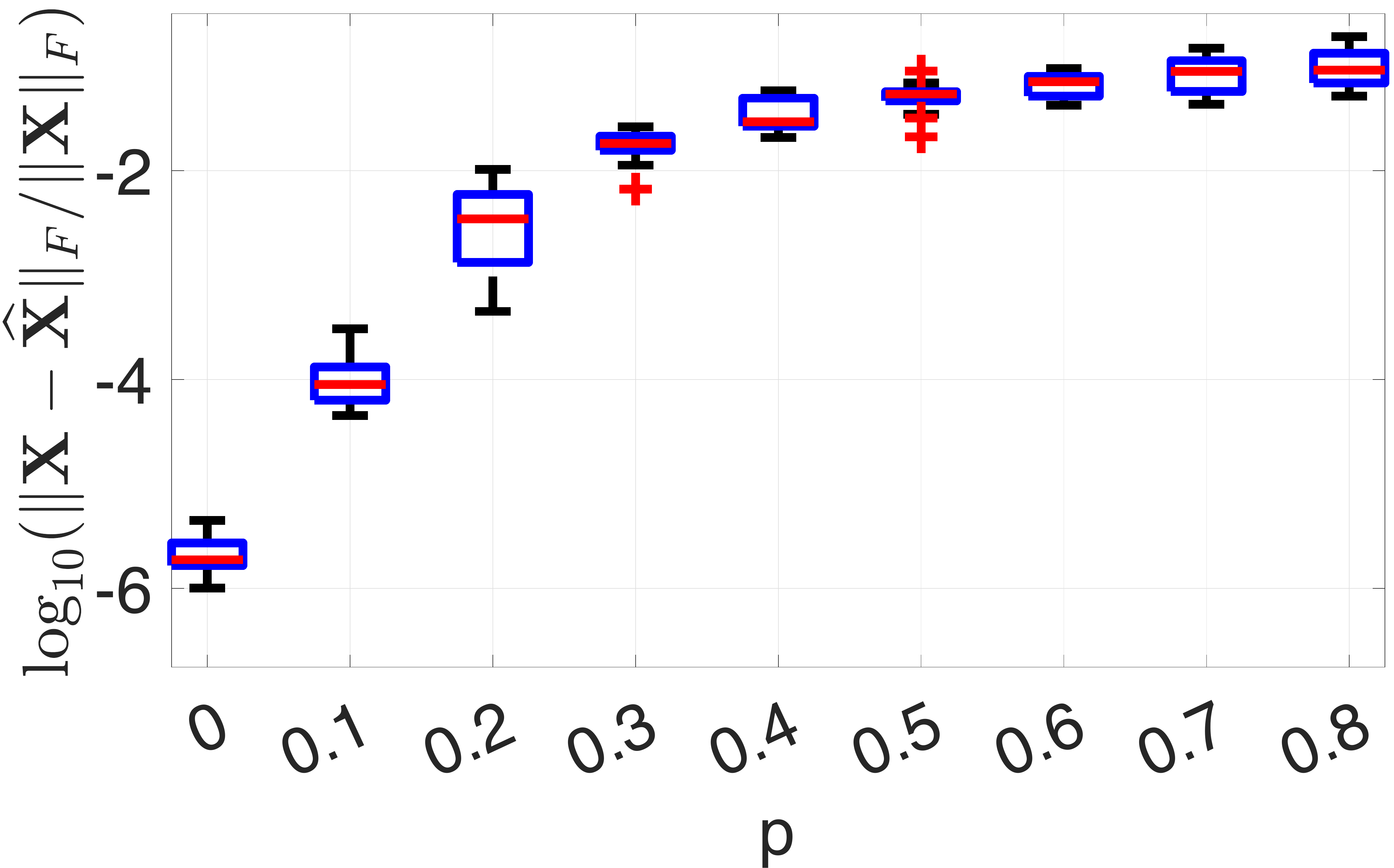}\label{fig:psdmf_SparseX_10x10_Xrand_MC20_K5_ra3_rb1_TolFun_1e13_TolFval_0_MaxCPUTime_600_MaxIter_1e6_ABG_LogErr}}
\hfil
\subfloat[ABG-P]{\includegraphics[width=0.48\columnwidth]{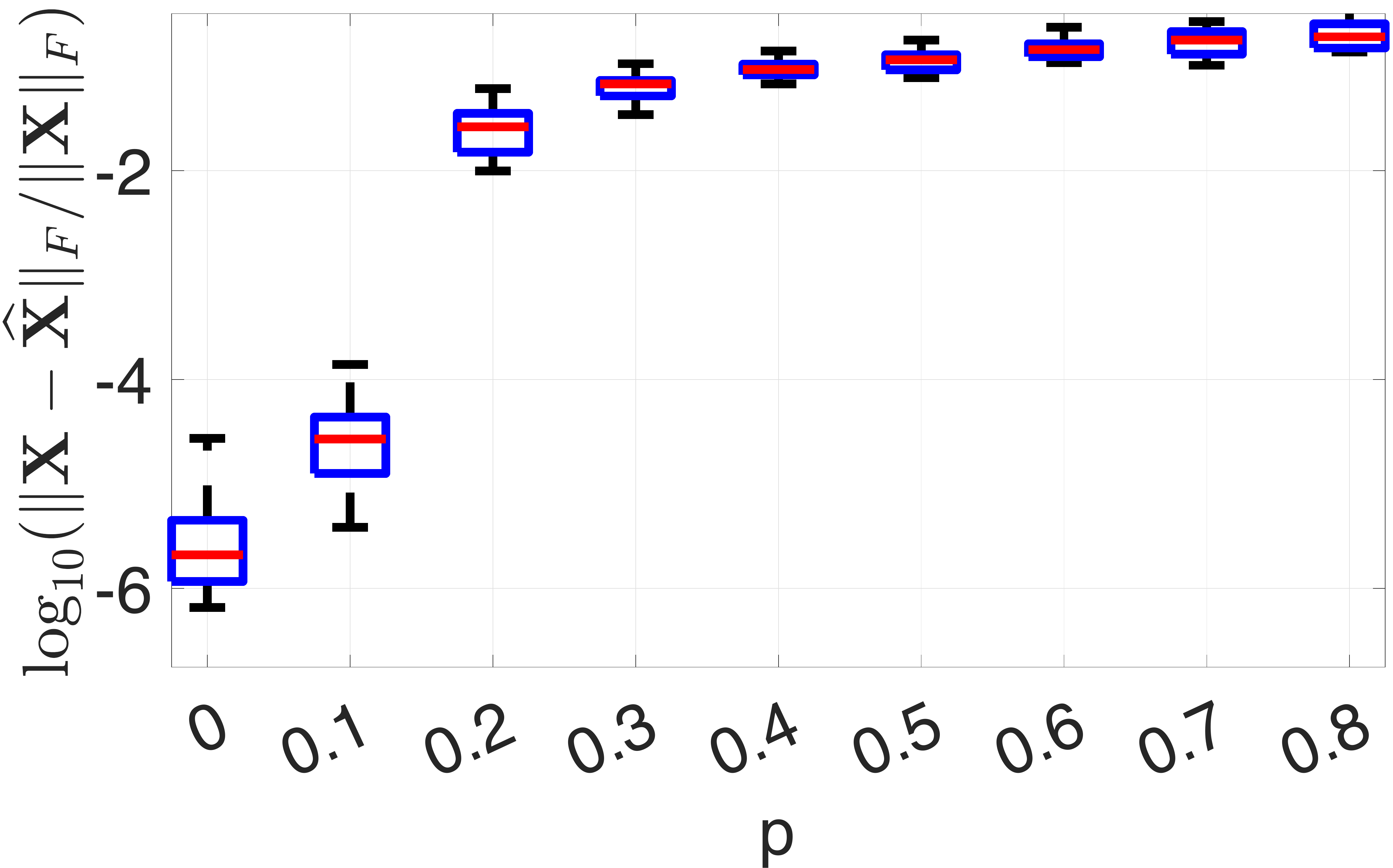}\label{fig:psdmf_SparseX_10x10_Xrand_MC20_K5_ra3_rb1_TolFun_1e13_TolFval_0_MaxCPUTime_600_MaxIter_1e6_ABGP_LogErr}}
\hfil
\subfloat[CGIHT]{\includegraphics[width=0.48\columnwidth]{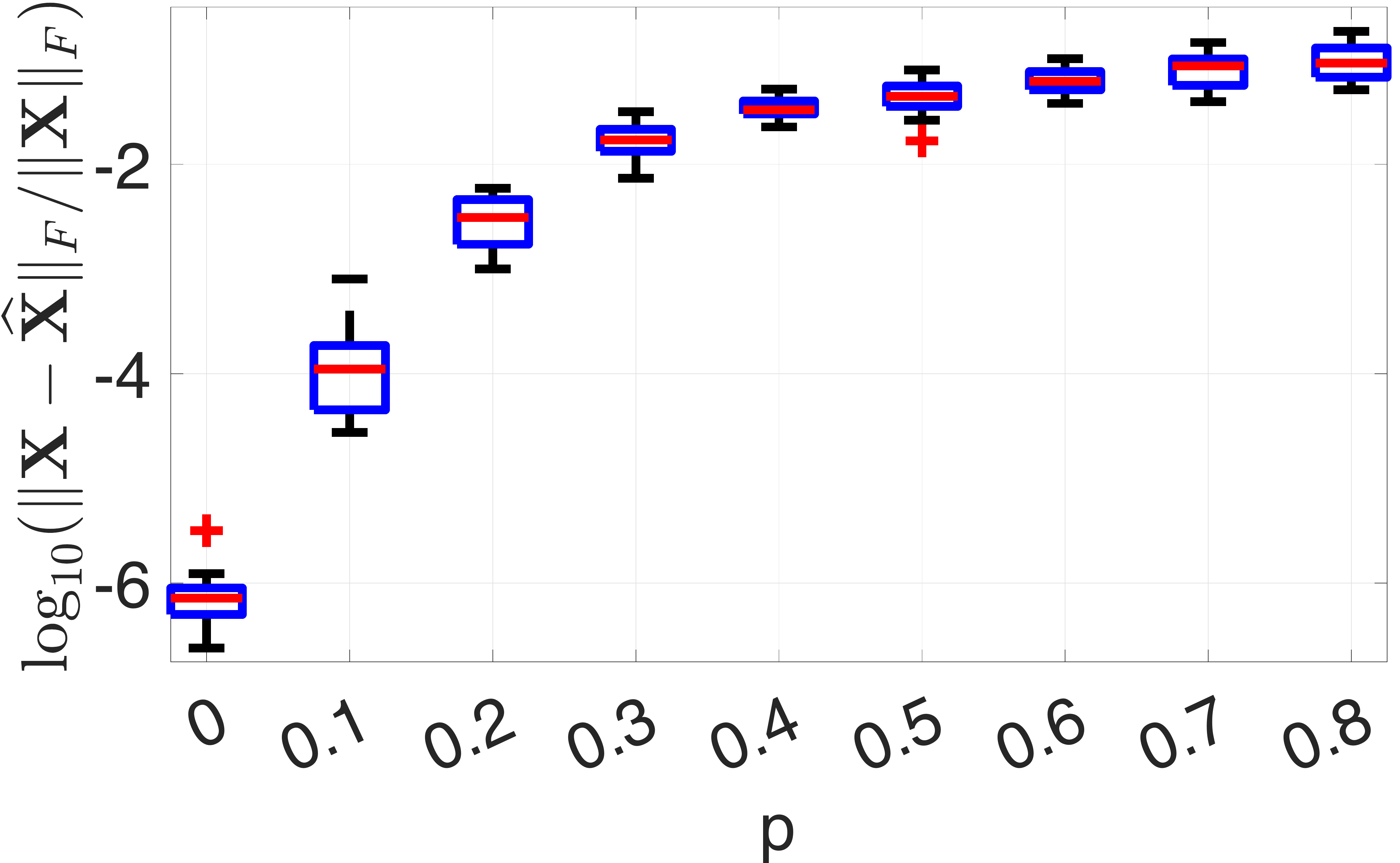}\label{fig:psdmf_SparseX_10x10_Xrand_MC20_K5_ra3_rb1_TolFun_1e13_TolFval_0_MaxCPUTime_600_MaxIter_1e6_CGIHT_LogErr}}
\hfil
\subfloat[CD cyc]{\includegraphics[width=0.48\columnwidth]{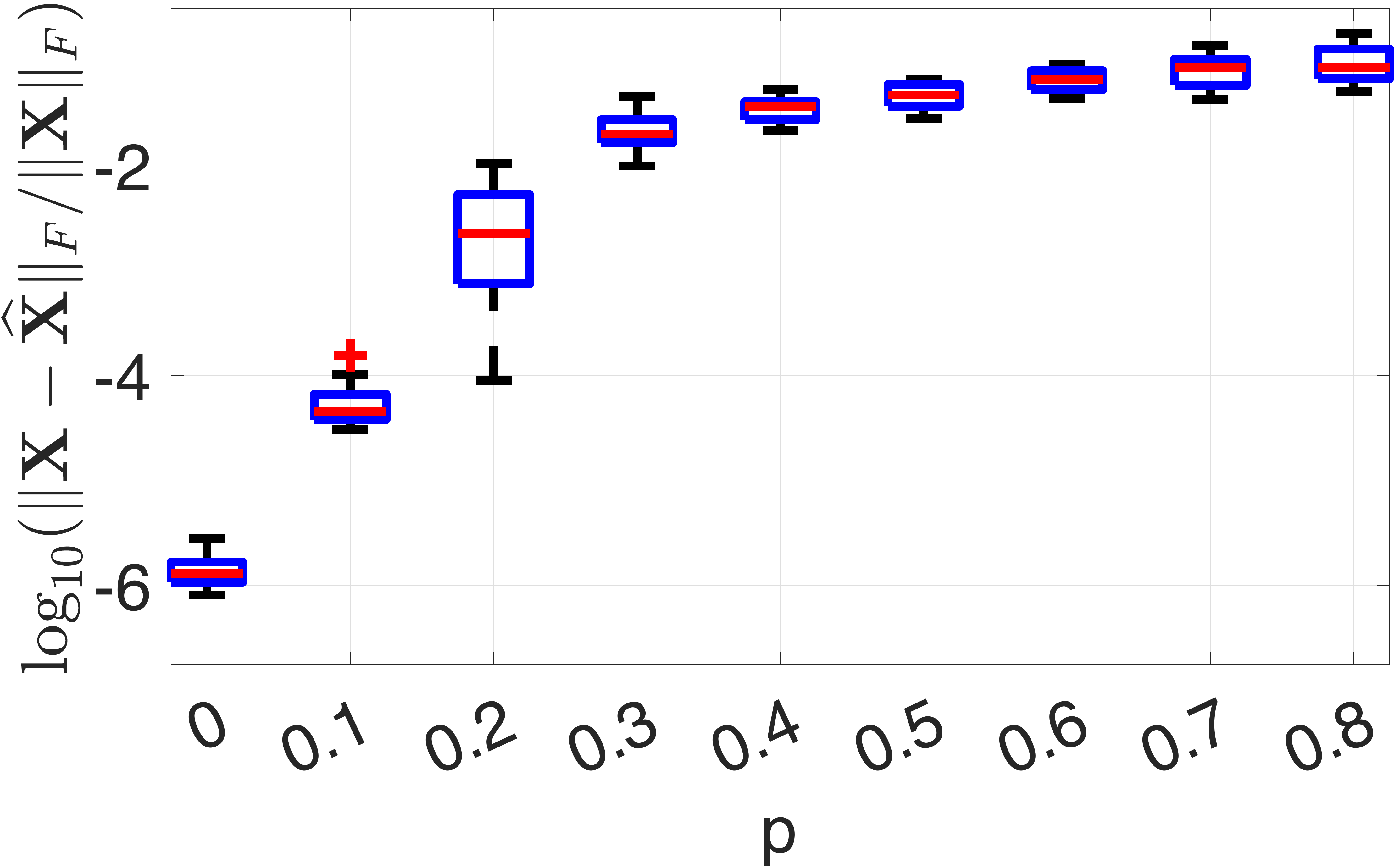}\label{fig:psdmf_SparseX_10x10_Xrand_MC20_K5_ra3_rb1_TolFun_1e13_TolFval_0_MaxCPUTime_600_MaxIter_1e6_CDcyc_LogErr}}
\hfil
\subfloat[CD GS]{\includegraphics[width=0.48\columnwidth]{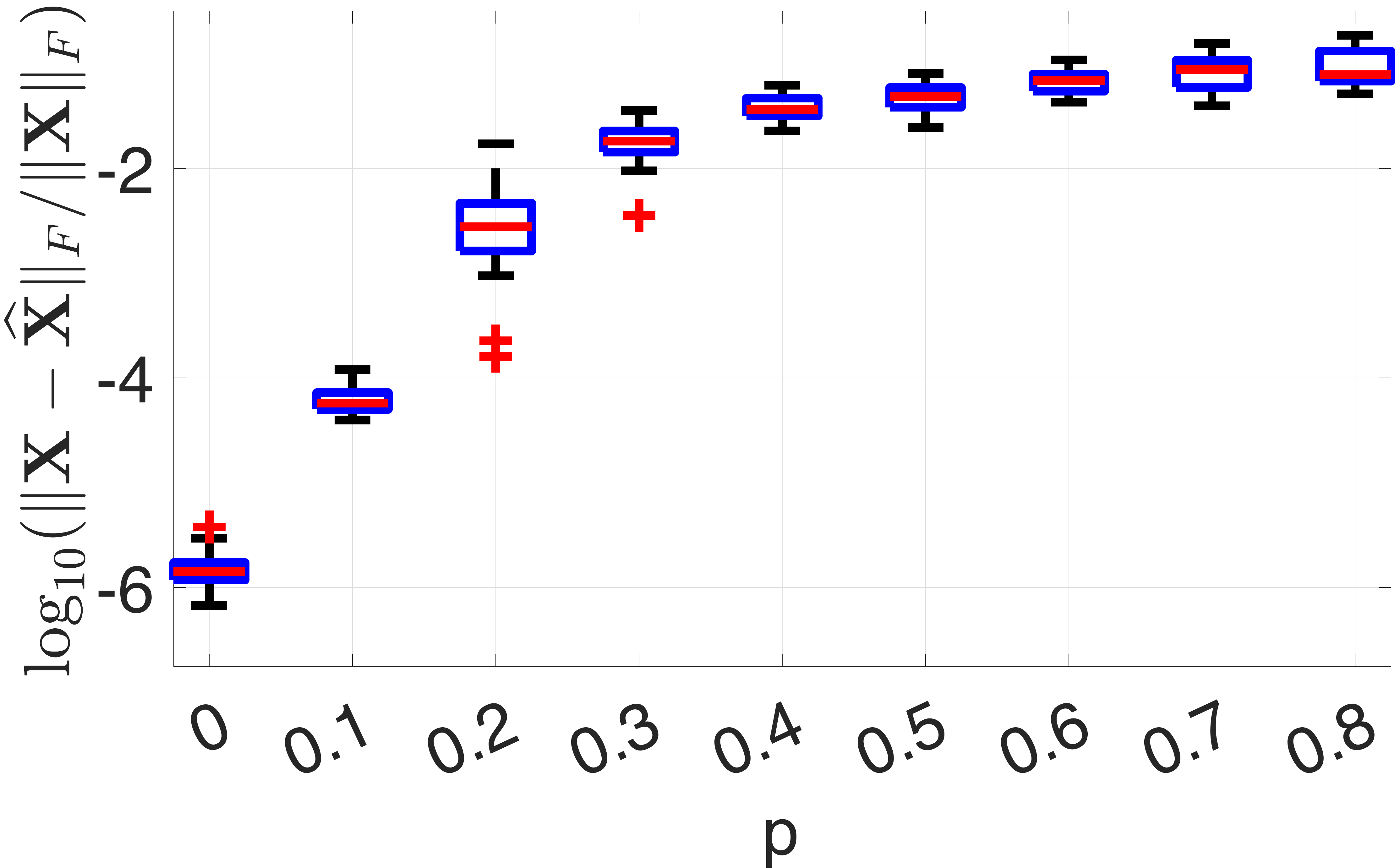}\label{fig:psdmf_SparseX_10x10_Xrand_MC20_K5_ra3_rb1_TolFun_1e13_TolFval_0_MaxCPUTime_600_MaxIter_1e6_CDgs_LogErr}}
\\
\subfloat[SVP]{\includegraphics[width=0.48\columnwidth]{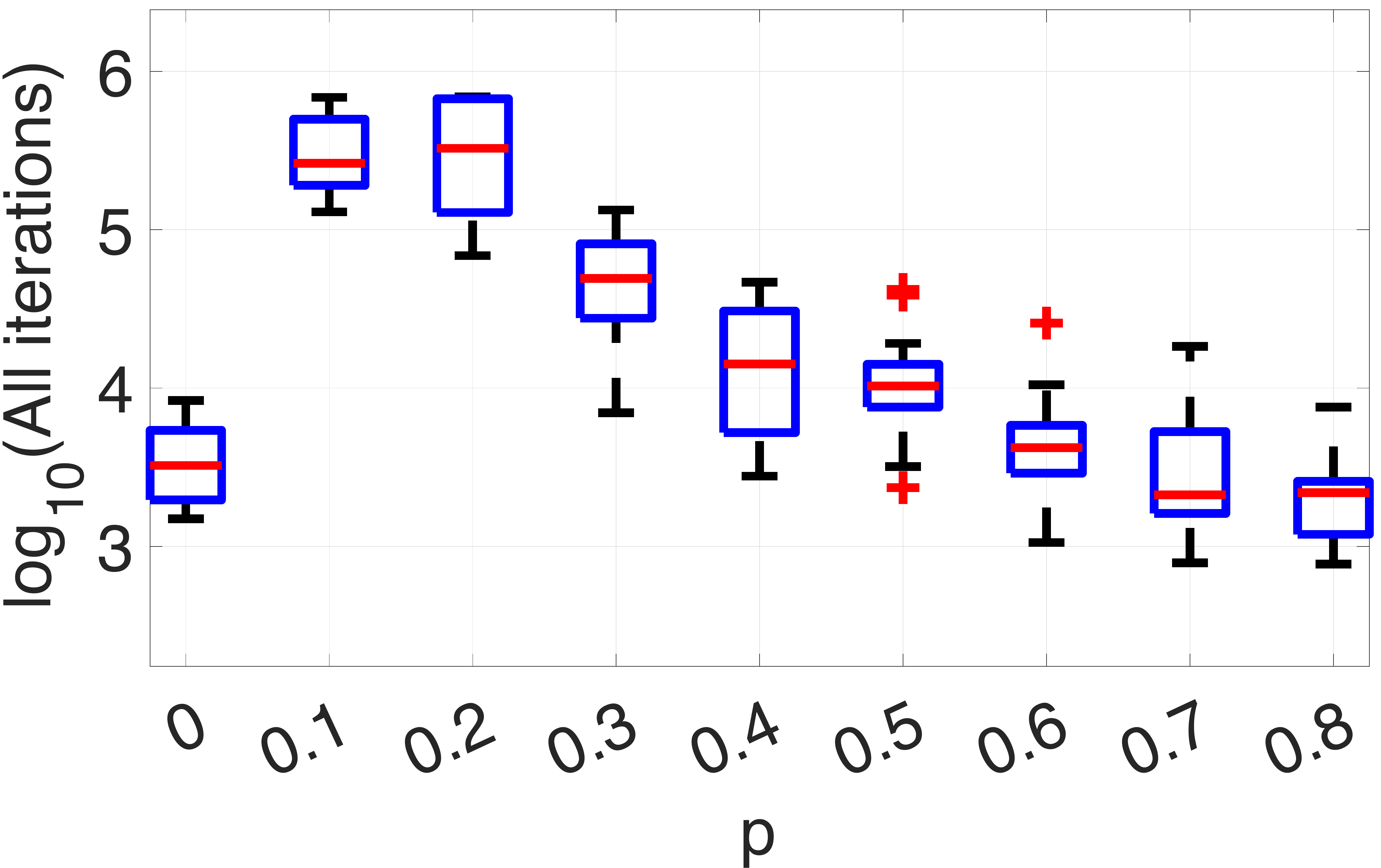}\label{fig:psdmf_SparseX_10x10_Xrand_MC20_K5_ra3_rb1_TolFun_1e13_TolFval_0_MaxCPUTime_600_MaxIter_1e6_SVP_LogAllIter}}
\hfil
\subfloat[FSVP]{\includegraphics[width=0.48\columnwidth]{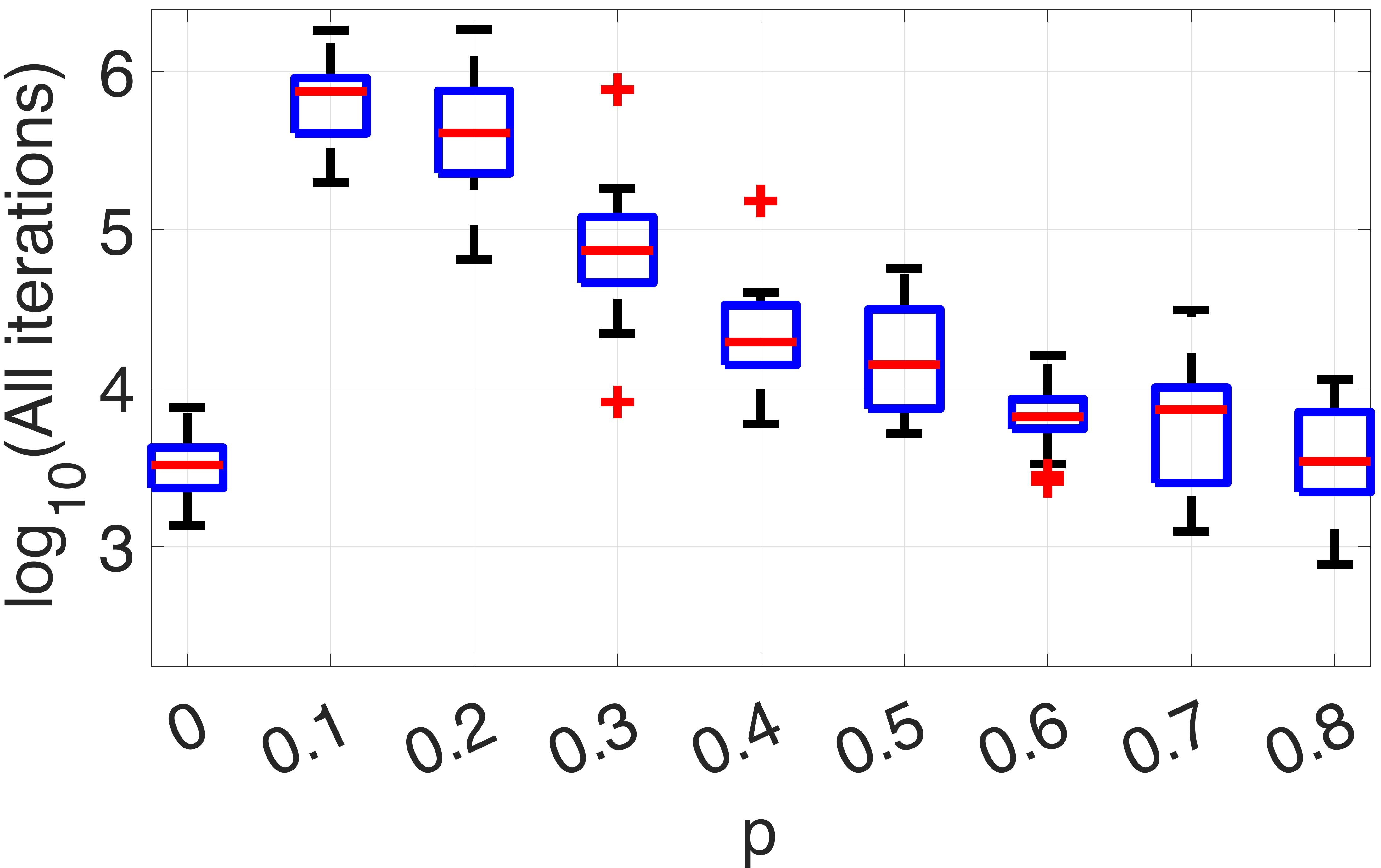}\label{fig:psdmf_SparseX_10x10_Xrand_MC20_K5_ra3_rb1_TolFun_1e13_TolFval_0_MaxCPUTime_600_MaxIter_1e6_FSVP_LogAllIter}}
\hfil
\subfloat[NIHT]{\includegraphics[width=0.48\columnwidth]{\gpath psdmf_SparseX_10x10_Xrand_MC20_K5_ra3_rb1_TolFun_1e13_TolFval_0_MaxCPUTime_600_MaxIter_1e6_NIHT_LogAllIter}\label{fig:psdmf_SparseX_10x10_Xrand_MC20_K5_ra3_rb1_TolFun_1e13_TolFval_0_MaxCPUTime_600_MaxIter_1e6_NIHT_LogAllIter}}
\hfil
\subfloat[ABG]{\includegraphics[width=0.48\columnwidth]{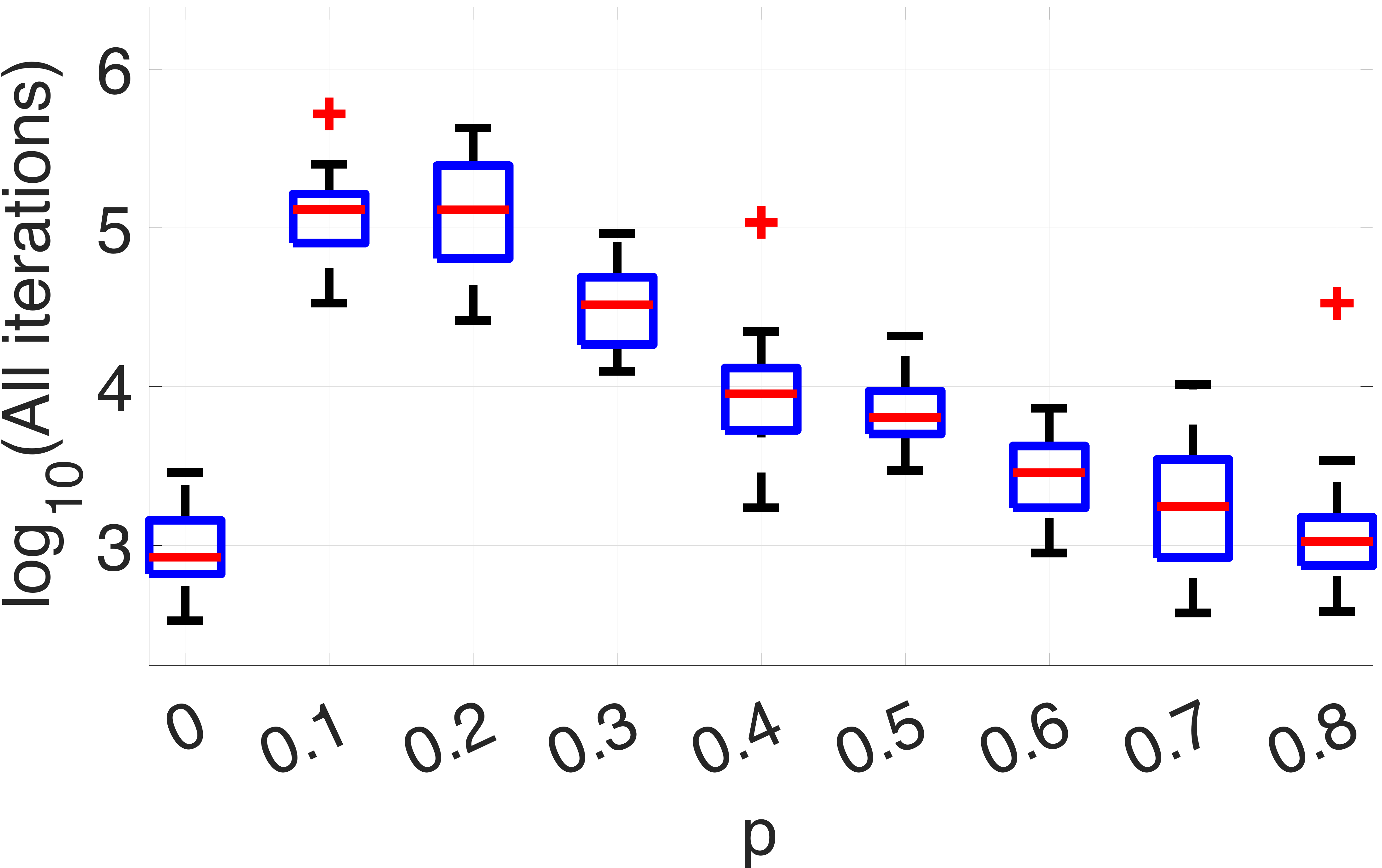}\label{fig:psdmf_SparseX_10x10_Xrand_MC20_K5_ra3_rb1_TolFun_1e13_TolFval_0_MaxCPUTime_600_MaxIter_1e6_ABG_LogAllIter}}
\hfil
\subfloat[ABG-P]{\includegraphics[width=0.48\columnwidth]{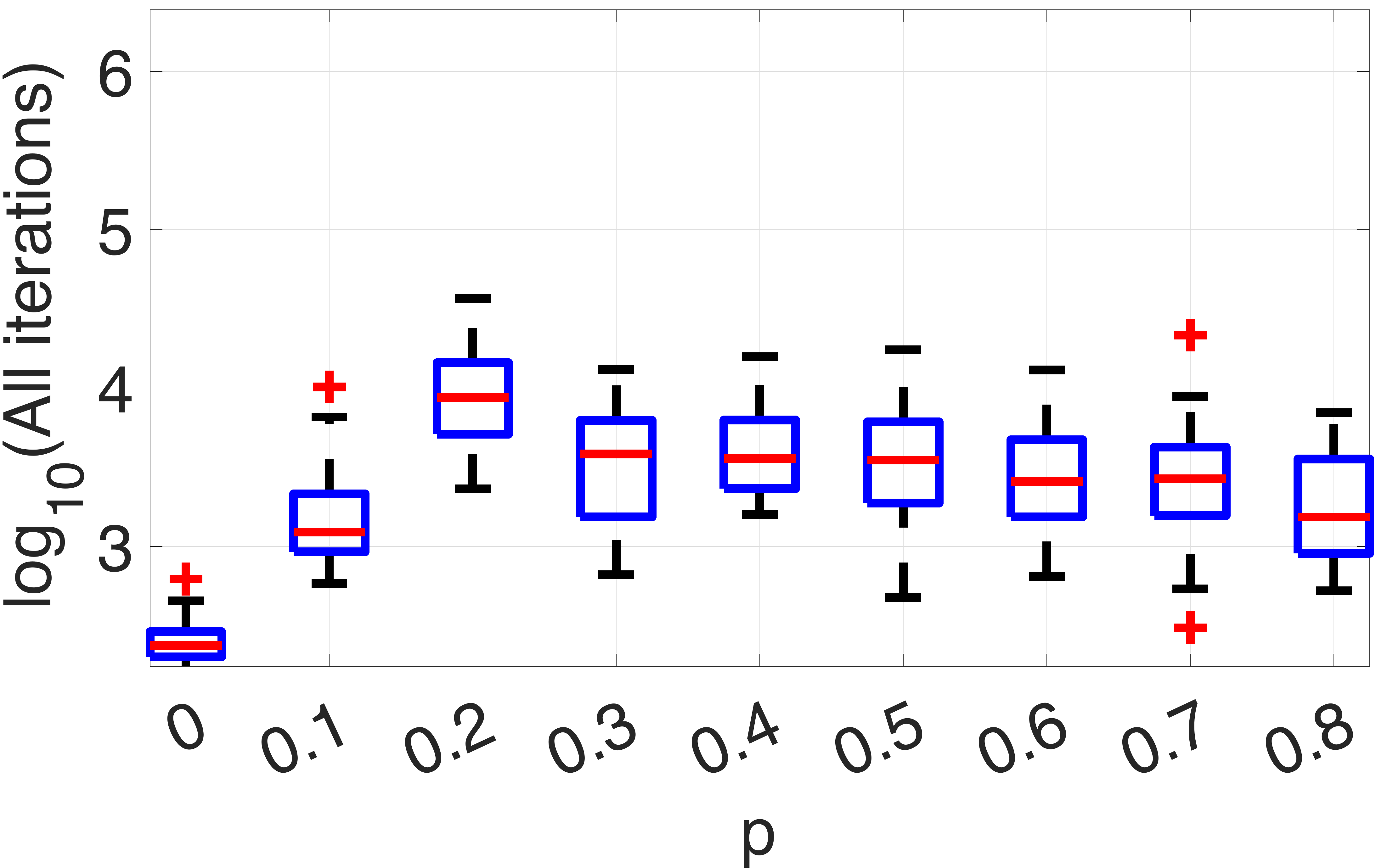}\label{fig:psdmf_SparseX_10x10_Xrand_MC20_K5_ra3_rb1_TolFun_1e13_TolFval_0_MaxCPUTime_600_MaxIter_1e6_ABGP_LogAllIter}}
\hfil
\subfloat[CGIHT]{\includegraphics[width=0.48\columnwidth]{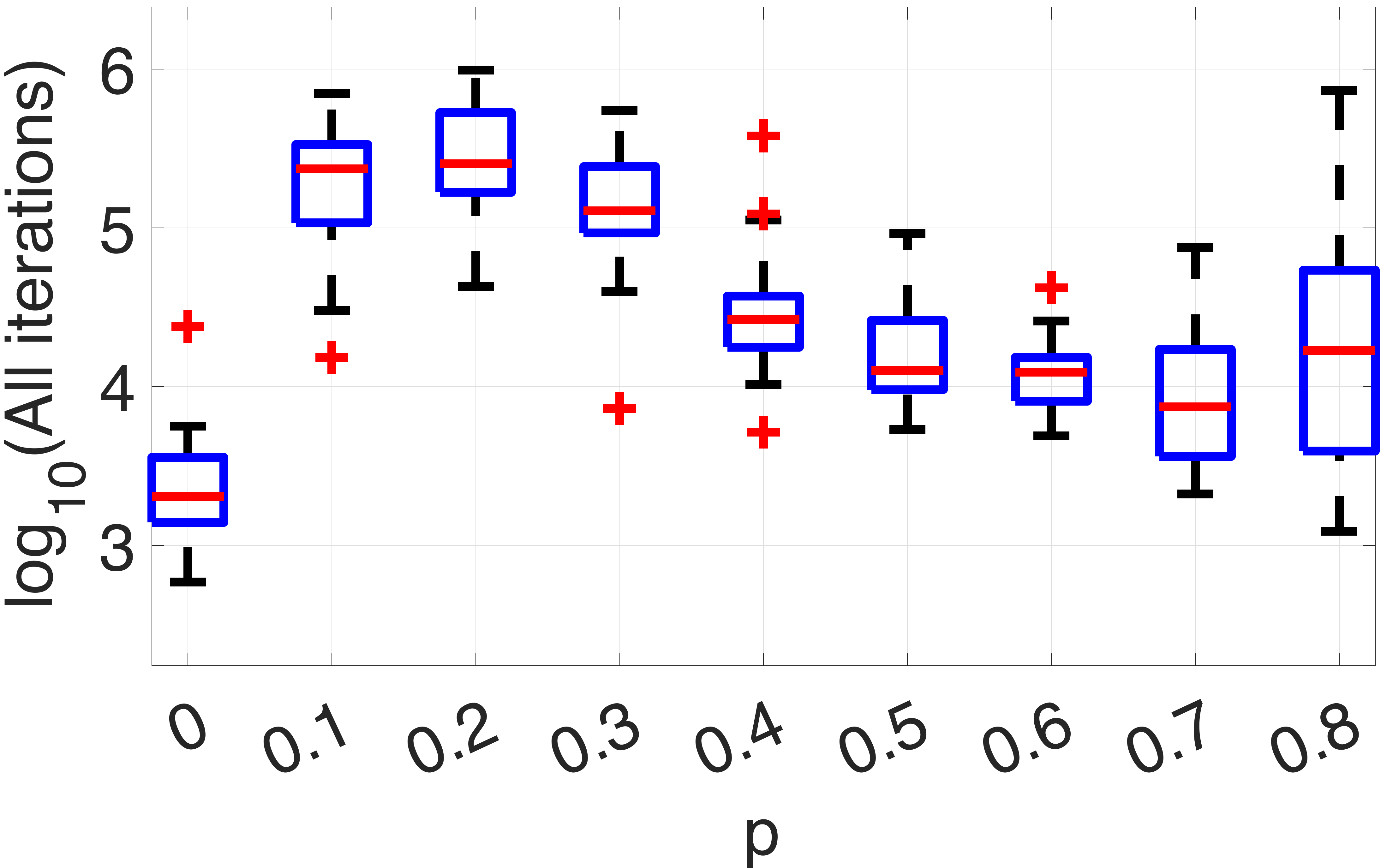}\label{fig:psdmf_SparseX_10x10_Xrand_MC20_K5_ra3_rb1_TolFun_1e13_TolFval_0_MaxCPUTime_600_MaxIter_1e6_CGIHT_LogAllIter}}
\hfil
\subfloat[CD cyc]{\includegraphics[width=0.48\columnwidth]{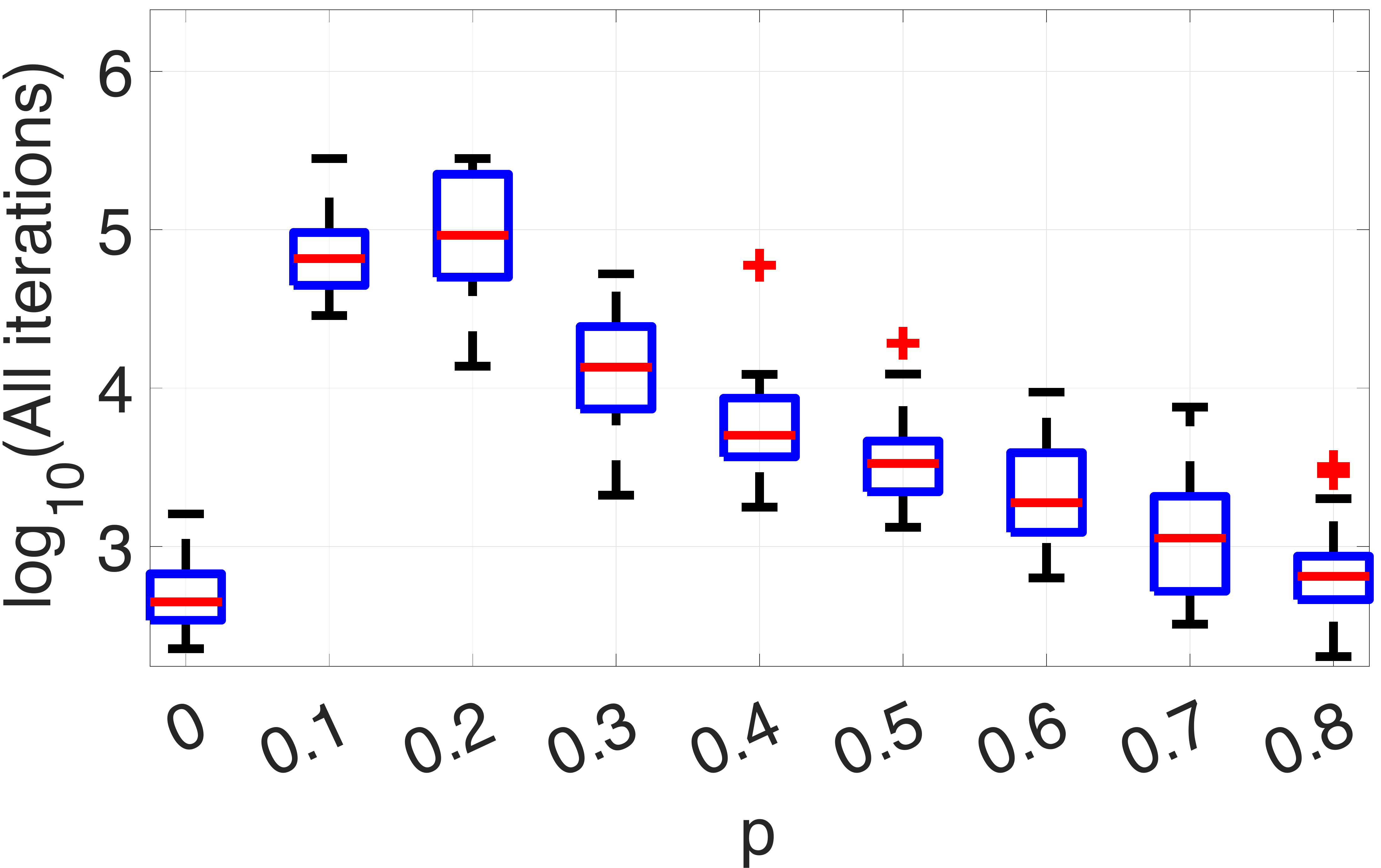}\label{fig:psdmf_SparseX_10x10_Xrand_MC20_K5_ra3_rb1_TolFun_1e13_TolFval_0_MaxCPUTime_600_MaxIter_1e6_CDcyc_LogAllIter}}
\hfil
\subfloat[CD GS]{\includegraphics[width=0.48\columnwidth]{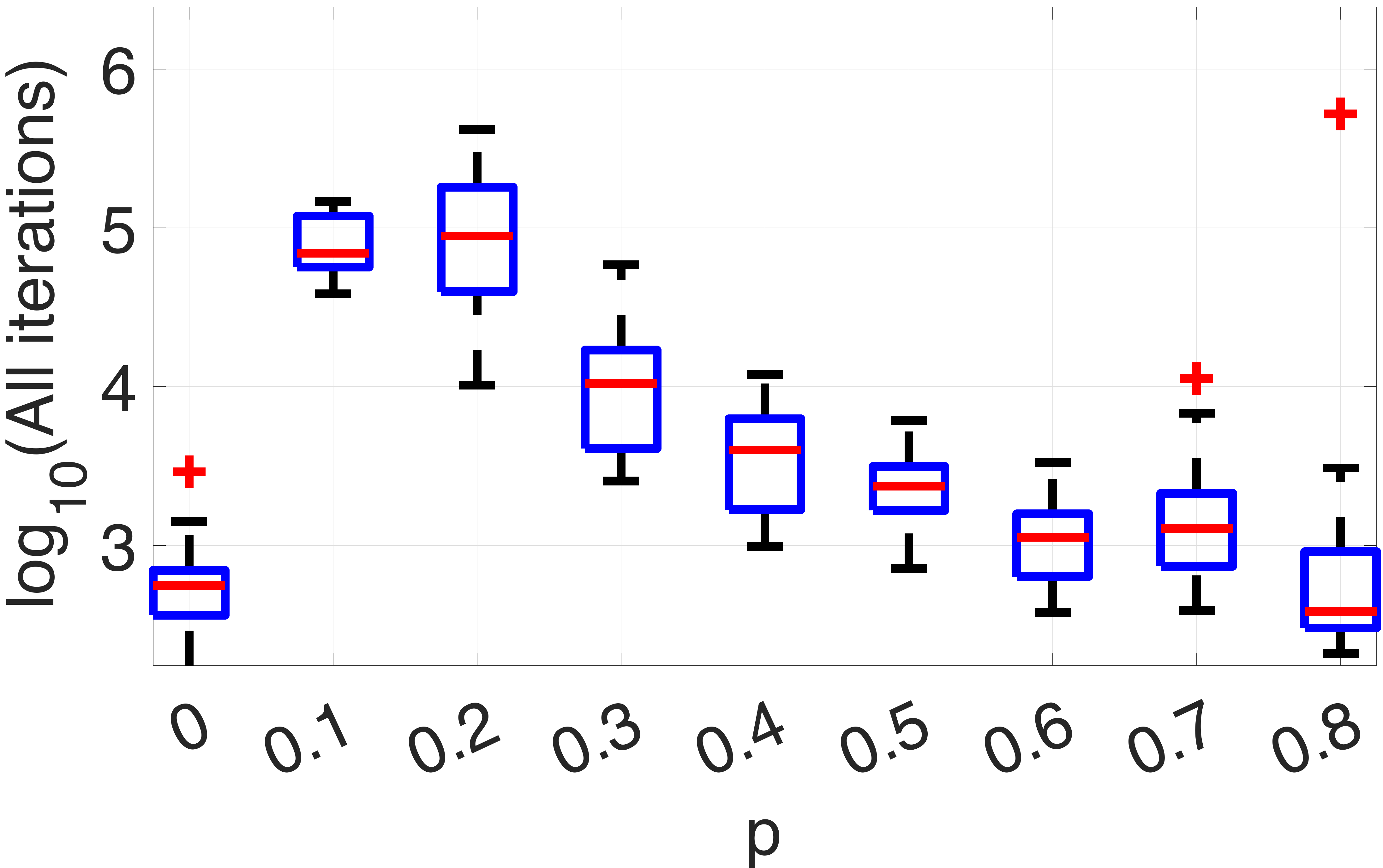}\label{fig:psdmf_SparseX_10x10_Xrand_MC20_K5_ra3_rb1_TolFun_1e13_TolFval_0_MaxCPUTime_600_MaxIter_1e6_CDgs_LogAllIter}}
\caption{Influence of the proportion of zeros $p$ in a $10\times 10$ random matrix on the model fit error (\cref{fig:psdmf_SparseX_10x10_Xrand_MC20_K5_ra3_rb1_TolFun_1e13_TolFval_0_MaxCPUTime_600_MaxIter_1e6_SVP_LogErr,fig:psdmf_SparseX_10x10_Xrand_MC20_K5_ra3_rb1_TolFun_1e13_TolFval_0_MaxCPUTime_600_MaxIter_1e6_FSVP_LogErr,fig:psdmf_SparseX_10x10_Xrand_MC20_K5_ra3_rb1_TolFun_1e13_TolFval_0_MaxCPUTime_600_MaxIter_1e6_NIHT_LogErr,fig:psdmf_SparseX_10x10_Xrand_MC20_K5_ra3_rb1_TolFun_1e13_TolFval_0_MaxCPUTime_600_MaxIter_1e6_ABG_LogErr,fig:psdmf_SparseX_10x10_Xrand_MC20_K5_ra3_rb1_TolFun_1e13_TolFval_0_MaxCPUTime_600_MaxIter_1e6_ABGP_LogErr,fig:psdmf_SparseX_10x10_Xrand_MC20_K5_ra3_rb1_TolFun_1e13_TolFval_0_MaxCPUTime_600_MaxIter_1e6_CGIHT_LogErr,fig:psdmf_SparseX_10x10_Xrand_MC20_K5_ra3_rb1_TolFun_1e13_TolFval_0_MaxCPUTime_600_MaxIter_1e6_CDcyc_LogErr,fig:psdmf_SparseX_10x10_Xrand_MC20_K5_ra3_rb1_TolFun_1e13_TolFval_0_MaxCPUTime_600_MaxIter_1e6_CDgs_LogErr}) and on the overall number of iterations (\cref{fig:psdmf_SparseX_10x10_Xrand_MC20_K5_ra3_rb1_TolFun_1e13_TolFval_0_MaxCPUTime_600_MaxIter_1e6_SVP_LogAllIter,fig:psdmf_SparseX_10x10_Xrand_MC20_K5_ra3_rb1_TolFun_1e13_TolFval_0_MaxCPUTime_600_MaxIter_1e6_FSVP_LogAllIter,fig:psdmf_SparseX_10x10_Xrand_MC20_K5_ra3_rb1_TolFun_1e13_TolFval_0_MaxCPUTime_600_MaxIter_1e6_NIHT_LogAllIter,fig:psdmf_SparseX_10x10_Xrand_MC20_K5_ra3_rb1_TolFun_1e13_TolFval_0_MaxCPUTime_600_MaxIter_1e6_ABG_LogAllIter,fig:psdmf_SparseX_10x10_Xrand_MC20_K5_ra3_rb1_TolFun_1e13_TolFval_0_MaxCPUTime_600_MaxIter_1e6_ABGP_LogAllIter,fig:psdmf_SparseX_10x10_Xrand_MC20_K5_ra3_rb1_TolFun_1e13_TolFval_0_MaxCPUTime_600_MaxIter_1e6_CGIHT_LogAllIter,fig:psdmf_SparseX_10x10_Xrand_MC20_K5_ra3_rb1_TolFun_1e13_TolFval_0_MaxCPUTime_600_MaxIter_1e6_CDcyc_LogAllIter,fig:psdmf_SparseX_10x10_Xrand_MC20_K5_ra3_rb1_TolFun_1e13_TolFval_0_MaxCPUTime_600_MaxIter_1e6_CDgs_LogAllIter}). Factorized with $K=5$, $R_A=3$, $R_B=1$. Stopping criterion: $\TolFun {}={}10^{-13}$. $D_{\textrm{FSVP}}{}={}14$, $D_{\textrm{CGIHT}}{}={}14$.
All plots have the same Y-axis for number of iterations, and similarly for the final \ac{RMFE}. 
We observe the same trend for all methods: as soon as $p>0$, the \ac{RMFE} cannot be reduced below a certain value. 
However, for $p=0$, further decreasing the stopping criterion \texttt{TolFun} will yield smaller \ac{RMFE} by several orders of magnitude.
 This matrix size and values of inner and outer ranks were chosen because they are the same as for $S_{10}$, a matrix that we also factorize in~\cref{sec:numerical_experiments}.} 
 \label{fig:psdmf_SparseX_10x10_Xrand_MC20_K5_ra3_rb1}
\end{figure*}

\Cref{fig:psdmf_SparseX_10x10_Xrand_MC20_K5_ra3_rb1} summarizes the plots generated using the same model and optimization parameters as in~\cref{fig:psdmf_10x10_MC20_K5_ra3_rb1_Xrand_SparseX} for all methods concerned in our numerical experiments.

\end{document}